\documentclass[12pt,a4paper]{article}
\pdfoutput=1
\usepackage{graphicx}

\usepackage{color}
\usepackage{subfig}
\usepackage{amsmath}
\usepackage{amsfonts}
\usepackage{amssymb}
\usepackage{caption}
\usepackage[hidelinks]{hyperref}
\usepackage{mathrsfs}
\usepackage{enumerate}
\usepackage{siunitx}

\usepackage{pgfplots}
\pgfplotsset{compat=newest}

\newcommand{\comment}[1]{}

\def\bea{\begin{eqnarray}}
\def\eea{\end{eqnarray}}
\def\be{\begin{equation}}
\def\ee{\end{equation}}

\newcommand{\mal}[1]{\mathcal #1}
\newcommand{\expect}[1]{\left\langle #1 \right\rangle}

\def\H{\mal{H}}

\def\d{\partial}
\def\ep{\epsilon}

\def\mpl{\ensuremath{M_{\text{Pl}}}}

\def\[{\left[}
\def\]{\right]}
\def\({\left(}
\def\){\right)}

\def\s{\sigma}

\newcommand{\half}{{\textstyle{\frac12}}}

\begin{document}

\begin{center}
\Large{\textbf{Light Particles with Spin in Inflation}} \\[0.5cm]
\end{center}
\vspace{0.1cm}

\begin{center}

\large{Lorenzo Bordin$^{\rm a,b}$, Paolo Creminelli$^{\rm c}$, Andrei Khmelnitsky$^{\rm c}$, \\  and  Leonardo Senatore$^{\rm d}$}
\\[0.5cm]

\small{
\textit{$^{\rm a}$ SISSA, via Bonomea 265, 34136, Trieste, Italy}}\\
\small{
\textit{$^{\rm b}$ INFN, National Institute for Nuclear Physics, Via Valerio 2, 34127 Trieste, Italy}}\\
\small{
\textit{$^{\rm c}$ Abdus Salam International Centre for Theoretical Physics\\ Strada Costiera 11, 34151, Trieste, Italy}}\\
\small{
\textit{$^{\rm d}$ SITP and KIPAC, Department of Physics and SLAC, Stanford University, Stanford, CA 94305}}
\end{center}

\vspace{.8cm}

\hrule \vspace{0.3cm}
\noindent \small{\textbf{Abstract}\\ 
The existence of light particles with spin during inflation is prohibited by the Higuchi bound. This conclusion can be evaded if one considers states with a sizeable coupling with the inflaton foliation, since this breaks the de Sitter isometries. The action for these states can be constructed within the Effective Field Theory of Inflation, or using a CCWZ procedure. Light particles with spin have prescribed couplings with soft inflaton perturbations, which are encoded in consistency relations. We study the phenomenology of light states with spin 2. These mix with the graviton changing the tensor power spectrum and can lead to sizeable tensor non-Gaussianities. They also give rise to a scalar bispectrum and trispectrum with a characteristic angle-dependent non-Gaussianity.
\vspace{0.3cm}
\noindent
\hrule
\def\thefootnote{\arabic{footnote}}
\setcounter{footnote}{0}

\section{Introduction}
The minimal set of light degrees of freedom during inflation is given by a scalar mode, the Goldstone of broken time-translations, and the graviton. The addition of extra light scalars has been studied for many years. They modify the predictions for the scalar spectrum and their presence gives unmistakeable signatures: isocurvature perturbations and non-Gaussianity of the local form~\cite{Maldacena:2002vr,Creminelli:2004yq}. The study of particles with spin is much more recent~\cite{1503.08043,Lee:2016vti,Delacretaz:2016nhw}. One of the reasons is that the mass $m$ of a particle with spin $s$ must satisfy $m^2 > s(s-1) H^2$, where $H$ is the Hubble rate during inflation. This inequality, called Higuchi bound~\cite{Higuchi:1986py}, is a consequence of the (approximate) de Sitter symmetries during inflation. It implies that particles with spin decay outside the horizon faster than $1/a$, where $a$ is the scale factor. This implies that they leave a small effect in the squeezed limit of correlation functions. The effect of even more massive particles, $m^2 > \left(s - \frac 1 2 \right)^2 H^2$, is exponentially suppressed but very peculiar, with an oscillatory pattern in the squeezed limit~\cite{1503.08043}. 

    The isometries of de Sitter are broken by the preferred foliation of the constant inflaton surfaces. This implies that states with a ``sizeable coupling" with the inflaton could violate the Higuchi bound. In this paper we show that this is indeed possible and study the physics of these light states with spin. The construction of the action for these states follows rather straightforwardly from the rules of the Effective Field Theory of Inflation (EFTI)~\cite{Cheung:2007st}, as we will discuss in Section~\ref{sec:EFTI}. The Higuchi bound implies that these states only exist in the presence of the preferred foliation induced by the time-dependent inflaton background, similarly to excitations of a fluid more than elementary particles. (In this sense what we propose is cosmological condensed matter rather than cosmological collider physics \cite{1503.08043}.)
 In the language on non-linearly realised symmetries, they are matter fields coupled to the Goldstones. As such their action can also be constructed following the usual Coleman-Callan-Wess-Zumino (CCWZ) rules for non-linearly realised spacetime symmetries, as we will discuss in Subection~\ref{sec:CCWZ}. These light states have prescribed couplings with the inflaton. Indeed the ``boost" isometries of de Sitter are spontaneously broken by the foliation and thus non-linearly realised. This gives rise to consistency relations: the 2-point function of light particles with spin is not de Sitter invariant (since it violates the Higuchi bound) and this variation fixes the coupling with the Goldstone of time-translations, i.e.~the inflaton fluctuations, in the squeezed limit. This relation will be explicitly verified in the simplest example of these theories: spin 1 (Section~\ref{sec:spin1}).
    
To study the phenomenological implications of these states, we are going to focus on the most interesting example: the one of helicity-2 states. The reason is two-fold. First of all, a simple parity argument suppresses the contribution of vectors in the squeezed limit which, as we discussed, is a most prominent signatures of light states. This makes the simplest case, the one of spin-1, not so interesting. Particles of spin-2 are unsuppressed in the squeezed limit and moreover they can mix with the graviton and modify the predictions related to tensor modes. These signatures are studied in Section~\ref{sec:pheno}, leaving some details of the calculations to the Appendices. A light particle of spin-2 modifies both the scalar and the tensor power spectrum. Depending of the parameters one of the spectra (or both) can be dominated by the exchange of the extra state.  We study the effect of the light spin-2 state on $\expect{\zeta\zeta\zeta}$ and $\expect{\zeta\zeta\zeta\zeta}$, where an angle dependent non-Gaussianity is induced, and on $\expect{\gamma\zeta\zeta}$ where the limit of soft graviton momentum shows a violation of the tensor consistency relation~\cite{Bordin:2016ruc,Dimastrogiovanni:2015pla}. 

The effect of these light states is enhanced when they have a small speed of propagation. In Section~\ref{sec: radiative constraints} we study the experimental and theoretical constraints on this speed of propagation. Conclusions are drawn in Section~\ref{sec:conclusions}.

Before starting, let us comment on the relation with other works in the literature. Light spin states may appear below the Higuchi bound in the form of partially massless states~\cite{Deser:2003gw}, whose possible phenomenology in inflation was studied in~\cite{Baumann:2017jvh, Franciolini:2017ktv}. In this paper we insisted on keeping the approximate shift-symmetry of the inflaton, which is behind the observed approximate scale-invariance of the scalar power-spectrum. A strong breaking of this symmetry can also efficiently violate the Higuchi bound as discussed in~\cite{Kehagias:2017cym}. Another way to get light states with spin is to start with a symmetry pattern of inflation that is different from the standard one: two examples are gauge-flation~\cite{Maleknejad:2011jw,Adshead:2013nka,Agrawal:2017awz} and gaugid inflation~\cite{Piazza:2017bsd}.

\section{\label{sec:EFTI}Particles with spin in the EFTI}

We want to understand how to describe matter fields, i.e.~fields in addition to the clock of inflation $\pi$, in the framework of the EFTI. For scalars, the procedure is straightforward: one just writes an action for an extra scalar $\sigma$ with the usual rules of the EFTI~\cite{Senatore:2010wk}. (For instance in unitary gauge a term of the form $(g^{00}+1) (g^{0\mu} \partial_\mu\sigma)$ describes the mixing between the inflaton and $\sigma$.) Things are somewhat different for particles with spin. Let us concentrate for concreteness on particles with spin-1. One might think to start with a four-vector $\Sigma^\mu$ and write an action with the usual rules of the EFTI; however this is not the most correct procedure. If we concentrate on scales much shorter than Hubble and forget about gravity, the inflaton background spontaneously breaks the Lorentz group to rotations.\footnote{Notice that the local breaking of the Lorentz symmetry will also break the de Sitter isometries and open the possibility to evade the Higuchi bound; however the construction works in a generic gravitational background.} The usual logic of non-linearly realised symmetries tells us that one should classify fields as representations of the unbroken group, in this case rotations.  
Therefore we should start from a 3-vector, not a 4-vector.\footnote{The breaking of the Lorentz symmetry induces a separation among the different helicities of a particle with spin. However, one cannot completely disentangle one helicity from the others since the selection of one helicity is an intrinsically non-local operation.} It is irrelevant to know to which Lorentz representation this 3-vector belongs: to build a Lagrangian which non-linearly realises the broken group one just needs to know the transformation properties under the unbroken group. Actually the question about the Lorentz representation is ill-defined: one in general will not be able to recombine the fields to form Lorentz multiplets in the same sense as in the chiral Lagrangian one cannot combine states under representations of the axial group. 
 
Let us see what is the procedure to build an action in terms of a 3-vector.
\begin{itemize}
\item In a generic gauge, the slices of constant inflaton are $\psi \equiv t+\pi(x^i,t) = c$. Given the preferred foliation there is a natural way to split the tangent space at each point, and thus tensors, in the projections parallel and orthogonal to the surface. Fields are classified as three-tensors.
\item It is also natural to parametrise the 3-surfaces of constant inflaton with the spatial coordinates $x^i$ that we use for the whole spacetime.  This gives a basis of the tangent space on the submanifold: $\partial/\partial x^i$. Objects living on the slice can be written in terms of this basis. A vector will have only three components: $\Sigma^i \cdot \partial/\partial x^i$. 
\item So far the fields live on the three-dimensional slices embedded in spacetime, but in order to describe their couplings to the four-dimensional fields including gravity one needs to ``push forward" them to objects living in four dimensions. Of course the mapping will depend on the particular configuration of the inflaton slices described by $\pi$
\be\label{3Dto4D}
\Sigma^\mu(\Sigma^i,\pi) = \left. \Sigma^i \, \frac{\partial x^\mu}{\partial x^i}\right|_{\psi}  \;.
\ee
This is a four-vector tangent to the surfaces of constant $\psi$ and it is specified by its three independent components $\Sigma^i$.
\item Given that
\be
\left.\frac{\partial t}{\partial x^i}\right|_\psi = - \frac{\partial\psi}{\partial x^i}\left/ \frac{\partial\psi}{\partial t} \right.= - \frac{\partial_i \pi}{1 + \dot \pi}
\ee
one can write an explicit expression for the components of the corresponding four-vector
\be\label{eq:Ai}
\Sigma^\mu(\Sigma^i,\pi) = \left(- \frac{\partial_i \pi \, \Sigma^i}{1 + \dot \pi} \; , \quad \Sigma^i \right) \;.
\ee
This is the object we will use to build the action: one can explicitly check that it transforms as a vector under all diffs.
\end{itemize}
This method can be applied to build an action for a field transforming in any tensor representation of three dimensional rotations, and therefore representing particles of arbitrary spin. In particular, a spin-2 particle would be described by a traceless rank-2 symmetric tensor $\Sigma^{ij}$ with five independent components. The diffeomorphism-invariant action for it can be written in terms of the traceless symmetric four-tensor $\Sigma^{\mu\nu}$ with the four extra components given in terms of $\pi$ and $\Sigma^{ij}$:
\begin{align}
\Sigma^{00} &= \frac{ \d_i \pi \d_j \pi}{(1 + \dot \pi)^2} \, \Sigma^{ij} \; , & \Sigma^{0j} &= - \frac{\d_i \pi }{1 + \dot \pi} \, \Sigma^{ij} \; .
\end{align}

The above construction can also be understood if one starts in unitary gauge, i.e. with $\pi =0$. In this gauge, since the slices of constant inflaton coincide with the ones of constant time, a vector on the surface has only spatial components. It transforms as a vector under time-dependent spatial diffs, so one can use it in this gauge provided spatial indexes are contracted.  If one goes out from the unitary gauge, doing a time diff.~, one has to perform the usual St\"uckelberg procedure introducing $\pi$. This gives the expression \eqref{3Dto4D} albeit from a somewhat different perspective.

\subsection{\label{sec:CCWZ}Matter fields in the CCWZ approach}

Since the presence of the foliation introduces a natural split of the tangent space, it breaks the local Lorentz invariance to the rotation subgroup. This suggests that one can employ the CCWZ approach~\cite{Coleman:1969sm, Callan:1969sn} for writing an action for the Goldstone and matter fields, which would non-linearly realises the broken symmetries. A formulation of single-clock inflation in the presence of an approximate shift-symmetry of the inflaton was presented in~\cite{Delacretaz:2014oxa} directly in the CCWZ language, including the coupling with gravity. The system has the same symmetries of a superfluid. The full symmetry group is taken to consist of the internal shift symmetry generated by $Q$ and local $ISO(1,3)$ Poincare group on the tangent space generated by translation $P_a$ and Lorentz transformation $J_{ab}$ operators.\footnote{Here we use the latin indices $a,b,\dots$ and $m,n,\dots$ for the four- and three-dimensional tangent spaces respectively in order to distinguish them from the coordinate indices $\mu,\nu,\dots$ and $i,j,\dots$.} The superfluid phase corresponds to a finite density of the global charge $Q$ with chemical potential $\mu$. The state breaks local boosts, time translations, and the global shift symmetry, but preserves a combination of time translations and shifts generated by $\bar P_0 = P_0 + \mu \, Q$. Moreover, in order to recover gravity in this approach one assumes that all the local Poincare shifts are non-linearly realised. The only linearly realised symmetries are local rotations generated by $J_{mn}$. The corresponding coset element is parameterised by eight Goldstone fields: $y^a$ for local translations, $\pi$ for the internal shift, and $\eta^m$ for boosts, and can be chosen to be
\begin{equation}\label{eq:coset}
\Omega = e^{i \, y^a \, \bar P_a} \, e^{i \, \pi \, Q} \, e^{i\, \eta^m J_{0m}} \; .
\end{equation}
All the building blocks that are allowed to be used in the action can be read off from the Maurer-Cartan form
\begin{equation}
\Omega^{-1} D_\mu \Omega \equiv  i \, \nabla_\mu y^a \, \left( \bar P_a + \nabla_a \pi \, Q + \nabla_a \eta^m \, J_{0m} + \frac12 {\mathcal{A}^{mn}}_a \, J_{mn}\right) \; .
\end{equation}
The coefficients in front of the broken generators correspond to the CCWZ covariant derivatives of the Goldstone fields and can be used in the action in the combinations that preserve unbroken local rotations. The coefficients of the unbroken generators give a connection that defines a covariant derivative needed to construct higher derivative terms and to act on non-Goldstone matter fields. All these objects are calculated in reference~\cite{Delacretaz:2014oxa} where it was also shown that they are the same as the building blocks of the EFTI action. In what follows we briefly review these building blocks and show that adding matter fields in the CCWZ language is equivalent to the EFTI construction introduced in the previous Section.

The covariant derivative $\nabla_\mu y^a$ of the translation Goldstones gives the ``coset vierbein"
\begin{equation}
\nabla_\mu y^a \equiv {E^a}_\mu = {\Lambda_b}^a \, {e^b}_\mu \; ,
\end{equation}
where we have introduced the boost matrix in the vector representation
\begin{equation}
{\Lambda^a}_b(\eta) \equiv {(e^{i \, \eta^m J_{0m}})^a}_b \; ,
\end{equation}
and ${e^b}_\mu$ is the space-time vierbein, which defines the metric as ${e^a}_\mu {e^b}_\nu \, \eta_{ab} = g_{\mu \nu}$. The coset vierbein transforms covariantly under the unbroken $SO(3)$ rotations: ${E^0}_\mu$ is a singlet and ${E^m}_\mu$ is a triplet. It can be used to construct an invariant integration measure $d^4x \, \det E = d^4x \, \det e$.

The covariant derivatives of the Goldstones $\pi$ and $\eta^m$ read
\begin{align}\label{eq:DpiDeta}
\nabla_a \pi &\equiv {e_b}^\mu \, {\Lambda^b}_a \, \d_\mu \psi - \mu \, \delta^0_a \; ,
& \nabla_a \eta^m &\equiv {e_b}^\mu \, {\Lambda^b}_a \, \left( {\Lambda_c}^{0} \d_\mu \Lambda^{cm} + {\omega^{c}}_{d\, \mu} {\Lambda_c}^0 {\Lambda}^{d m}\right) \; ,
\end{align}
where we have introduced the field $\psi \equiv \pi + \mu \, y^0$, and ${\omega^{c}}_{d \, \mu}$ is the spin-connection that corresponds to the vierbein ${e^b}_\mu$. These covariant derivatives can also be used to impose additional constraints on the effective theory. If these constraints allow one to reduce the number of the Goldstone fields by expressing some of the Goldstone fields in terms of the others in a local manner then they are called inverse Higgs constraints. In the case at hand we can impose the condition $\nabla_m \pi = 0$, which allows to express $\eta^m$ in terms of the derivatives of $\pi$:
\begin{equation}
\nabla_m \pi = {\Lambda^a}_m \, {e_a}^\mu \d_\mu \psi = 0 \; .
\end{equation}
It means that the transformation ${\Lambda^a}_b$ corresponds to a boost with velocity 
\begin{equation}\label{eq:inverse higgs}
\beta_m \equiv \frac {\eta_m} \eta \, \tanh \eta = - \frac{{e_m}^\mu \d_\mu \psi}{{e_0}^\mu \d_\mu \psi} \,,
\end{equation}
where $\eta \equiv \sqrt{\eta_m \eta_n \delta_{mn}}$.
Therefore, the first row of the boost matrix ${\Lambda^a}_b$ gives the unit normal to the constant $\psi$ slices in the orthonormal basis and the other three rows are three orthonormal vectors lying in the tangent space to a slice:
\begin{align}
{\Lambda^a}_0 &= n^a \equiv - \frac{{e^a}_\mu \d^\mu \psi}{\sqrt{-\d_\nu\psi \d^\nu \psi}} \; , & n_a {\Lambda^a}_m &= 0 \; , &\eta_{ab}\,{\Lambda^a}_m {\Lambda^b}_n &= \delta_{mn} \; .
\end{align}
Four vectors ${\Lambda^a}_m$ give an embedding of the space tangent to the slice in the tangent space of the space-time. In particular, any vector tangent to a constant $\psi$ slice can be written in terms of three components $\Sigma^m$ in this orthonormal basis as
\begin{equation}\label{eq:4dAm}
\Sigma^\mu(\Sigma^m, \pi) = {e_a}^\mu \, {\Lambda^a}_m \, \Sigma^m = \Sigma^m \, {E_m}^\mu \; ,
\end{equation} 
and the normal to a slice is given by the remaining forth tetrad $n^\mu = {E_0}^\mu$. Comparing this embedding to equation~\eqref{3Dto4D} we can write an explicit transformation between the three components of a tangent vector in the coordinate and orthonormal bases respectively:
\begin{equation}\label{eq:AmtoAi}
\Sigma^m = {E^m}_\mu \, \frac{\partial x^\mu}{\partial x^i}\Big|_{\psi} \, \Sigma^i \; .
\end{equation}
The transformation matrix is nothing else but a dreibein for the induced metric on constant $\psi$ slices.

After fixing the boost Goldstones $\eta$, the remaining covariant derivatives~\eqref{eq:DpiDeta} give the familiar objects of the EFTI:
\begin{align}
\nabla_0 \pi &= \sqrt{- \d_\mu \psi \d^\mu \psi} - \mu = - \mu (\sqrt{-g^{00}} + 1) \; , \\
\nabla_0 \eta_m & = {E_m}^\mu \, n^\nu \nabla_\nu n_\mu = -{e_m}^\mu \d_\mu \log{\sqrt{-g^{00}}}\; ,\\
\nabla_n \eta_m & = {E_m}^\mu \, {E_n}^\nu \nabla_\nu n_\mu  = {e_m}^\mu \, {e_n}^\nu \, K_{\nu\mu} \; ,
\end{align}
where $\nabla_\mu$ is the usual four-dimensional covariant derivative associated with the metric $g_{\mu\nu}$ and the last equalities give the corresponding unitary gauge objects. Higher order terms in the derivative expansion are obtained by acting on the above terms by the CCWZ covariant derivative.

Additional matter fields in the CCWZ language should belong to the representations of the unbroken subgroup, the $SO(3)$ rotations in our case. Under the broken symmetry transformations these fields transform with Goldstone-dependent, and thus space-time dependent rotations.  In order to preserve the non-linearly realised symmetries one has to use the CCWZ covariant derivatives with the Goldstone-dependent connection:
\begin{equation}
{\mathcal A^m}_{n\, a} = {E_a}^\mu \, \left(  {\Lambda_c}^{m} \d_\mu {\Lambda^c}_n + {\omega^{c}}_{d \, \mu} {\Lambda_c}^m {\Lambda^d}_n \right)\; ,
\end{equation}
with parameters $\eta_m$ of the boost matrix $\Lambda$ fixed in terms of the time translations Goldstone field $\pi$ by equation~\eqref{eq:inverse higgs}.
For a vector field under $SO(3)$ rotations $\Sigma^m$ the covariant derivative is thus
\begin{align}
\nabla_a \Sigma^m &\equiv {E_a}^\mu \d_\mu \Sigma^m + {\mathcal{ A}^m}_{n\, a} \, \Sigma^n \notag\\
&= {E_a}^\mu \, \left( \d_\mu \Sigma^m + 
( {\Lambda_c}^{m} \d_\mu {\Lambda^c}_n + {\omega^{c}}_{d\,\mu} {\Lambda_c}^m {\Lambda^d}_{n}) \,\Sigma^n \right) \notag\\
&= {E_a}^\mu \, {\Lambda_c}^{m} \, \left( \d_\mu ({\Lambda^c}_n \Sigma^n ) + {\omega^{c}}_{d\,\mu}  {\Lambda^d}_{n} \,\Sigma^n \right) \notag\\
&=  {E_a}^\mu \, {E^m}_\nu \nabla_\mu (\Sigma^m \, {E_m}^\nu) \; .
\end{align}
It is nothing else but a projection to the tangent space of a slice of the usual four-dimensional covariant derivative of corresponding four-vector $\Sigma^\mu(\Sigma^m, \pi)$ defined in equation~\eqref{eq:4dAm}. This can be readily extended to a matter field that transforms in any other representation of the unbroken rotations. We have thus shown that the CCWZ construction is equivalent to the natural embedding prescription introduced before up to a $\pi$-dependent field redefinition~\eqref{eq:AmtoAi}. In what follows we will use the latter since it can be easily combined with the usual EFT of Inflation routine.

\section{\label{sec:spin1} Spin-one example}

In order to warm up let us consider the theory of a massive spin-1 particle in inflation and discuss the connection to the relativistic Proca theory. The reader interested in the phenomenology of light spinning particles can skip straight to Section~\ref{sec:pheno}. In order to write a general quadratic action for the vector field $\Sigma^i$ we apply the EFTI prescription to the four-vector $\Sigma^\mu$ . The most general quadratic Lagrangian for $\Sigma^\mu$ can be written as\footnote{Here and in the following we assume parity invariance. Otherwise we could have added a term $\epsilon_{\mu\nu\lambda\rho} n^\mu \Sigma^\nu \nabla^\lambda \Sigma^\rho$, which would split opposite helicities. This operator would give parity-odd observables like a TB correlation. We thank the Referee for pointing this out.}
\begin{equation}\label{eq:LagBdS}
\mathcal L_2 =  \frac12 (1 - c_1^2) \, n^\nu n^\lambda \nabla_\nu \Sigma^\mu \nabla_\lambda \Sigma_\mu 
- \frac12 c_1^2 \, \nabla_\mu \Sigma^\nu \nabla^\mu \Sigma_\nu
- \frac12 (c_0^2 - c_1^2) \, \nabla_\mu \Sigma^\mu \nabla_\nu \Sigma^\nu  
 - \frac12 (m^2 + c_1^2 H^2)\Sigma^\mu \Sigma_\mu\; .
\end{equation}
There are three independent kinetic terms one can write. We have fixed the overall normalisation to have a canonical time kinetic term and chosen the remaining two parameters $c_0^2$ and $c_1^2$ to be the propagation speeds for the helicity-0 and -1 modes respectively. Given that $\Sigma^\mu \, n_\mu \equiv 0$,  all the possible kinetic terms where $n_\mu$ is contracted with the vector field index contribute to the quadratic action only by changing the mass term: $n_\mu \nabla_\nu \Sigma^\mu = - \Sigma^\mu \nabla_\nu n^\mu$. They do, however, contribute to the interactions with $\pi$ and have to be considered for the phenomenological applications. The quadratic action for $\Sigma^i$ then reads
\begin{equation}\label{eq:S2A}
S_2 = \frac12 \int d^3 x \, dt \, a^3 \, \Big( (\dot \sigma^i)^2  - c_1^2 \, a^{-2} \, (\d_j \sigma^i)^2 - (c_0^2 - c_1^2) \, a^{-2} \, (\d_i \sigma^i)^2  - m^2 \, (\sigma^i)^2 \Big) \;,
\end{equation}
where we have defined a new field $\sigma^i \equiv a \, \Sigma^i$ which has a time kinetic term of a canonical scalar field. The Latin indices here and in what follows are summed with $\delta_{ij}$. Note that, at variance with the Lorentz-invariant Proca action for the massive spin-1 field the sign of the helicity-0 kinetic term is not related to the mass parameter. It shows that the scaling dimension of the spin-1 field in the inflationary background is not bounded by unitarity requirements (Higuchi Bound) and can be chosen to be arbitrary, even when the space-time metric is exactly de Sitter.

In Fourier space we decompose $\s_i$ in terms of its helicites,
$\s_{i}(\vec k, \eta) \equiv \sum_{h} \s^{(h)}_k (\eta)\, \epsilon^{(h)}_{i}(\hat k)\,,$
where $\sum_{i}\epsilon^{(h)}_{i}(\hat k)\ ({\epsilon^{(h')}_{ i } (\hat k)})^*= \delta^{h h'}$. 
The mode functions for a given comoving momentum $k$ and helicity $h$ that correspond to the Minkowski-like vacuum for the deep subhorizon modes are given by
\begin{equation}\label{eq:Bsol}
\sigma_{k}^{(h)} (\eta) = H \, \frac {\sqrt{\pi}} 2 \, e^{\half i \pi (\nu + \half)} \, (-\eta)^{3/2} \, H_\nu^{(1)} (- c_h \, k \, \eta) \;,
\qquad \text{with} \qquad \nu = \sqrt{\frac 94 - \frac {m^2}{H^2}} \; .
\end{equation}
The two-point function at late times takes the form
\begin{equation}\label{eq:A2pt}
\langle \sigma^i(\eta, \vec k) \sigma^j(\eta, -\vec k)\rangle' \simeq \frac{2^{2\nu - 2}  \, \Gamma^2(\nu) \, H^2 }{\pi \, k^{2 \nu} \, (-\eta)^{2\nu - 3}} \,
 \left( (\delta_{ij} - \hat k_i \hat k_j) \, c_1^{-2\nu} + \hat k_i \hat k_j \, c_0^{-2\nu}\right) \;,
\end{equation}
which corresponds to an operator with scaling dimension $\Delta = \frac 3 2 - \nu$.
Notice that the prime on the correlation functions indicates that the momentum conserving delta function has been removed.
The scaling dimension is determined by the mass parameter in the quadratic action and can be made to lie below the Higuchi bound $0 \le \Delta \le 1$.  Negative scaling dimensions for a massive field would mean that the stress energy tensor of the fluctuations is growing at late time signalling the breakdown of perturbation theory.

Apart from the quadratic action~\eqref{eq:S2A} the Lagrangian~\eqref{eq:LagBdS} leads also to cubic interactions of order $\mathcal O (\sigma^2 \, \pi)$ given by the following Hamiltonian:
\begin{equation}\label{eq:HpiBB}
H_{\text{int}}^{\sigma\sigma\pi} = - \int d^3 x \, a^2 \, \Big( 2 c_1^2 \, H \, \d_i \pi \, \dot \sigma^i \, \d_j \sigma^j 
 - (c_0^2 - c_1^2) \,  \d_i \pi \, \sigma^i \, (\d_t - H) \d_j \sigma^j 
 + (c_1^2 - 1) \, \d_i \pi \, \d_i \sigma^j \, \dot \sigma^j  \Big) \;.
\end{equation}
The scale suppressing these operators becomes lower and lower as one turns off the time-dependence of the inflaton background, since in canonical normalization $\pi_c = \sqrt{2 \epsilon} H \mpl \,\pi$ and $\epsilon$ goes to zero in this limit. To perform one more check that this geometric construction reproduces the correct couplings to the Goldstone field $\pi$ in the next subsection we shall check the conformal consistency relation for the vector field $\sigma^i$. We calculate the $\pi\sigma\sigma$ three-point function sourced by this interactions in the limit of the soft $\pi$ mode and confirm that is related to the action of the special conformal transformation on the two-point function of $\sigma$. But before we proceed let us comment on the relation of the developed formalism to the standard Proca action for a massive vector field.

{\bf Proca field.} Since in our formalism the spin-1 field is embedded in a four-vector with only three independent components, it might not be obvious how, in the same formalism, we can discuss the ordinary Proca theory, which has four components. Let us see how to do this. Let us consider a Proca four-vector, $B^\mu$, and let us decompose it in a component perpendicular to the slicing of uniform physical clock and three components parallel to it:
\be\label{eq:procafield}
B^\mu=\Sigma^\mu+n^\mu \;\phi\ ,\quad{\rm with}\quad n_\mu\, \Sigma^\mu=0\ .
\ee
To describe a Proca field, $B^\mu$, we need an additional scalar field, $\phi$, on top of $\Sigma^\mu=\Sigma^\mu(\Sigma^i,\pi)$, which is the spin-1 field we defined in~(\ref{3Dto4D}). $B^\mu$ is a combination of $\Sigma^i,\phi$ and $\pi$:
\begin{equation}
B^\mu = \left(
\begin{array}{c}
\phi-\d_j\pi \, \Sigma^j \\
\Sigma^i - a^{-2} \, \d_i \pi \, \phi \\
\end{array} \right)\ .
\end{equation}
The dependence on $\pi$ appears because our splitting in $\Sigma^i$ and $\phi$ is done with reference to the time-slicing induced by the physical clock, while the splitting $B^{\{0,1,2,3\}}$ is done with respect to a clock-independent slicing. One can check that the dependence on $\Sigma^i,\phi$ and $\pi$ indeed guarantees that $B^\mu$ transforms as a four-vector (as obvious from~(\ref{eq:procafield})). 

Now, a simplification occurs if $B^\mu$ appears in the unitary-gauge Lagrangian always contracted in a fully diff. invariant way, for example as it appears in the combination of a diff. invariant mass term: $B^\mu B_\mu$. In this case, we expect that we should not need $\pi$ to describe the field $B^\mu(\Sigma^i,\phi,\pi)$. Indeed, the problem is solved by performing a somewhat obvious field redefinition
 \begin{equation}\label{eq:componentsproca}
 \left(
\begin{array}{c}
\Sigma^i \\
\phi\\
 \pi\\
\end{array} \right)\qquad\to\qquad  \left(
\begin{array}{c}
B^\mu\\
 \pi\\
\end{array} \right)\ .
\end{equation} 
 If $B^\mu(\Sigma^i,\phi,\pi)$ appears in the action only in such a way that $B^\mu$ is contracted in a diff. invariant way, then the action in the Proca sector will depend only on the four components $B^\mu$, reproducing the familiar approach. However, as it made clear by our formalism, given the spontaneous breaking of time diff.s provided by inflation, there is no need to introduce a scalar component $\phi$ to describe a spin-1 field.  

We notice that in the case of Proca, the component $B^0$ of the vector field, which in our formalism is described by $\phi$, does not have a time-kinetic term. Therefore, a Proca Lagrangian describes the propagation of three helicities, i.e. the same number as in a theory described solely in terms of $\Sigma^i$. However, the physics associated to the two Lagrangians is completely different, as it is evident from the fact that the spectrum of the theory built with just $\Sigma^i$ can violate the Higuchi bound.

This observation highlights a freedom in our construction: the possibility of having auxiliary fields.  
Lacking a time-kinetic term, these fields do not propagate additional degrees of freedom. 
They can therefore be integrated out by solving for them and plugging back the resulting solution. This will lead to a non-local-looking Lagrangian, as it will contain several factors of inverse-Laplacians. 
This can lead to instantaneous propagation which is not compatible with a standard Lorentz invariant UV completion. Therefore one should impose microcausality, i.e.~the commutativity of field outside the lightcone, as a restriction on the model parameters. Due to the abundance of auxiliary fields and terms that we can add, the process can become quite cumbersome, and it would be nice to find a straightforward way to add these auxiliary fields automatically preserving microcausality.\footnote{Working directly in terms of the field $B^\mu$ does not seem to help. While in the Lorentz invariant case of Proca one can easily verify that the Lagrangian obtained after integrating out $B^0$ is local, notwithstanding the non-local-looking factors, microcausality is not generically preserved as we move away from the Lorentz invariant case by adding couplings of $B^\mu$ to the foliation.}

\subsection{Consistency relation for special conformal transformations}

The Higuchi bound \cite{Higuchi:1986py} is a consequence of the de Sitter isometries, or equivalently of the conformal symmetry of late-time correlators. The 2-point function is fixed by these symmetries and for small masses the longitudinal component becomes a ghost. The coupling with the foliation breaks the de Sitter isometries\footnote{The foliation breaks both the dilation and the boost isometries of de Sitter. We are here interested in the boosts since these relate the different helicities and therefore must be broken to violate the Higuchi bound. Furthermore in this paper we assume that the inflaton is endowed with an approximate shift symmetry, so that a residual diagonal symmetry---dilation combined with an inflaton shift---is linearly realised. This is the usual slow-roll assumption and it is the origin of the observed approximate scale-invariance of the scalar power spectrum.}: the 2-point functions of the various helicities are not anymore related to each other and thus the Higuchi bound can be avoided. However, the breaking induced by the inflaton background is spontaneous and not explicit, so that the symmetry is still there albeit non-linearly realised. The 2-point function of a particle with spin is now not invariant under the symmetries, but its variation is related to the coupling with soft Goldstone modes. The consistency relations associated with special conformal transformations (or boost isometries of de Sitter) were studied in~\cite{1203.4595,Hinterbichler:2012nm,Goldberger:2013rsa}.

These relations state that the effect of the gradient of the soft mode $\d_i \pi_L$ is equivalent to a special conformal transformation with parameter $b_i = - \half \, H \, \d_i \pi$:
\begin{equation}\label{CCC}
\langle \pi(\vec{q}) \, \sigma^i(\vec{k}) \, \sigma^j(-\vec{k} - \vec{q}) \rangle' 
\underset{\vec q \to 0}{=} \frac12 \, H \, P_\pi(q) \, \left(\vec{q} \cdot \vec{K}\right) \, \langle \sigma^i(\vec{k}) \, \sigma^j(-\vec{k}) \rangle' 
+ \mathcal O (q/k )^2 \; ,
\end{equation}
where $\vec K$ is the generator of the special conformal transformations in momentum space.\footnote{An explicit expression for $\vec K$ can be found in the equation~(A.157) of reference~\cite{1503.08043}.} Particles with spin with masses below the Higuchi bound do not have a conformal invariant 2-point function: the RHS of the equation above is therefore non-zero and it implies a prescribed coupling with the soft inflaton fluctuations. It is worthwhile stressing that this consistency relation implies a coupling with $\pi$ that remains perturbative, i.e.~suppressed by $\Delta_\zeta$, even when we are well below the Higuchi bound and the 2-point function is very far from conformal invariance. This remains true even when one gives very different speeds of propagations to the various helicities. 

Let us now verify eq.~\eqref{CCC}. In order to calculate the special conformal transformation of the two-point function it is useful to contract the tensor indices of the vectors $\sigma^i$ with the polarization vectors $\epsilon_i, \, \tilde\epsilon_j$. The late time limit of the two-point function~\eqref{eq:A2pt} then takes the form:
\begin{equation}\label{eq:2pteps}
\langle \epsilon_i \, \sigma^i(\eta, \vec k) \, \tilde \epsilon_j \,\sigma^j(\eta, -\vec k)\rangle' \underset{\eta \to 0}{=}
 \frac{2^{2\nu - 2}\, \Gamma^2(\nu) \, H^2  }{\pi \, k^{2 \nu} \, (-\eta)^{2\nu - 3}}  \left( \frac1{c_1^{2\nu}} \, (\vec \epsilon \cdot \vec { \tilde\epsilon}) + \left(\frac1{c_0^{2\nu}} - \frac1{c_1^{2\nu}}\right) \, (\vec\epsilon \cdot \hat k) (\vec{\tilde \epsilon} \cdot {\hat k}) \right) \;.
\end{equation}
For a particular choice of polarisation vectors and the soft momentum:
\begin{align}
\vec{q} \cdot \vec{k} &= \vec{q} \cdot \vec \epsilon = 0 \;, &\vec q \cdot \vec{\tilde\epsilon} \neq 0 \; ,
\end{align}
the special conformal transformation of a two-point function was calculated in~\cite{1503.08043} and for the 2-point function~\eqref{eq:2pteps} reads
\begin{equation}
\(\vec q \cdot \vec K\) \, \langle \epsilon_i \, \sigma^i(\eta, \vec k) \, \tilde \epsilon_j \, \sigma^j(\eta, -\vec k)\rangle' = 
 -  \( \frac{2\nu +1} {c_1^{ 2\nu}} +  \frac{2\nu - 1}{c_0^{ 2\nu}} \) \,
  \frac{H^2 \, 2^{2\nu-2} \, \Gamma^2(\nu) \, (\vec \epsilon \cdot {\hat k}) (\vec{\tilde \epsilon} \cdot \vec{q})}{ k^{2\nu+1} (-\eta)^{2\nu - 3}} \; .
\end{equation}
Notice that the 2-point function is not conformal invariant even for masses above the Higuchi Bound. 
This is compatible with the fact our formalism is intrisically related  to the foliation and does not simply interpolate with that de Sitter invariant Proca action.
The perturbative contribution to the three point function $\langle \pi \, \sigma \, \sigma \rangle$ due to the interaction~\eqref{eq:HpiBB} is given by
\begin{equation}
\langle \pi(\vec{q}) \, \epsilon_i \, \sigma^i(\vec k) \, \tilde \epsilon_j \, \sigma^j(\vec p) \rangle_{\eta_0} =
-i \int_{-\infty}^{\eta_*} d\eta
\langle \left[  \pi(\vec{q}) \, \epsilon_i \, \sigma^i(\vec k) \, \tilde \epsilon_j \, \sigma^j(\vec p), \, H_\text{int}^{\pi \sigma\sigma}(\eta)\right] \rangle \; .
\end{equation}
For the consistency relation we need only the leading order in $q/k$, and it is possible to obtain an explicit expression for the three-point function at late times $\eta_* \to 0$:
\begin{multline}
\langle \pi(\vec{q}) \, \epsilon_i \, \sigma^i(\vec k) \, \tilde \epsilon_j \, \sigma^j(\vec p) \rangle'_{\eta_*\to 0} = \\
-i \, H \, P_\pi(q) \left[ (\vec \epsilon \cdot \vec{k}) (\vec{\tilde \epsilon} \cdot \vec{q}) - (\vec\epsilon \cdot \vec{q}) (\vec{\tilde \epsilon} \cdot \vec{k}) \,\right] 
\big( \sigma_k^{(0)}(\eta_*) \, \sigma_k^{(1)}(\eta_*) \big)_{\eta_* \to 0} \\
  \int_{-\infty}^0 d\eta 
\left\{ \left( (c_1^2 - c_0^2) \, (-\eta) {\sigma'_k}^{(0)}(\eta) \sigma_k^{(1)}(\eta) + 2 c_0^2 \, \sigma_k^{(0)}(\eta) \sigma_k^{(1)}(\eta)  \right) - \text{c.c.} \right\} \; ,
\end{multline}
where the mode functions are given in~\eqref{eq:Bsol}. Taking the integral one can check that the consistency relation holds for all values of the sound speeds and scaling dimension.

\section{Minimal spin-$s$ theory}

As discussed, a minimal description of a spin-$s$ particle during inflation is given by a traceless\footnote{The trace of $\Sigma^{i_1 \dots i_s}$ should be taken using the induced metric on the constant inflaton slices: $h_{ij} = g_{\mu\nu} \frac{\d x^\mu}{\d x^i}\Big|_\psi \frac{\d x^\nu}{\d x^j}\Big|_\psi$. This implies that the four-dimensional field $\Sigma^{\nu_1 \dots \nu_s}$ is also traceless.} rank-$s$ tensor field $\Sigma^{i_1 \dots i_s}$. The action that explicitly preserves all space-time symmetries can be written by using its four-dimensional version $\Sigma^{\nu_1 \dots \nu_s}$, and contracting with the vector $n^\mu$:
\begin{multline}\label{eq:spins}
S = \frac1{2 s!} \int a^3 \, d^3x\, d t \left( (1- c_s^2) \, n^\mu n^\lambda \, \nabla_\mu \Sigma^{\nu_1 \dots \nu_s} \nabla_\lambda \Sigma_{\nu_1 \dots \nu_s} -  \, c_s^2 \,\nabla_\mu \Sigma^{\nu_1 \dots \nu_s} \nabla^\mu \Sigma_{\nu_1 \dots \nu_s} \right. \\
\left. -  \delta c_s^2 \, \nabla_\mu \Sigma^{\mu\nu_2 \dots \nu_s} \nabla_\lambda {\Sigma^\lambda}_{\nu_2 \dots \nu_s} -  (m^2 + s \, c_s^2 H^2) \, \Sigma^{\nu_1 \dots \nu_s} \Sigma_{\nu_1 \dots \nu_s}
\right)  \;.
\end{multline}
For a canonically normalized field there are only three free parameters in the quadratic action at leading order in derivatives, independently of the spin: two speed parameters $c_s^2$ and $\delta c_s^2$ fixing the propagation speeds for all the helicity modes, and the mass $m$. 

Expanding the action in powers of the Goldstone field one obtains a free action for the spinning particle and its leading interactions with $\pi$. The quadratic action reads
\begin{equation}\label{eq:S2}
S_2 = \frac1{2 s!} \int a^3 \, d^3x\, d t \Big(  \, (\dot \sigma^{i_1 \dots i_s})^2 -  \, c_s^2 \, a^{-2} (\d_j \sigma^{i_1 \dots i_s})^2 - \, \delta c_s^2 \, a^{-2} (\d_j \sigma^{j i_2 \dots i_s})^2
-  \, m^2 \, (\sigma^{i_1 \dots i_s})^2 
\Big)  \;,
\end{equation}
where we have defined a new field $\sigma^{i_1 \dots i_s} \equiv a^s \, \Sigma^{i_1 \dots i_s}$, which has the same time-kinetic term as a canonical scalar field with mass $m^2$. Note that there are only two independent operators that constitute the spatial kinetic term for a traceless tensor field. This means that although the action propagates $2 s + 1$ different helicity modes, all propagation speeds $c_h^2$ for $h = 0, \dots, s$ can be expressed as some linear combinations of $c_s^2$ and $\delta c_s^2$. The coefficients in these combinations are not universal and depend on the spin. It is easy to see, however, that the speed of the highest helicity mode $h = s$ is given by $c_s$ for any spin.\footnote{One can also show that $c_h^2 = a \, c_s^2 + (1 - a) \, c_0^2$ with $0 \leq a \leq 1$ for any helicity $h$.} In particular, for a spin-1 field $c_1^2 = c_s^2$ and $c_0^2 = c_s^2 + \delta c_s^2$, and for a spin-2 field
\begin{align}
c_2^2 &= c_s^2 \;, & c_0^2 &=  c_s^2  + \frac 23 \delta c_s^2\; , & c_1^2 &= c_s^2 + \frac12 \delta c_s^2 = \frac14 c_2^2 + \frac34 c_0^2 \; .
\end{align}
In particular, in order to avoid gradient instabilities and superluminal propagation for all the modes of a speed-2 particle, the speed parameters have to be in the range $0 \le c_s^2 \le 1$ and $0 \le c_s^2 + \frac 23 \delta c_s^2\le 1$.
Notice that the mass term is the same for all the helicities and this implies that the time dependence on super-horizon scales is common to the whole multiplet.

Given that the transformation of $\sigma$ under boosts is $\pi$-dependent, the action~\eqref{eq:spins} has to contain interactions between $\sigma$ and the Goldstone field $\pi$. The structure of these minimal interactions is completely fixed by symmetries and all the couplings are determined in terms of the parameters of the free Lagrangian. In order to study the phenomenological and theoretical consequences of these interactions, we need to consider the structure of the cubic $\sigma^2 \pi$ and quartic $\sigma^2 \pi^2$ interactions. The cubic action is given by
\begin{multline}\label{eq:S3}
S_3 = \frac 1 {s!} \int dt \, d^3x\, a^3 \Big( (c_s^2 - 1)  \, \d_j \pi \, \d_j \sigma^{i_1 \dots i_s}  \, \dot \sigma^{i_1 \dots i_s} \\
+ 2 c_s^2 \, s H  \, \d_i \pi \, \sigma^{i i_2 \dots i_s}  \, \d_j \sigma^{j i_2 \dots i_s}
 - \delta c_s^2 \,  \d_i \pi \, \sigma^{i i_2 \dots i_s}  \, (\d_t - s H) \, \d_j \sigma^{j i_2 \dots i_s} \Big) \;,
\end{multline}
while the quartic interactions are given by the following action:
\begin{multline}\label{eq:S4}
S_4 = \frac 1 {2 s!} \int d^3x\, d t \, a^3 \Big(  \big(s \, m^2 - (5 s + (5 s^2 - 4 s ) \, c_s^2 + (s-1)(s + 4) \, \delta c_s^2 )\, H^2  \big) \, a^{-2} \, ( \d_i \pi \, \sigma^{i i_2 \dots i_s})^2\\
+ 2 (2 c_s^2 + \delta c_s^2) \, s H \, a^{-2} \dot \pi \, \d_i \pi \, \sigma^{i i_2 \dots i_s} \, \d_j \sigma^{j i_2 \dots i_s}
+ 2 (1 - c_s^2) \, s H \, a^{-2} \, \d_i \dot \pi \, \d_j \pi \, \sigma^{i i_2 \dots i_s} \, \sigma^{j i_2 \dots i_s}
\\
- (s + \delta c_s^2 (s-1) ) \,  a^{-2} \, (\d_t ( \d_i \pi \, \sigma^{i i_2 \dots i_s} ))^2
+ \delta c_s^2 \, s H \, a^{-2} \, \dot \pi \, \d_i \pi \, \sigma^{i i_2 \dots i_s} \, \d_j \dot \sigma^{j i_2 \dots i_s}
\\
+ c_s^2 \, s \, a^{-4} \, (\d_j (\d_i \pi \, \sigma^{i i_2 \dots i_s}))^2 + \delta c_s^2 \, (s - 1) \, a^{-4} \, (\d_j (\d_i \pi \, \sigma^{i j i_3 \dots i_s}))^2
\\
+ (1 - c_s^2) \, \big [ a^{-4} \, (\d_j \pi \, \d_j \sigma )^2 + 2 \, a^{-2} \, \dot \pi \, \d_j \pi \, \dot \sigma \, \d_j \sigma
+ a^{-2} \, (\d_j \pi)^2 \, (\dot \sigma)^2
\big ]
\Big)
 \; .
\end{multline}
As expected, the fact that the matter field $\sigma$ have to realise non-linearly the full Poincare symmetry fixes the form and the coefficients of its interactions with the Goldstone field $\pi$.\footnote{Some of these interaction terms are degenerate with other interactions that one is allowed to add to the effective field theory action and that start at higher order in Goldstone fields. 
The coefficients of such terms are not fixed by symmetries.} 
These interaction will induce non-Gaussianities in $\pi$ correlation functions that are inevitable consequences of the presence of the extra field $\sigma$ and its non-relativistic structure. 
In Section~\ref{sec: radiative constraints} we provide an estimate for these non-Gaussianities, where we also take into account the constraints induced by the radiative stability of the theory.
We will find that radiative stability and observational constraints do not qualitatively limit the observational consequences associated to the presence of spinning particles coupled to the inflaton, that we elaborate in detail in the next section for the particularly interesting case of a spin-2 particle.

\section{Phenomenology of a light spin-two}\label{sec:pheno}

In this section we focus on the phenomenology of  a light spin-2 field $\Sigma^{ij}$ during inflation. 
There are two main reasons to skip the simpler spin-1 case.
First, the contributions to the squeezed limit of the bispectra $\langle{\zeta_{\vec q \to 0} \zeta_{\vec k} \zeta_{-\vec k}}\rangle$ and $\langle{\gamma^{(s)}_{\vec q \to 0} \, \zeta_{\vec k} \, \zeta_{- \vec k}}\rangle$ mediated by a field with an odd spin have an extra $q/k$ suppression~\cite{Lee:2016vti}. 
Second, the helicity-2 mode of the spin-2 field can mix with the tensor metric fluctuations: this affects the phenomenology of gravitational waves (besides the scalar sector).

In our construction, a spin-2 field is embedded in a four-dimensional tensor $\Sigma^{\alpha\beta}$ which is traceless and orthogonal to $n^\alpha$.
In this case the general action \eqref{eq:spins} reads
\begin{align}\label{cov_action}
S[\Sigma] &= \frac14 \int \!\! d^4x \, \sqrt{-g} \Big( (1-c_2^2) n^\mu n^\nu\nabla_\mu \Sigma^{\alpha\beta} \, \nabla_\nu \Sigma_{\alpha\beta} - c_2^2 \nabla_\mu \Sigma^{\alpha\beta} \, \nabla ^\mu \Sigma_{\alpha\beta} \notag \\
&- \frac32 (c_0^2 -c_2^2)\, \nabla_\mu\Sigma^{\mu\alpha}\, \nabla^\nu\Sigma_{\nu\alpha}  - (m^2 + 2 \, c_2^2 H^2)\,  \Sigma^{\alpha\beta}\Sigma_{\alpha\beta}  \Big) \; .
\end{align}
The parameters $c_0$ and $c_2$ give the sound speeds for the helicity-0 and helicity-2 modes respectively. The sound speed for the helicity-1 mode $c_1^2 = \frac 14 ( 3 c_0^2 + c_2^2)$ is always positive and less then unity given that $0 \le c_0^2, \, c_2^2 \le 1$. The quadratic action reads
\be\label{toy_model}
S[\sigma] = \frac14 \int \!\! d t \, d^3 x \, a^3 \Big( (\dot\sigma^{ij})^2 - c_2^2 \, a^{-2} \, (\d_i \sigma^{jk})^2 - \frac32 (c_0^2 -c_2^2) \, a^{-2} \,  (\d_i\sigma^{ij})^2 - \, m^2\, (\sigma^{ij})^2 \Big) \; ,
\ee
where, as before, we have defined $\sigma^{ij} \equiv a^2 \, \Sigma^{ij}$. Apart from the quadratic part, the covariant action~\eqref{cov_action} includes also interactions of $\sigma^{ij}$ with $\pi$ and $\gamma_{ij}$ dictated by the non-linearly realised diffeomorphism invariance. For a systematic study of the phenomenology of the spin-2 field we also have to include other interaction and mixing terms allowed by the symmetries. At the leading order in fields and derivatives one has the following operators\footnote{For simplicity we omit the operator $\tilde m^2 \delta g^{00} \Sigma^2$. This induces a $\pi \s\s$ coupling that is subdominant to  the ones in of eq.~\eqref{eq:S3} in the regime $\tilde m^2 \ll H^2$. }
\begin{equation}\label{interactions}
S_{{ \rm int}} =  \int \!\! d^4x \, \sqrt{-g} \Big( \mpl \, \rho \, \delta K_{\alpha\beta} \Sigma^{\alpha\beta} + \mpl \,\tilde\rho \; \delta g^{00} \delta K_{\alpha\beta} \Sigma^{\alpha\beta} - \mu \, \Sigma^{\alpha\beta} {\Sigma_\alpha}^\gamma \Sigma_{\gamma\beta} \Big) \; ,
\end{equation}
where $\delta K_{\alpha\beta} \equiv K_{\alpha\beta} - a^2 H h_{\alpha\beta}$ is the fluctuation of the extrinsic curvature of constant $\psi$ surfaces and $\rho, \tilde\rho$, and $\mu$ are coupling constants with mass dimension one. The term proportional to $\rho$ is responsible of the mixing of $\sigma$ both with scalar and tensor perturbations. Going to the decoupling limit and in terms of the canonically normalized fields, $\pi_c \equiv (2 \epsilon H^2 \mpl^2)^{1/2} \pi$ and $\gamma_{ij}^{(c)} \equiv \mpl \gamma_{ij}$ one has, up to cubic order:
\begin{multline}\label{intercanonical}
S_{\rm int} = \int \!\! d^4x \, \sqrt{-g} \left[ - \frac{\rho}{\sqrt{2\epsilon} \, H} \, a^{-2} \, \partial_i\partial_j \pi_c\, \sigma^{ij}+\frac12 \, \rho \, \dot\gamma_{c\ ij} \sigma^{ij} \right.\\
 \left. - \frac{\rho}{2 \epsilon H^2 \mpl} \, a^{-2} \, \left( \partial_{i}\pi_c\partial_{j} \pi_c \, \dot \sigma^{ij} + 2 H \, \partial_i\pi_c\partial_j\pi_c \, \sigma^{ij} \right) + \frac{\tilde\rho}{\epsilon H^2 \mpl}\, a^{-2} \, \dot\pi_c \partial_i\partial_j \pi_c \sigma^{ij} - \mu (\sigma^{ij})^3\right] \;.
\end{multline}

In the following we are going to study the phenomenology associated with the action above when $\sigma$ is light, $m \ll H$. 
Notice that a background for the field $\s$ would induce a certain amount of anisotropy  \cite{Franciolini:2018eno,Bartolo:2015dga}. 
In the presence of a small mass for $\s$ the background (slowly) redshifts away. In this paper we assume that the field has a negligible background value.

Let us comment on the radiative stability of our setup. Since the interactions involve $\sigma$ without derivatives, one expects loop corrections to generate a mass for $\sigma$. The interactions $\rho$ and $\tilde \rho$ would not generate a $\sigma$ mass in flat space since they preserve {the} shift symmetry in $\sigma$ up to the terms proportional to $H$. One can therefore estimate the radiative corrections to the mass {induced by the operator in $\rho$} to be
\be\label{eq:massrho}
\delta m^2_\rho \sim H^2 \, \(\frac{\rho}{\sqrt \epsilon\, H}\)^2 \frac{\Lambda^4}{\epsilon \, H^2 \, \mpl^2} ,
\ee
where $\Lambda$ is a cutoff at which the loop is cut. 
As we will discuss later $\rho/(\sqrt{\epsilon} H) \lesssim 1$ is the condition for the stability of the system.
In this case, requiring $\delta m^2 \ll H^2$ gives $\Lambda^4 \ll \ep H^2 M_{\rm Pl}^2$. The loop must be cut at a scale somewhat below {$\ep H^2\mpl^2$}, which is the unitarity cut-off associated to the $\rho$ interaction itself. The same estimate eq.~\eqref{eq:massrho} works for the $\tilde\rho$ interaction.
The cubic $\mu$ interaction gives a logarithmic divergent contribution to the mass, so that for naturalness one needs $\mu \ll H$.
Notice that the smallness of the couplings $\rho,\,\tilde\rho$ and $\mu$ is radiatively stable since these couplings are odd for $\s\to-\s$.
In conclusion, it is technically natural to have a light $\s$ with sizeable mixing with $\gamma$ and $\zeta$.

\subsection{Estimates of the effects}
In this Section we will estimate the effects of the $\sigma$ field, described by {the} action \eqref{toy_model}-\eqref{interactions}, while in the following we will make explicit calculations in some specific cases. The spin-2 field $\sigma$ affects the observables {in particular} through the term in the action $\propto \rho$; this induces a mixing of $\sigma$ both with scalar and tensor perturbations. In terms of the canonically normalized scalar and tensor perturbations $\pi_c$ and $\gamma_c$ the mixings are schematically of the form
\be
\sim \frac{\rho}{\sqrt{\epsilon} H} \, \partial\partial\pi_c \, \sigma_c {\, ,} \qquad \sim \rho \, \dot \gamma_c  \, \sigma_c \;.
\ee
The mixing with the scalar perturbations is enhanced with respect to the tensor one since $\epsilon \ll 1$. This suggests that the effects of $\sigma$ should be searched for in the statistics of scalar perturbations only. 
However, as we discussed, the different helicity components of $\sigma$ will have in general different propagation speeds.\footnote{In this paper we assume, for simplicity, that the speed of the scalar perturbations $\pi$ is unity. 
The speed of propagation of tensor perturbations can always be taken to be one without loss of generality~\cite{Creminelli:2014wna}.}
(One expects the speeds to be also different from the speed of light, since $\sigma$ is an intrinsically non-Lorentz-invariant object.) If the speed of propagation of the helicity-2 {component}, $c_2$, is smaller than the one of the helicity-0, $c_0$, then the helicity-2 power spectrum is boosted and this can easily overcome the smaller mixing. 
{Additionally}, the mixing between $\gamma$ and $\sigma$ does not turn off outside the horizon: as we will see, this induces, for sufficiently small mass of $\sigma$, an enhancement of the {effect of the} mixing of order $N^2$, where $N$ is the number of e-folds of observable inflation. 
We assume $\rho/(\sqrt{\epsilon} H)\ll c_0$. In the opposite case the speed of propagation of $\sigma$ is dominated by the mixing term, this leads to gradient instabilities.

{\bf Power spectra.} It is easy to realise that the contribution of $\sigma$ to the scalar and tensor power spectra {is} of the form (see figure \ref{Fig_power_spectrum})
\be\label{power_estimates}
P_\zeta \sim \frac{H^2}{\epsilon \mpl^2} \frac{1}{k^3}\left(\frac{\rho}{\sqrt{\epsilon}H}\right)^2 \frac{1}{c_0^3} {\, ,} \qquad  P_\gamma \sim \frac{H^2}{\mpl^2} \frac{1}{k^3}\left(\frac{\rho}{H}\right)^2 \frac{N^2}{c_2^3}\;,
\ee
where we assumed $\sigma$ massless for simplicity.
Depending on the parameters, one can get sizeable modification of either the tensor or scalar power spectrum, or both. Notice also that, for sufficiently small {sound-}speeds $c_{0,2}$, both power spectra may be dominated by the $\sigma$ exchange, while remaining in the weak mixing regime $\rho/(\sqrt{\epsilon} H) \ll 1$. 

\begin{figure}[h] 
\centering
\subfloat[][\label{Fig_power_zeta}]{\includegraphics[scale=.15]{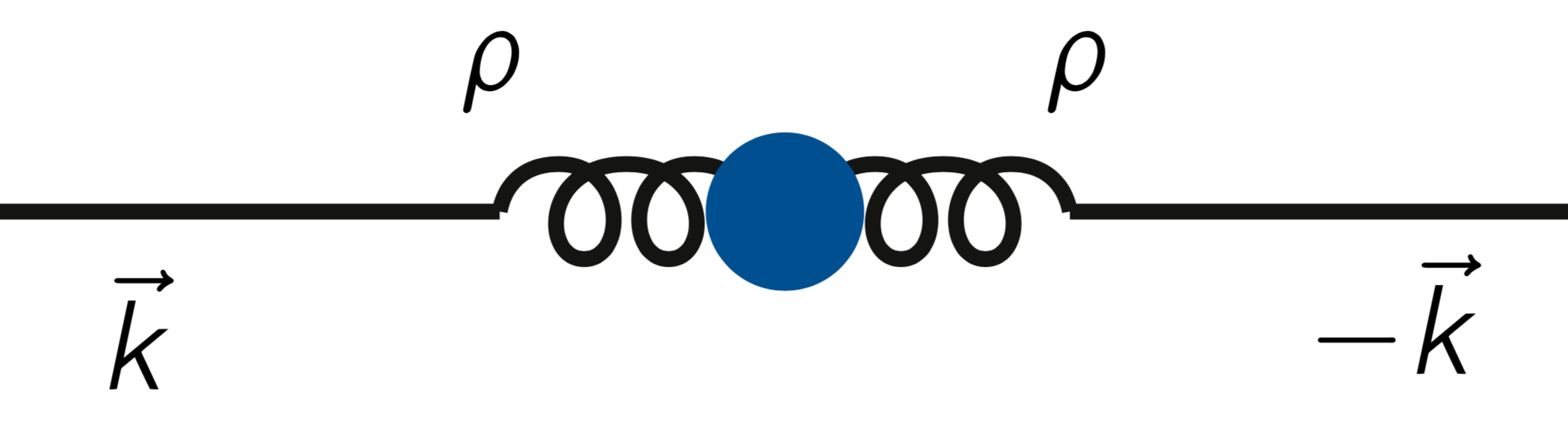}} \quad\quad\quad
\subfloat[][\label{Fig_power_gamma}]{\includegraphics[scale=.15]{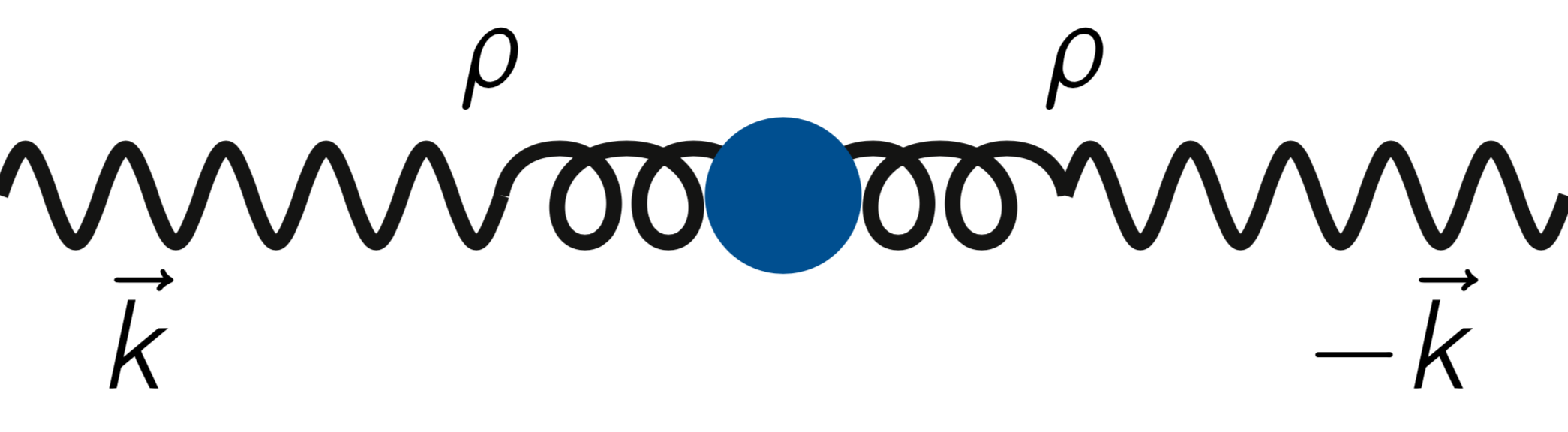}} \quad
\caption{\small Contributions to the scalar and tensor power spectra due to the exchange of a $\sigma$ field. Solid lines correspond to $\pi$, wavy ones to $\gamma$ and curly ones to $\sigma$.
The dots indicate a contraction between a pair of free fields, i.e.~the insertion of a power spectrum. One has to put the minimal number of dots in such a way that external lines cannot be connected without going through a dot (contraction), and each dot is connected to external lines from both sides, see for instance \cite{Mirbabayi:2014zpa}.
\label{Fig_power_spectrum}}
\end{figure} 

{\bf 3-point functions.} {For the purpose of estimates, we will focus on the squeezed limit of the 3-point function}. As it is well-known, this limit is sensitive to the content of light fields during inflation. Since we now understand that fields with masses below the Higuchi bound can exist, it is natural to look for their signatures in the squeezed limit. By symmetry arguments one can easily write the behaviour in the squeezed limit of the 3-point functions up to an overall factor
\be
\expect{\zeta_{\vec q \to 0} \zeta_{\vec k} \zeta_{-\vec k}}' = B_\zeta \(\frac{q}{k}\)^{\frac 3 2 - \nu} P_{\zeta}(q)\, P_{\zeta}(k)  \((\hat q \cdot \hat k)^2-\frac{1}{3} \)\,,
\ee
\be
\expect{\gamma^{(s)}_{\vec q \to 0} \, \zeta_{\vec k} \, \zeta_{- \vec k}}' = B_\gamma \(\frac{q}{k}\)^{\frac{3}{2}-\nu} P_{\gamma}(q) \ P_\zeta (k) \ \epsilon_{ij}^{(s)}\hat k_i \hat k_j\,,
\ee
where $\nu \equiv \sqrt{\frac{9}{4}-\frac{m^2}{H^2}}$ {and where $\epsilon_{ij}^{(s)} (\hat q)$ is the polarization tensor relative to the $s$-th helicity of $\s$}. (The present experimental limit on $B_\zeta$, for $\nu \simeq \frac 3 2$, from Planck is: $|B_\zeta| = |-\frac{144}{5} f_{\rm NL}^{L=2}| \lesssim 200$ at $2 \sigma$ \cite{Ade:2015ava,Shiraishi:2013vja}. 
$B_\zeta$ induces a scale-dependent tidal alignment of galaxies which could be observed in future surveys \cite{Schmidt:2015xka,Chisari:2016xki}. 
For other future constraints see \cite{MoradinezhadDizgah:2017szk, MoradinezhadDizgah:2018ssw,Franciolini:2018eno,MoradinezhadDizgah:2018pfo}.) 
It is also quite easy to give a parametric estimate of the prefactors in the various cases
\be\label{scalar_estimates}
\begin{split}
\rho \rho\;{\rm interaction.\;fig~\ref{XXX_zeta}} & \qquad B_\zeta \sim \(\frac{\rho}{H \sqrt{\epsilon}}\)^2 \frac{1}{c_0^{2\nu}}\,, \\
\rho \rho\;{\rm interaction.\;fig~\ref{YYY_zeta}} & \qquad B_\zeta \sim \(\frac{\rho}{H \sqrt{\epsilon}}\)^2 \frac{1}{c_0^{4\nu}}\,, \\
\rho \tilde\rho\;{\rm interaction.\;fig~\ref{XXX_zeta}} & \qquad B_\zeta \sim \(\frac{\rho}{H \sqrt{\epsilon}}\)\(\frac{\tilde\rho}{H \sqrt{\epsilon}}\) \frac{1}{c_0^{2\nu}}\,, \\
\mu\;{\rm interaction.\;fig~\ref{WWW_zeta}} & \qquad B_\zeta \sim \frac{\mu}{H}\(\frac{\rho}{H \sqrt{\epsilon}}\)^3 \Delta_\zeta^{-1} \frac{1}{c_0^{4\nu}}\,. \\
\end{split}
\ee
The dependence of $c_0$ is obtained looking at the $\s$ power spectra in the graphs, taking into account that $P_{\s^{(s)}} \propto 1/(c_s\,q)^{2\nu}\,,$ and that $\s$ freezes before $\pi$ crosses the Hubble horizon.

\begin{figure}[h] 
\centering
\subfloat[][\label{XXX_zeta}]{\includegraphics[scale=.18]{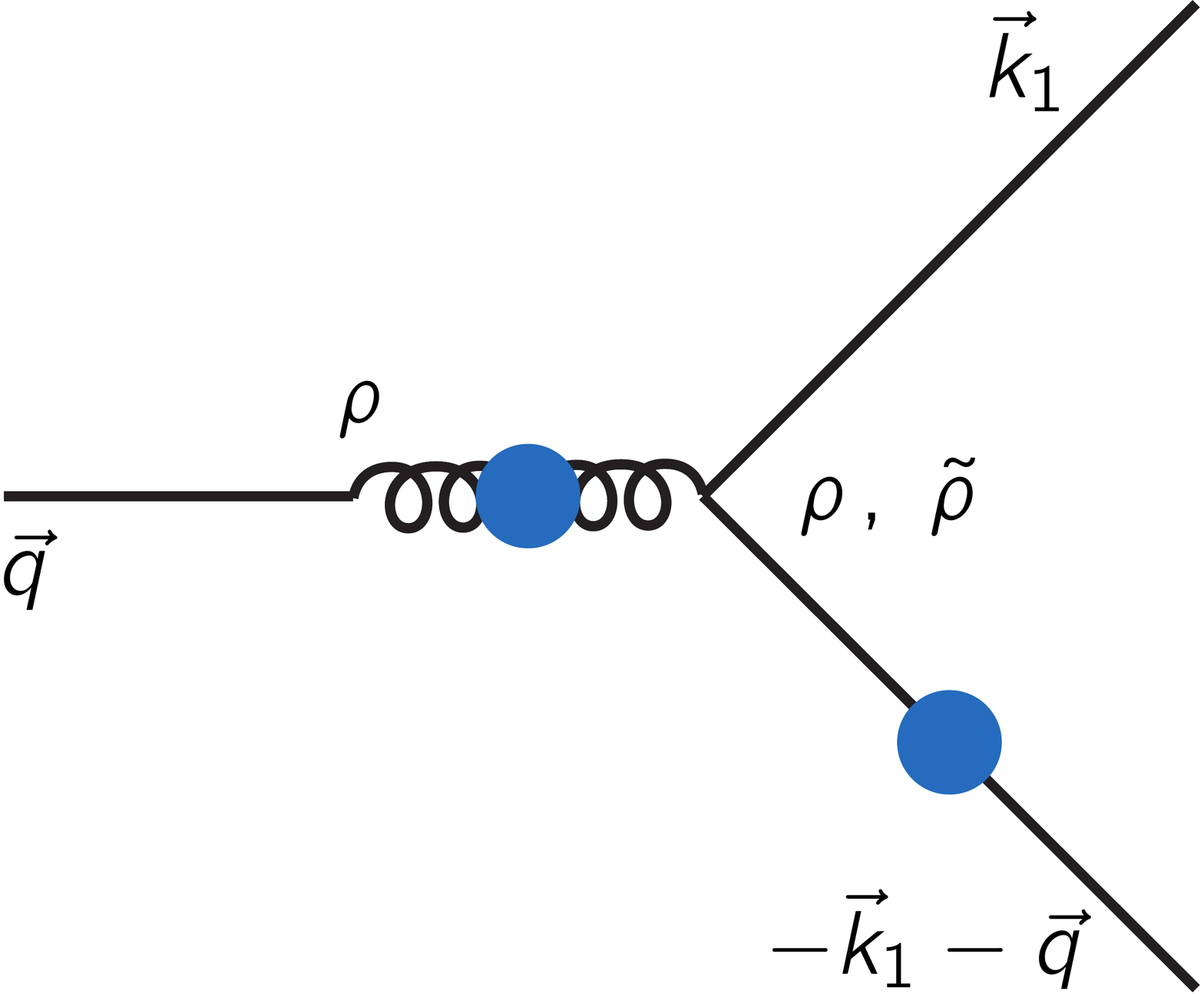}} \quad
\subfloat[][\label{YYY_zeta}]{\includegraphics[scale=.18]{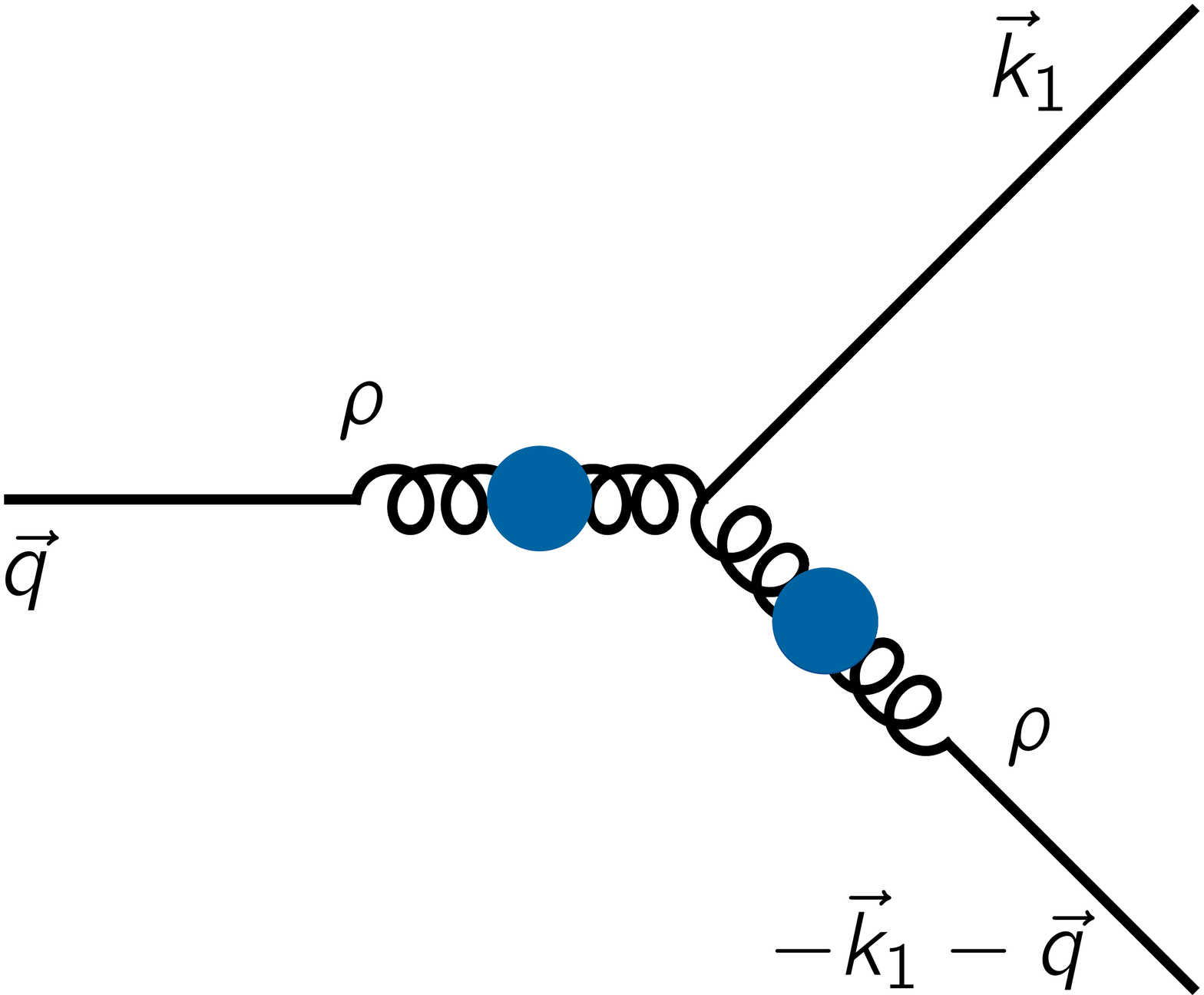}} \quad
\subfloat[][\label{WWW_zeta}]{\includegraphics[scale=.18]{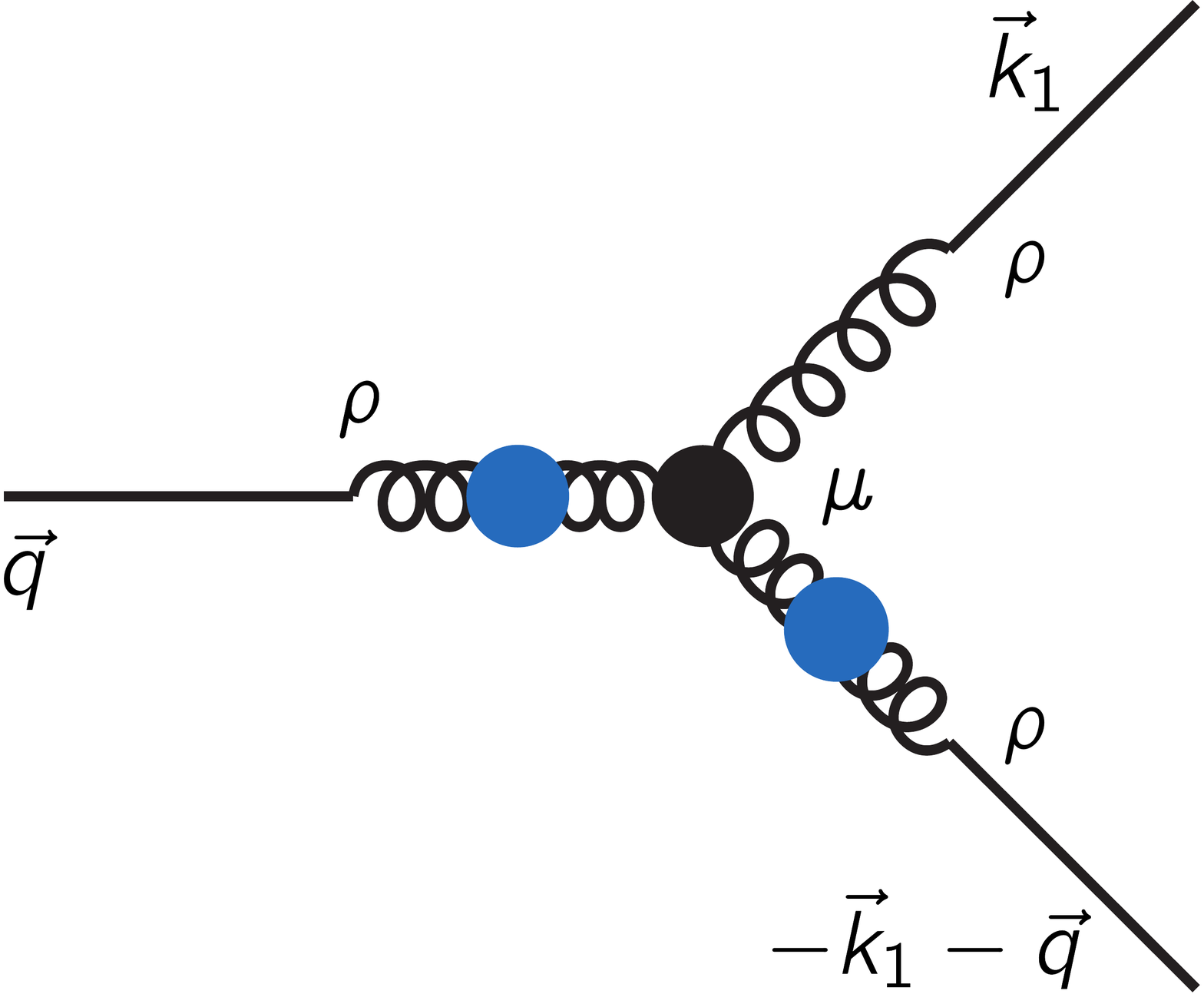}} \quad
\caption{\small Leading contributions to the $\expect{\zeta \zeta \zeta}$ 3-point function.  Diagrams where the dots are located in different places are subdominant for small $c_0$.
}
\end{figure} 

The correlator $\expect{\gamma^{(s)}_{\vec q} \, \zeta_{\vec k} \, \zeta_{- \vec k}} '$ can be estimated\footnote{Notice that the estimates computed in eqs.~\eqref{scalar_estimates} and \eqref{tensor_estimates} of $B_\zeta$ and $B_\gamma$   are valid even away from the squeezed limit with the only possible exception of the last line of eq.~\eqref{tensor_estimates}, fig~\ref{WWW_gamma}. 
Away from the squeezed limit there is another contribution to $\expect{\gamma\zeta\zeta}$ besides fig~\ref{WWW_gamma}. It is proportional to $c_0^{- 4\nu}$ instead of $(c_0\,c_2)^{-2\nu}$.} 
to be
\be\label{tensor_estimates}
\begin{split}
\rho \rho\;{\rm interaction.\;fig~\ref{XXX_gamma}} & \qquad B_\gamma \sim \(\frac{\rho}{H \sqrt{\epsilon}}\)^2 \frac{1}{c_2^{2\nu}} \,,  \\
\rho \rho\;{\rm interaction.\;fig~\ref{YYY_gamma}} & \qquad B_\gamma \sim \(\frac{\rho}{H \sqrt{\epsilon}}\)^2 \frac{1}{c_0^{2\nu} c_2^{2\nu}}\,, \\
\rho \tilde\rho\;{\rm interaction.\;fig~\ref{XXX_gamma}} & \qquad B_\gamma \sim \(\frac{\rho}{H \sqrt{\epsilon}}\)\(\frac{\tilde\rho}{H \sqrt{\epsilon}}\) \frac{1}{c_2^{2\nu}} \,, \\
\mu\;{\rm interaction.\;fig~\ref{WWW_gamma}} & \qquad B_\gamma \sim \frac{\mu}{H}\(\frac{\rho}{H \sqrt{\epsilon}}\)^3 \Delta_\zeta^{-1} \frac{1}{c_0^{2\nu} c_2^{2\nu}} \,.\\
\end{split}
\ee

\begin{figure}[h] 
\centering
\subfloat[][\label{XXX_gamma}]{\includegraphics[scale=.18 ]{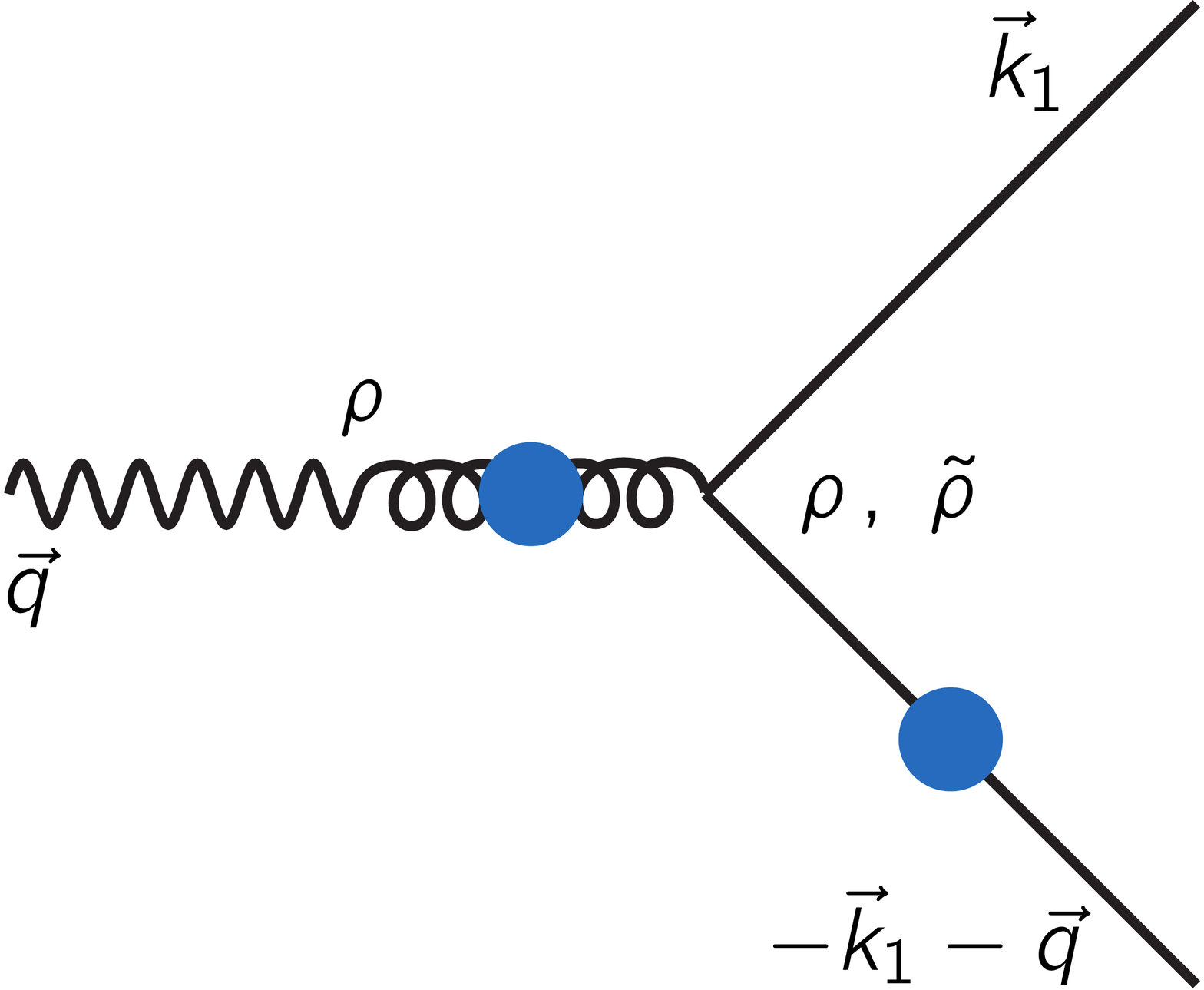}} \quad
\subfloat[][\label{YYY_gamma}]{\includegraphics[scale=.18]{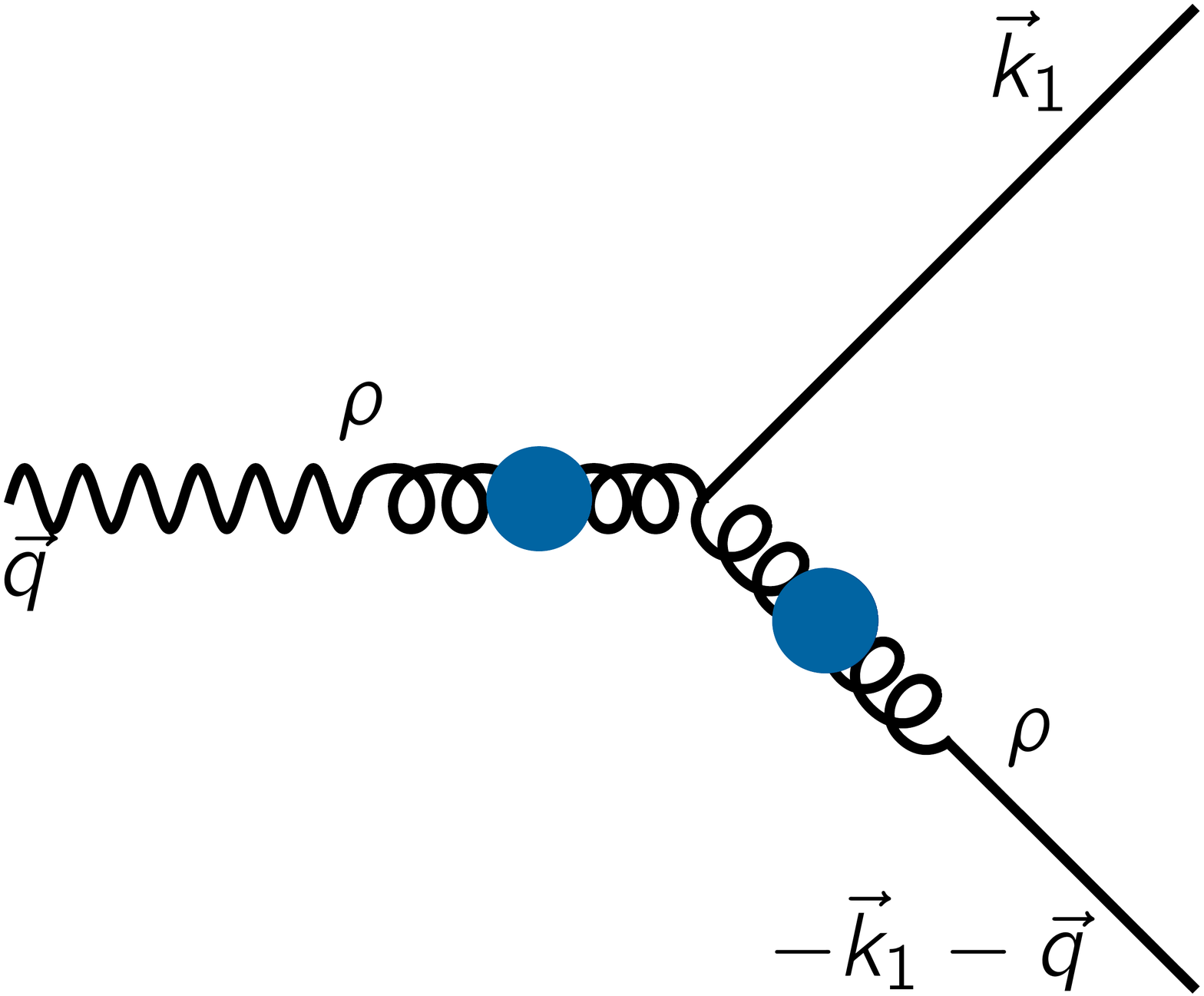}} \quad
\subfloat[][\label{WWW_gamma}]{\includegraphics[scale=.18]{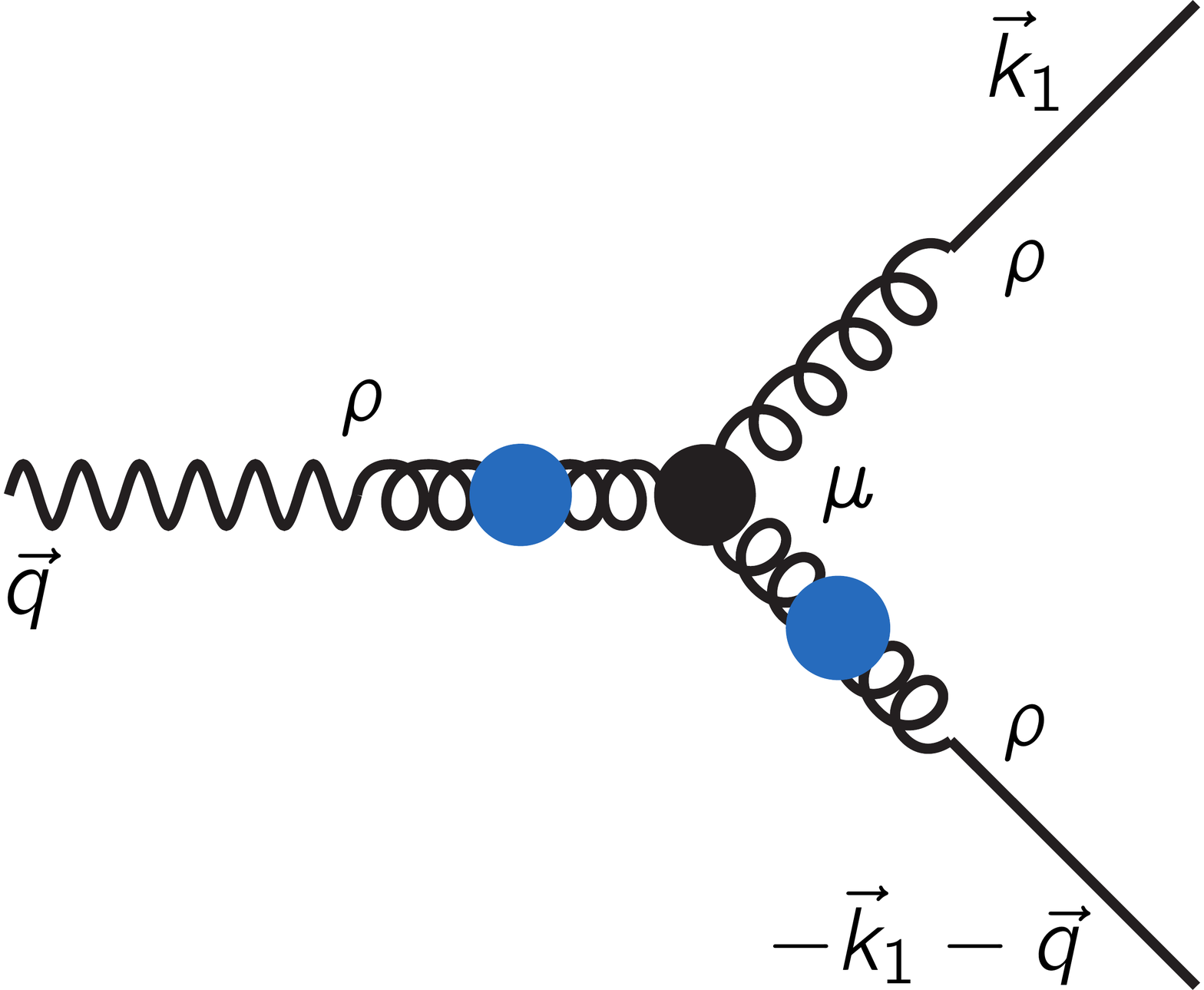}} \quad
\caption{\small Leading contributions to the $\expect{\gamma \zeta \zeta}$ 3-point function.
}
\end{figure} 

{\bf Consistency relations and isocurvature perturbations.}
In our setup, besides the inflaton, one has the additional field $\sigma_{ij}$. Since this field is traceless, it cannot mix with scalar perturbations in the long wavelength limit, i.e.~when derivatives are negligible.\footnote{At quadratic order $\sigma_{ij}\sigma_{ij}$ is scalar so the mixing is possible.} One can thus conclude, even without knowing the details of reheating, that isocurvature perturbations cannot be generated. Following the same logic one can conclude that the usual Maldacena consistency relation holds after an angular average over the orientation of the long mode:
\be
\label{averagedCR}
\int \!\frac{d^2\hat q}{4\pi}\expect{\zeta_{\vec q} \zeta_{\vec k} \zeta_{-\vec k-\vec q}}'_{q \ll k} = -\frac{d \log k^3 P_\zeta(k)}{d\log k}  P_\zeta(q) P_\zeta(k) \,.
\ee
(This is similar to what happens in the case of Solid Inflation \cite{Endlich:2012pz,Bordin:2017ozj}.) Notice that this consistency relation will hold even when the power spectrum of $\zeta$ is dominated by the $\sigma$ exchange.
Eq.~\eqref{averagedCR} implies that there is no way to generate angle-independent local non-Gaussianities in this setup. There is no sense in which tensor consistency relations are preserved. Moreover, since  $\sigma$ mixes with $\gamma$ also in the long-wavelength limit, one in general expects that this mixing also occurs during reheating: both the tensor power spectrum and ``local" tensor non-Gaussianities can be generated at reheating, similarly to what happens for scalar perturbations when one has more than one scalar field. It would be interesting to understand the most general tensor bispectrum generated in this way.

{\bf 4-point functions.} The 4-point function simplifies in the countercollinear limit, when couples of external momenta are almost equal and opposite. It receives independent contributions from all helicities of $\sigma$. In the countercollinear limit the behaviour of the correlator is fixed by symmetries. The exchange of the helicity states   gives
\be\label{tris_shape}
\expect{\zeta_{\vec k_1 - \vec q} \zeta_{-\vec k_1} \zeta_{\vec k_2 + \vec q} \zeta_{-\vec k_2}}' = T_0 \(\frac{q}{\sqrt{k_1 k_2}}\)^{3/2-\nu} P_{\zeta}(q)\, P_{\zeta}(k_1)\, P_{\zeta}(k_2)  \sum_{s\,s' = -2}^{+2} \(\epsilon_{ij}^{(s)} (\hat q) \, \hat k_{1\,,i} \hat k_{1\,j}\) \(\epsilon_{ij}^{(s')} (\hat q) \, \hat k_{2\,,i} \hat k_{2\,j}\) \,,\ee
where $\epsilon_{ij}^{(s)} (\hat q)$ is the helicity-$s$ polarization tensor. 
It is easy to estimate the prefactors, for example
\be
\begin{split}
\mu\mu\;{\rm interaction.\;fig~\ref{AAA}} & \qquad T_0 \sim \(\frac{\mu}{H}\)^2 \(\frac{\rho}{H \sqrt{\epsilon}}\)^4 \Delta_\zeta^{-2} \frac{1}{c_{i}^{2\nu} \, c_0 ^{4\nu}}\,, \\
\tilde\rho \tilde\rho\;{\rm interaction.\;fig~\ref{BBB}} & \qquad T_0 \sim \(\frac{\tilde\rho}{H \sqrt{\epsilon}}\)^2 \frac{1}{c_i^{2\nu}} \;,
\end{split}
\ee
where the $c_i$ stands for the speed of propagation of the helicity exchanged in the horizontal propagator.
Notice that the $\tilde\rho\tilde\rho$ diagram is present even in the absence of the mixing $\propto\rho$.

\begin{figure}[h] 
\hspace{\fill}
\subfloat[][\label{AAA}]{\includegraphics[scale=.19]{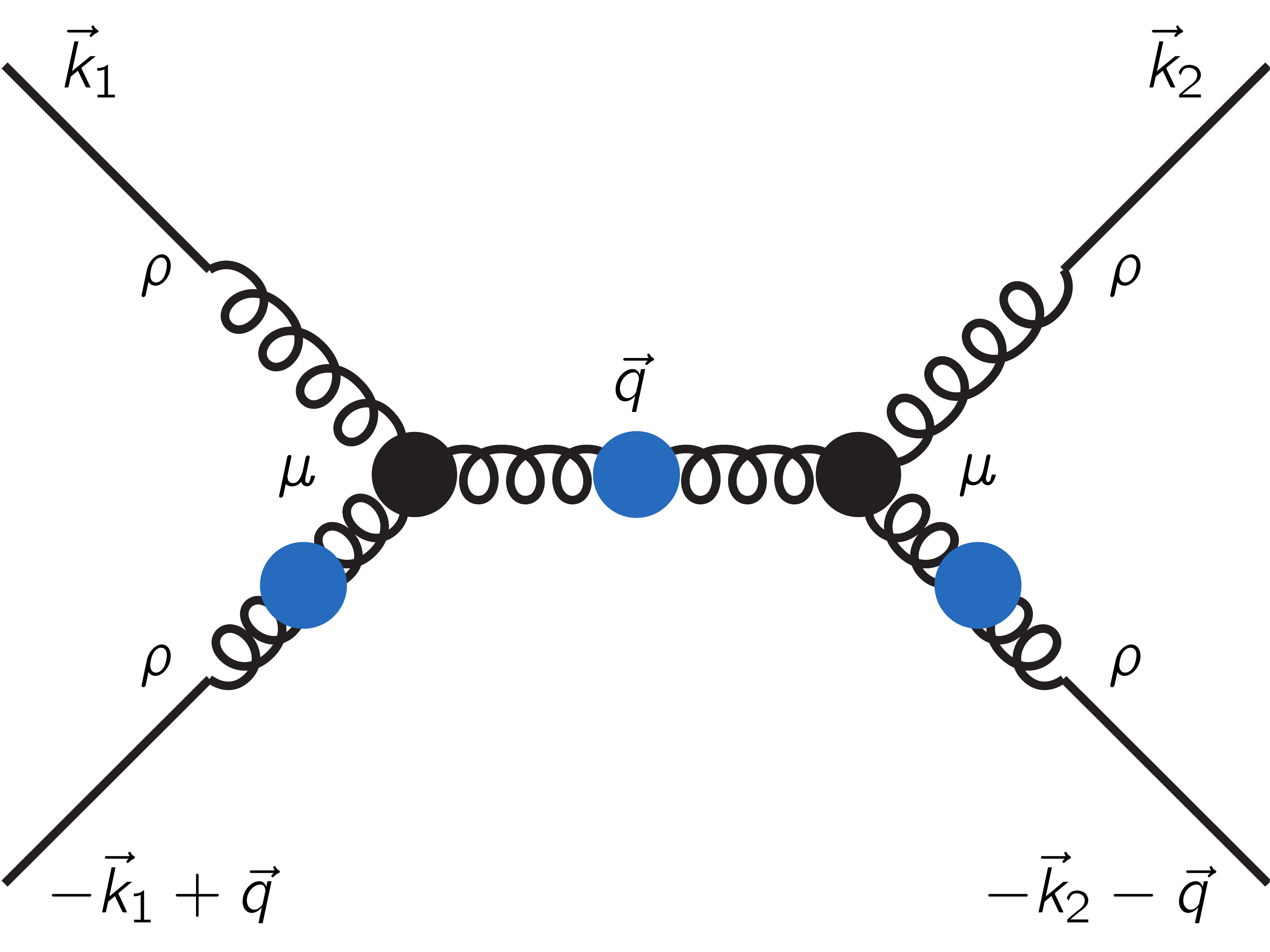}} \hspace{\fill}
\subfloat[][\label{BBB}]{\includegraphics[scale=.19]{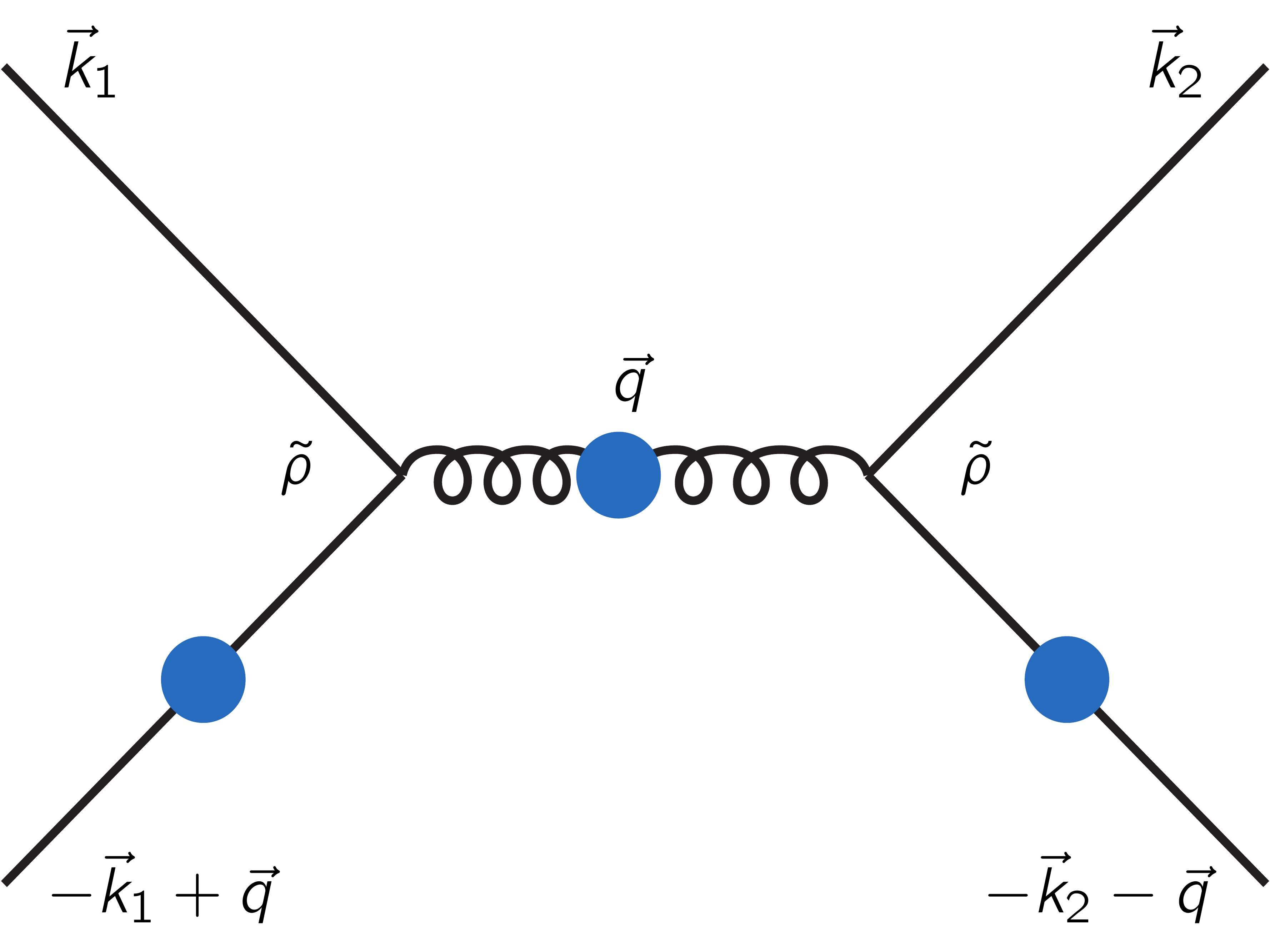}} \hspace{\fill}
\caption{\small Leading contribution to the $\expect{\zeta\zeta \zeta \zeta}$. }
\end{figure}

{\bf Large tensor non-Gaussianities.} The presence of a light helicity-2 state during inflation changes the prediction for tensor modes. In particular one can consider a regime where the tensor spectrum is dominated by the mixing with $\sigma$, while the correction to the scalar power spectrum remains small. 
If the $\sigma$ sector is quite non-Gaussian, one will have large non-Gaussianity for tensors, while scalar perturbations may remain close to Gaussian, as required by experiments. For concreteness let us focus on the cubic term proportional to $\mu$ and see how it affects the various 3-point functions. Schematically, the deviations from a Gaussian statistics are given by the following dimensionless estimates:
\be
\begin{split}
\frac{\expect{\gamma\gamma\gamma}}{\Delta_\gamma^3} & \sim \frac{\mu}{H c_2^{3/2}} \; ,\\
\frac{\expect{\gamma\zeta\zeta}}{\Delta_\gamma \, \Delta_\zeta^2} & \sim \frac{\mu}{H c_2^{3/2}} \left(\frac{\rho}{\sqrt{\epsilon} H}\right)^2 \frac{1}{c_0^3} \; , \\
\frac{\expect{\zeta\zeta\zeta}}{\Delta_\zeta^3} & \sim \frac{\mu}{H c_2^{3/2}} \left( \left(\frac{\rho}{\sqrt{\epsilon} H}\right)^2 \frac{1}{c_0^3} \right)^{3/2} \left(\frac{c_2}{c_0}\right)^{3/2}  \;.
\end{split}
\ee
In these estimates we assumed that the tensor power spectrum is dominated by the $\sigma$ exchange (while in eq.~\eqref{tensor_estimates} we assumed that the $\sigma$ exchange was only a correction to the standard prediction). When $\mu/(H c_2^{3/2})$
approaches unity, tensor fluctuations become strongly non-Gaussian. This can happen keeping the other correlators involving $\zeta$ close to a Gaussian statistics. Indeed $(\rho/\sqrt{\epsilon} H)^2 \cdot 1/c_0^3 \ll 1$ if one wants the scalar mixing with $\sigma$ not to change significantly the scalar power spectrum (see eq.~\eqref{power_estimates}) and this makes the non-Gaussianity in $\expect{\gamma\zeta\zeta}$ subdominant. The same will happen for $\expect{\zeta\zeta\zeta}$ in the regime $c_2 \lesssim c_0$. 

In the rest of this Section we are going to confirm some of these estimates by explicit calculations.

\subsection{Power Spectra}
\label{PPS}

In this section we will compute explicitly the leading corrections to both the power spectra of curvature perturbation and of gravitational waves. 
For, simplicity we will do the computations only in the limit $c_0,\,c_2\ll1$. 
As we will show this is the most interesting limit since in both cases the corrections can dominate the power spectra, being proportional to $c_0^{-{2\nu}}$ and $c_2^{-{2\nu}}$ respectively.
Denoting with $X$ the perturbation $\zeta$ or $\gamma$, the leading correction to their power spectra is given, using the in-in formalism, by
\bea\label{power_spectrum_corr}
\expect{X_{\vec k} \ X_{-\vec k}} \!\!\! &=& \int_{-\infty}^0 \!\!\!\!\! d\eta \!\! \int_{-\infty}^\eta \!\!\!\!\! d\eta' \, \expect{\[ H_{\rm int}^{X\s}(\eta), \, \[ H_{\rm int}^{X\s}(\eta'), \, X_{\vec k}(0) \, X_{-\vec k}(0) \]\]} \nonumber \\
&=& \int_{-\infty}^0 \!\!\!\!\! d\eta \!\! \int_{-\infty}^0 \!\!\!\!\! d\eta' \, \expect{H_{\rm int}^{X\s}(\eta) \, X_{\vec k} X_{-\vec k} H_{\rm int}^{X\s}(\eta')} 
-2 {\rm Re} \[\int_{-\infty}^0 \!\!\!\!\! d\eta \!\! \int_{-\infty}^\eta \!\!\!\!\! d\eta' \, \expect{X_{\vec k} X_{-\vec k} H_{\rm int}^{X\s}(\eta) \,  H_{\rm int}^{X\s}(\eta')}\] \nonumber\,.\\
\eea
$H_{\rm int}^{X\s}(\eta)$ denotes the interaction Hamiltonian between $X$ and $\s$. This interaction is proportional to $\rho$ both for scalars and tensors: see the first line of eq.~\eqref{intercanonical}. 

\paragraph{Power spectrum of curvature perturbations.}
Let us compute the contribution to $\expect{\zeta\zeta}$. Substituting the wavefunctions in eq.~\eqref{power_spectrum_corr} we find
\be
P_{\zeta}(k) = \frac{H^2}{4 \mpl^2\epsilon \ k^3} \(1 + \frac{\mathcal C_\zeta(\nu)}{\epsilon \, {c_0}^{2\nu} } \(\frac{\rho}{H}\)^2\)\,
\ee
where,
\bea
\mathcal C_\zeta(\nu) &\equiv& \mathcal C_{\zeta,\,1}(\nu) + \mathcal C_{\zeta,\,2}(\nu) \,, \\
\label{C_1_zeta} \mathcal C_{\zeta,\,1}(\nu) &\equiv& \frac{\pi}{6} \, {c_0}^{2\nu} \ \left| \int_{0}^\infty \!\! dx\  \frac{e^{-i x}\, (1+i x) \, H^{(2)}_\nu (c_0\, x) }{\sqrt x}  \right|^2\,, \\ 
\label{C_2_zeta} \mathcal C_{\zeta,\,2}(\nu) &\equiv& -\frac{\pi}{3} \, {c_0}^{2\nu} \  {\rm Re}\[\int_{0}^\infty \!\! dx\  \frac{e^{-i x}\, (1+i x) \, H^{(1)}_\nu (c_0\, x) }{\sqrt x} \!\! \int_{x}^\infty \!\! dy\  \frac{e^{-i y}\, (1+i y) \, H^{(2)}_\nu (c_0\, y) }{\sqrt y} \]\,. \nonumber \\
\eea
For $c_0 \ll 1$ the integrals can be computed analytically. We get
\be\label{C_zeta_nu}
\mathcal C_\zeta  = \frac{2^{2\nu -3} \, (3-2 \nu )^2 \, \Gamma \left(\frac{1}{2}-\nu \right)^2 \Gamma (\nu )^2  \, (1-\sin (\pi  \nu ))}{3 \pi  }\,.
\ee
\begin{figure}[h] 
\centering
\includegraphics[scale=.8]{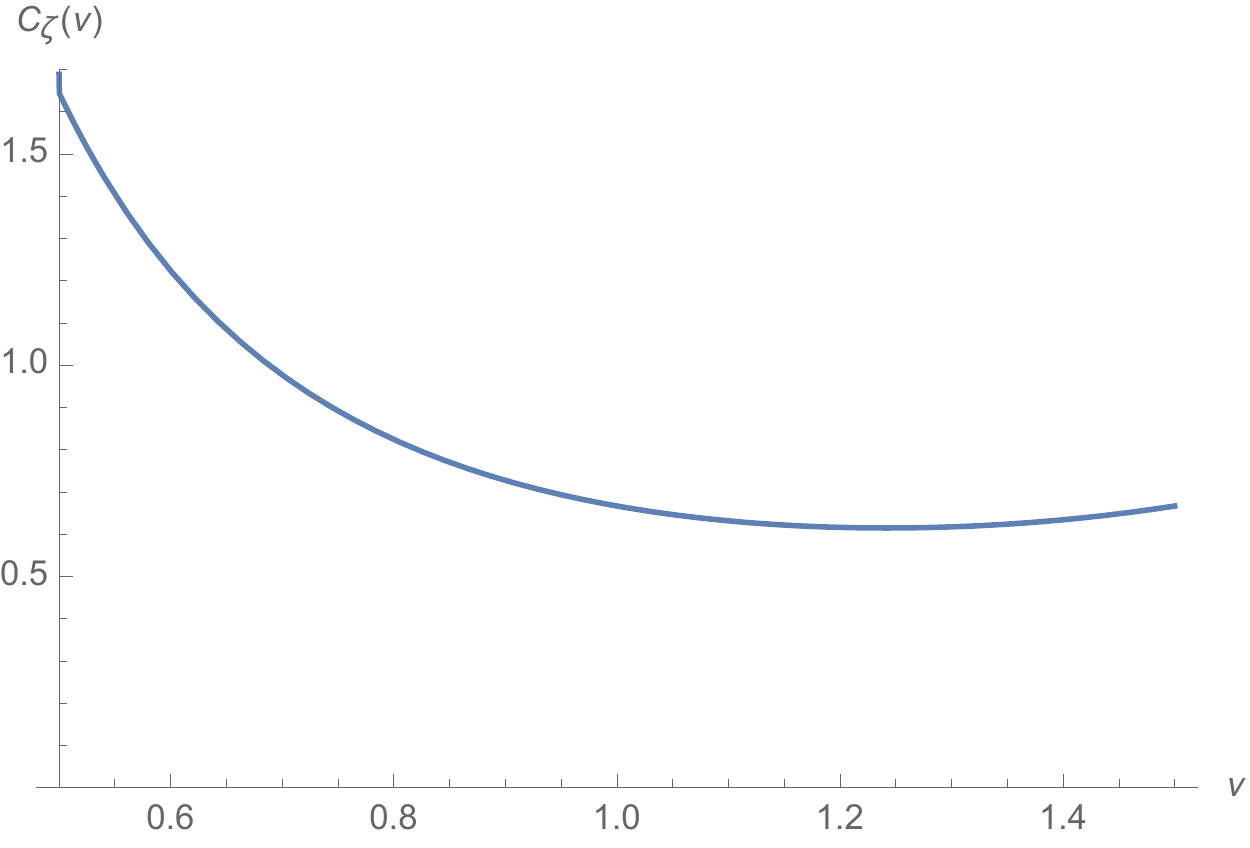}
\caption{\small $\mathcal C_\zeta$ as a function of $\nu = \sqrt{\frac 9 4 - \left( \frac m H \right)^2}$, in the range of masses below the Higuchi bound: $\nu \in [\frac 1 2, \frac 3 2]$. \label{Fig_C_zeta}}
\end{figure} 

\noindent
figure \ref{Fig_C_zeta} shows the plot of $C_\zeta$ as a function of $\nu$ in the mass range below the Higuchi bound.
In the massless case one gets $\mathcal C_\zeta(\nu=3/2) = 2/3$.
The expression in eq.~\eqref{C_zeta_nu} receives relative corrections of order $c_0^{2\nu}$ so it should not be trusted for small $\nu$.   
More details about the computation can be found in App.~\ref{power_spectra}.

\paragraph{Power spectrum of tensor perturbations.}
Let us move now to the computation of the tensor power spectrum. The contribution to $\expect{\gamma\gamma}$ is due to an exchange of a helicity-2 mode $\sigma^{(2)}_{ij}$. Substituting the wavefunctions in eq.~\eqref{power_spectrum_corr} we find
\be\label{p_gamma_c}
P_{\gamma}(k) = \frac{4 H^2}{ \mpl^2 \ k^3} \(1 + \frac{ \mathcal C_\gamma(\nu)}{{c_2}^{2\nu}} \(\frac{\rho}{H}\)^2\)\,,
\ee
where,
\bea
\mathcal C_\gamma(\nu) &\equiv&  \mathcal C_{\gamma,\,1}(\nu) + \mathcal C_{\gamma,\,2}(\nu) \,, \\
\label{C_1_gamma}\mathcal C_{\gamma,\,1}(\nu) &\equiv&  \frac \pi 2 \, {c_2}^{2\nu} \ \left| \int_{0}^\infty \!\! dx\  \frac{e^{-i x} \, H^{(2)}_\nu (c_2\, x) }{\sqrt x}  \right|^2\,, \\ 
\label{C_2_gamma} \mathcal C_{\gamma,\,2}(\nu) &\equiv& - \pi \, {c_2}^{2\nu} \ {\rm Re}\[\int_{0}^\infty \!\! dx\  \frac{e^{-i x} \, H^{(1)}_\nu (c_2\, x) }{\sqrt x} \!\! \int_{x}^\infty \!\! dy\  \frac{e^{-i y} \, H^{(2)}_\nu (c_2\, y) }{\sqrt y}\] \,. 
\eea
Even in this case the integrals can be computed analytically in the limit $c_2\ll1$. For some details about the computation we refer the reader to App.~\ref{power_spectra}. The final result is
\be\label{c_gamma}
\mathcal C_\gamma  = \frac{2^{2 \nu -1} \, \Gamma \left(\frac{1}{2}-\nu \right)^2  \, \Gamma (\nu )^2 (1-\sin(\pi \nu))}{\pi} \,.
\ee
\begin{figure}[h] 
\centering
\includegraphics[scale=.8]{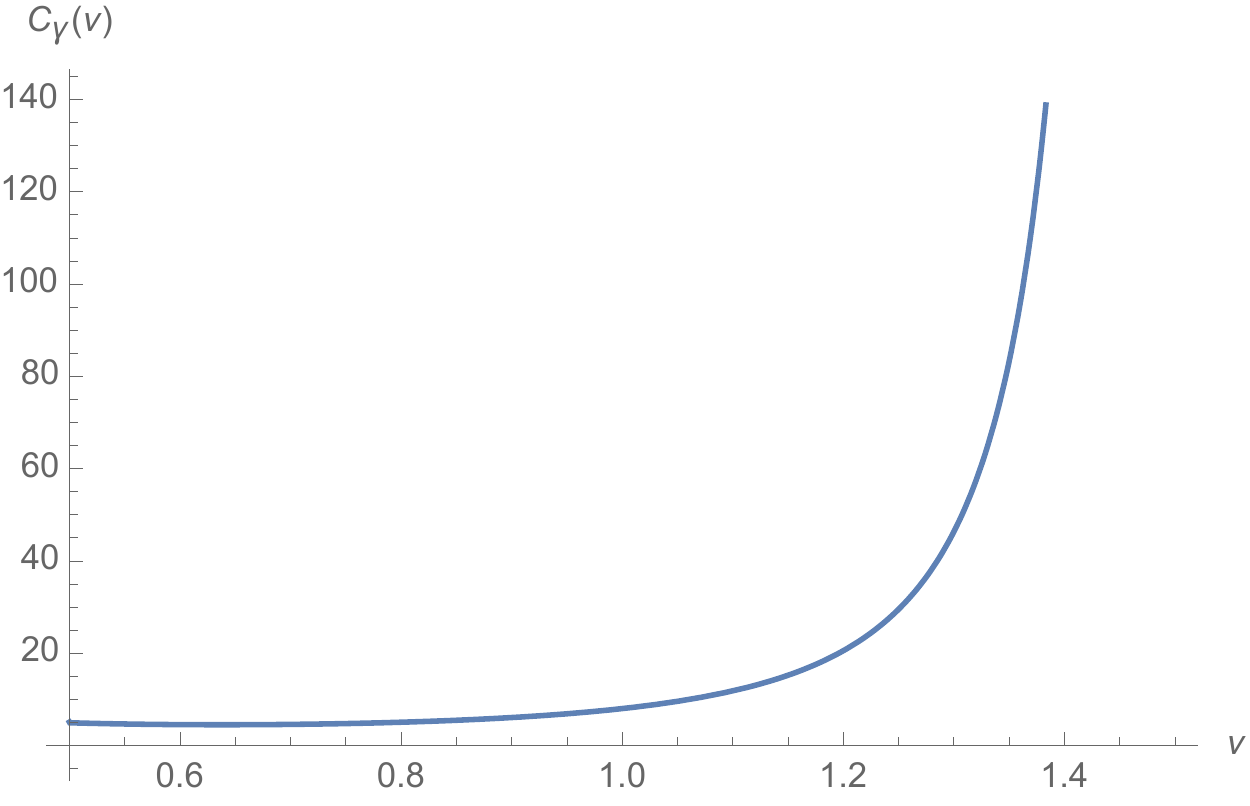}
\caption{\small $\mathcal C_\gamma$ as a function of $\nu = \sqrt{\frac 9 4 - \left( \frac m H \right)^2}$, in the range of masses below the Higuchi bound: $\nu \in [\frac 1 2, \frac 3 2]$.  \label{Fig_C_gamma}}
\end{figure} 

\noindent
In figure \ref{Fig_C_gamma} we plot $C_\gamma$ as function of $\nu$ in the mass range below the Higuchi Bound.
Notice that the result in this case diverges in the massless limit since the mixing does not turn off outside the horizon.
The divergence will be eventually regulated by the finite duration of inflation. 
For small mass the time evolution of $\s$ outside the horizon is $\s \propto (-\eta)^{m^2/3H^2} \sim e^{- N m^2/3H^2}$, where $N$ is the number of e-folds to the end of inflation. Therefore the result \eqref{c_gamma} is accurate only for $(m/H)^2 \gg 1/N$.
One can take into account the finite duration of inflation cutting off the integrals eqs.~\eqref{C_1_gamma} and  \eqref{C_2_gamma}.
In the massless case $\nu=3/2$ one gets
\be\label{eq:p_gamma}
P_{\gamma}(k) = \frac{4 H^2}{ \mpl^2 \ k^3} \(1 + \frac{2}{{c_2}^{3}} \(\frac{N\, \rho}{H}\)^2\)\,.
\ee
The enhancement in the massless limit can be quite sizeable: $2N^2\sim 10^4$.

If $P_\gamma$ is dominated by the $\s$ exchange, while the correction to the scalar spectrum is small, the tensor-to-scalar ratio $r$ is given by\footnote{In the regime in which also $P_\zeta$ is dominated by the $\s$ exchange we have
\be
r \simeq 48 \(\ep N\)^2 \( \frac{c_0}{c_2} \)^3\,.
\ee }
\begin{equation}
r \equiv \frac{P_\gamma}{P_\zeta} \simeq 32 \, \left(\frac{c_0}{c_2}\right)^3 \, \left( \left(\frac{\rho}{\sqrt{\epsilon} H}\right)^2 \frac{1}{c_0^3} \right)  \, \left(\epsilon \, N \right)^2 \; .
\end{equation}
Notice that the tilt of the tensor power spectrum is given by the usual formula, $n_t = -2 \epsilon$, in the regime $(m/H)^2 \gg 1/N$ (although the tensor to scalar ratio is now not fixed in term of $\epsilon$). In the regime of smaller masses one has the extra $N$ dependence of eq.~\eqref{eq:p_gamma}, which gives an additional negative contribution to the tilt:
\be
n_t = -2 \epsilon - \frac{2}{N} \;.
\ee
This formula neglects the possible time dependence of $c_2$: this is arbitrary and can make tensor modes large on the short scales observed by interferometers. This opens the possibility of studying the statistics of primordial gravitational waves on short scales, see e.g.~\cite{Bartolo:2018qqn}. 

As it is discussed in Appendix~\ref{gamma_sigma_model}, one can exactly solve the coupled equations of $\gamma$ and $\s$ instead of treating the mixing pertubatively. The mixing term effectively gives rise to an additional mass term of order $\rho^2$, however this effect is never relevant since we are always in the regime $(\rho/H)^2 \ll \ep \lesssim 1/N$.

\subsection{Bispectra}

\subsubsection{Squeezed limit}

In this section we compute the squeezed limit of $\expect{\zeta \zeta \zeta}$ and $\expect{\gamma\zeta \zeta}$.
For simplicity we assume $\rho \ll \tilde \rho$ and take $\mu =0$. 
In this way we can just focus on the contributions of fig.~\ref{XXX_zeta} for $\expect{\zeta \zeta \zeta}$ and of fig.~\ref{XXX_gamma} for $\expect{\gamma\zeta \zeta}$. 
In the squeezed limit, the long mode leaves the horizon much earlier than the short ones. 
We can then split the computation of the bispectrum in two:  first we look at the effect of the long mode on the small scales and then we compute the mixing between $\s$ and the soft field, $\zeta$ or $\gamma_{ij}$.

Let $X_{\vec q}$ be the long mode that mixes with $\sigma$. 
Leaving the horizon much earlier than the other two modes, $X_{\vec q}$ acts on the smaller scales as a classical background. 
Working in spatially flat gauge we have
\be
\expect{X_{\vec q} \, \pi_{\vec k} \, \pi_{-\vec k - \vec q}}_{\, \vec q\to0} = 
\expect{X_{\vec q}\  \expect{\pi_{\vec k} \, \pi_{-\vec k}}_{\s_b} }\,,
\ee
where  $\langle\pi_{\vec k} \, \pi_{-\vec k}\rangle_{\s_b}$  is the power spectrum of $\pi$ modulated by the long $\s$ mode. It can be expanded in a power series in terms of the long mode,
\be\label{modul_curv_power}
\expect{\pi_{\vec k} \, \pi_{-\vec k}}_{\s_b} \simeq \expect{\pi_{\vec k} \, \pi_{-\vec k}} + \sum_s \s^{(s)}_{\vec q} \,\frac{\expect{\s^{(s)}_{\vec q} \, \pi_{\vec k} \, \pi_{-\vec k}}}{P_{\s^{(s)}}(q)}\,.
\ee
Bracketing eq.~\eqref{modul_curv_power} with the soft $X_{\vec q}$ we get the leading contribution to the squeezed 3-point function,  
\be\label{squeezed_bis}
\expect{ X_{\vec q} \, \zeta_{\vec k} \, \zeta_{-\vec k - \vec q}}'_{\vec q \to 0} = H^2  \frac{\expect{X_{\vec q} \, \s^{(s)}_{- \vec q}} '\expect{\sigma^{(s)}_{\vec q} \, \pi_{\vec k} \, \pi_{- \vec k}}'}{P_{\s^{(s)}}(q)}\,.
\ee
where we used the fact that $\zeta = - H \pi\,.$
We use the in-in formalism to compute the correlator $\langle X\s^{(s)} \rangle$ and the 3-point function $\langle \s^{(s)}\pi\pi \rangle$, we get
\bea\label{spp_bispectrum}
\expect{\s^{(s)}_{\vec q} \, \pi_{\vec k} \, \pi_{- \vec k}} \!\!\! &=& \!\!\! -i \int_{-\infty}^0 \!\! d\eta' \langle \[ \hat\s^{(s)}_{\vec q} \, \hat\pi_{\vec k} \, \hat\pi_{- \vec k}\,,\ H^{\pi\pi\s}_{\rm int}(\eta') \] \rangle \\
&=& \!\!\! -4 \ \tilde \rho \, \mpl \, k^2 \ \(\s^{(s)}_q (\eta_*) \, \pi_k (\eta_*) \, \pi_k (\eta_*)\) {\rm Im}\!\[ \int_{-\infty}^{\eta_*} \!\! d\eta \, a\  {\sigma_q^{(s)}}^*(\eta) \, {{\pi_k}^*}'(\eta) \, {\pi_k}^*(\eta) \] \epsilon^{(s)}_{ij}(\hat q) \, \hat k_i \hat k_j\,. \nonumber
\eea
and
\bea\label{sx_corr}
\expect{X_{-\vec q} \, \s^{(s)}_{\vec q}} \!\!\! &=& \!\!\! -i \int_{-\infty}^{\eta_*} \!\! d\eta \langle \[ \hat\s^{(s)}_{\vec q} \, \hat\pi_{-\vec q} \,,\ H^{X\s}_{\rm int}(\eta) \] \rangle \,.
\eea
{Here $\eta_*$ is the conformal time at the end of inflation.}
In the next two subsections we report the expression for both the scalar and the tensor squeezed bispectra as a function of the mass and the speed of propagation of $\s$. 
The details of the calculations can be found in App.~\ref{bispectra}.

\paragraph{Scalar bispectrum in the squeezed limit.}

Only the helicity-0 component of $\s$ can mix with the soft $\zeta$.
Therefore we need to evaluate the mixing $\langle\pi_{\vec q} \, \s^{(0)}_{-\vec q}\rangle$ and  then the  3-point function $\langle \s^{(0)}_{\vec q} \, \pi_{\vec k} \, \pi_{-\vec k-\vec q} \rangle$.

The expression for the mixing $\langle\pi_{\vec q} \, \s^{(0)}_{-\vec q}\rangle$ cannot be written analytically for generic $c_0$. In the limit $c_0\ll1$ one has 
\be\expect{\s^{(0)}_{\vec q} \, \pi_{-\vec q}}' = \frac{d_\pi(\nu)}{c_0^{2\nu}}_{\vec q \to 0} \ \mpl \ \rho \ (-q \, \eta_*)^{3/2-\nu} \ P_\pi (k)\,,
\ee
with $d_\pi(\nu)$ given in eq.~\eqref{d_pi_nu}.
The 3-point correlation function $\langle\s^{(0)}_{\vec q} \, \pi_{\vec k} \, \pi_{-\vec k-\vec q}\rangle$ is given by
\be \label{squeezed_sigma_0_pi_pi}
\expect{\s^{(0)}_{\vec q} \, \pi_{\vec k} \, \pi_{- \vec k}} ' _{\vec q \to 0}= \frac{\sqrt 3 \, c(\nu)}{\epsilon \ \mpl} \ \frac{\tilde \rho}{H} \ (-k\,\eta_*)^{-\frac{3}{2}+\nu} \ P_{\sigma^{(0)}}(q) \, P_\pi (k) \((\hat q \cdot \hat k)^2-\frac{1}{3} \)\,.
\ee
The coefficient $c(\nu)$ is given in eq.~\eqref{c_nu} and the power spectrum of $\s$ is (see eq.~\eqref{late_sigma})
\be\label{sigma_power_spectrum}
P_{\sigma^{(s)}}(q) = \frac{2^{2\nu-2} \, \Gamma(\nu)^2 \, H^2}{\pi} \  \frac{(-\eta_*)^{3-2\nu}}{(c_s \, q)^{2\nu}}\,, \ \ \ \ \ s=0\,,1\,,2\,.
\ee 
The final expression of the curvature bispectrum is then,
\be\label{zeta_bis_gen_mass}
\expect{\zeta_{\vec q} \zeta_{\vec k} \zeta_{-\vec k}}' _{\vec q \to 0}= \sqrt 3 \ \frac{ \mathcal F_\pi(\nu)}{c_0^{2\nu}}\ \frac{\rho \, \tilde \rho}{\epsilon \, H^2} \ \(\frac{q}{k}\)^{\frac 3 2-\nu} P_{\zeta}(q)\, P_{\zeta}(k)  \((\hat q \cdot \hat k)^2-\frac{1}{3} \)\,,
\ee
where again we used the leading relation among $\zeta$ and $\pi$, $\zeta = -H \pi$. 
Notice that, even if $\langle\s\pi\rangle$ and $\langle\s\pi\pi\rangle$ decay in time, their product has the exact time dependence of $P_{\s^{(0)}}$ so that the contribution due to an exchange of $\s$ to the squeezed scalar bispectrum is time-independent, as expected.
The function $\mathcal F_\pi(\nu)$ is plotted in fig.~\ref{Fig_F_pi} in the mass range below the Higuchi bound. Its expression is given by
\be
\mathcal F_\pi(\nu) = d_\pi(\nu)\, c(\nu)\,.
\ee
\begin{figure}[h] 
\centering
\includegraphics[scale=.8]{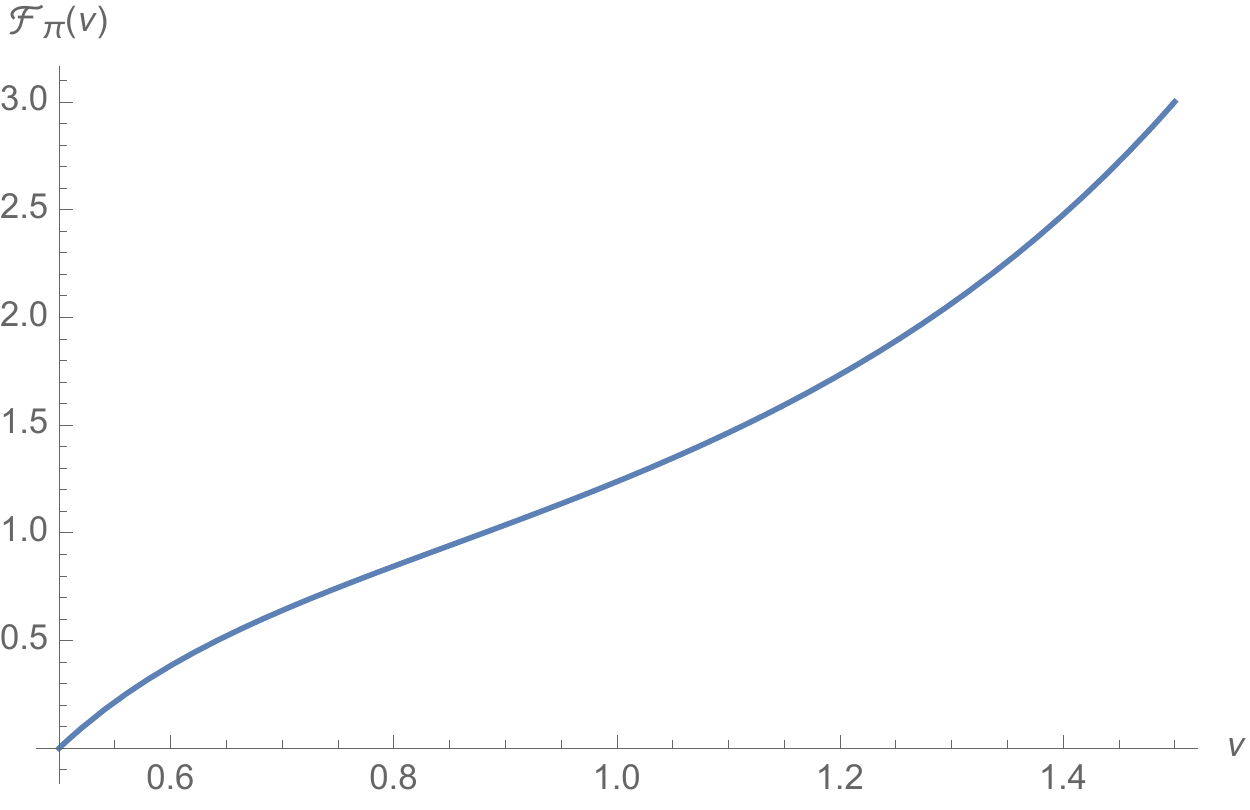}
\caption{\small $\mathcal F_\pi$ as a function of $\nu = \sqrt{\frac 9 4 - \left( \frac m H \right)^2}$, in the range of masses below the Higuchi bound: $\nu \in [\frac 1 2, \frac 3 2]$. \label{Fig_F_pi}}
\end{figure} 
\noindent
The bispectrum $\expect{\zeta_{\vec q \to 0}\zeta\zeta}$ can be exactly calculated when $\sigma$ is massless for any value of the sound speed  $c_0$: 
\be
\expect{\zeta_{\vec q} \, \zeta_{\vec k} \, \zeta_{-\vec k}}' _{\vec q \to 0}=  3 \, \frac{1+c_0 + c_0^2}{c_0^3 \, (1+c_0)} \,  \frac{\rho \, \tilde \rho}{\epsilon \, H^2} \, P_{\zeta}(q) P_{\zeta}(k) \ \((\hat q \cdot \hat k)^2-\frac{1}{3} \)\,.
\ee

\paragraph{Tensor-scalar-scalar bispectrum in the squeezed limit.}

Let us now compute the squeezed tensor-scalar-scalar bispectrum.
The expression of $\langle\s\gamma\rangle$, in the limit $c_2\ll1$ is
\be\label{s-g_corr_gen_mass}
\langle \s^{(2)}_{\vec q} \, \gamma_{-\vec q} \rangle' = \frac{d_\gamma(\nu)}{c_2^{2\nu}}\ \mpl \ \frac{\rho}{H}\ (-q\,\eta_*)^{\frac{3}{2}-\nu} \ P_\gamma(q)\,.
\ee
The coefficient $d_\gamma(\nu)$ is given in eq.~\eqref{d_gamma_nu}.
Notice that, although the computation of the above expression has been carried out explicitly   assuming $c_2\ll1$, one can check that eq.~\eqref{s-g_corr_gen_mass} is  a good approximation {i.e. ({$\mathcal O (1)$)} even if $c_2\simeq 1$ in the range $1/2 < \nu < 3/2\,,$ i.e. the most interesting mass range, since it corresponds to the masses below the Higuchi Bound.
The correlator $\expect{\s^{(\pm2)}\pi\pi}$ is, 
\be\label{squeezed_sigma_2_pi_pi}
\expect{\s^{(\pm 2)}_{\vec q} \, \pi_{\vec k} \, \pi_{- \vec k}} ' _{\vec q \to 0}=  \frac{c(\nu)}{\epsilon \ \mpl} \ \frac{\tilde \rho}{H} \ (-k\,\eta_*)^{-\frac{3}{2}+\nu} \ P_{\sigma^{(2)}}(q) \, P_\pi (k) \ \epsilon_{ij}^{(\pm2)} (\hat q) \, \hat k_i \hat k_j \,,
\ee 
with $c(\nu)$ given in eq.~\eqref{c_nu}.
The expression of the squeezed $\langle \gamma\zeta\zeta\rangle$ bispectrum for a generic mass of the field $\sigma_{ij}$ is
\be
\expect{\gamma^{(s)}_{\vec q} \, \zeta_{\vec k} \, \zeta_{- \vec k}} ' _{\vec q \to 0}=  \frac{\mathcal F_\gamma (\nu)}{c_2^{2\nu}}\ \frac{\rho\,\tilde \rho}{\epsilon \, H^2}\ \(\frac{q}{k}\)^{\frac{3}{2}-\nu} P_{\gamma}(q) \ P_\zeta (k) \ \epsilon_{ij}^{(s)} (\hat q) \, \hat k_i \hat k_j\,,
\ee
being,
\be
\mathcal F_{\gamma}(\nu) = - d_\gamma(\nu)\,c(\nu)\,.
\ee
\begin{figure}[h] 
\centering
\includegraphics[scale=.8]{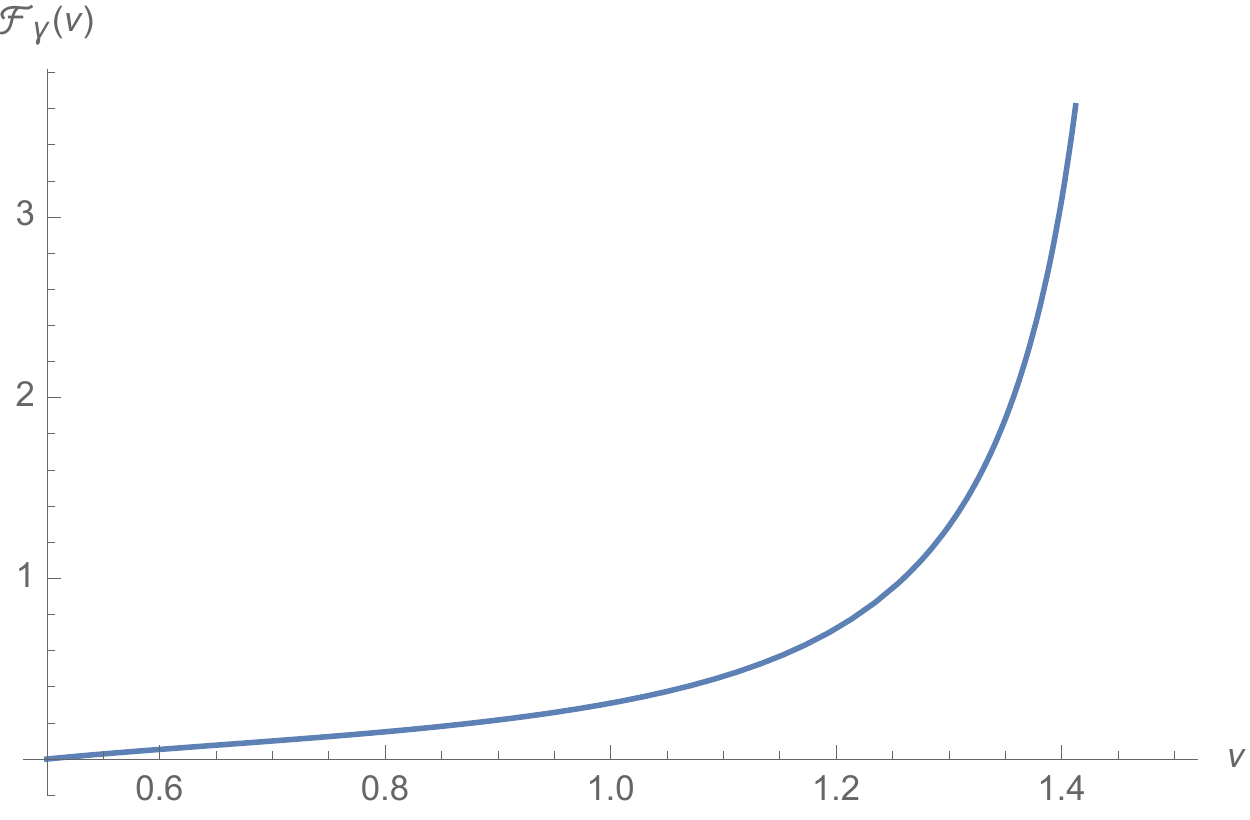}
\caption{\small $\mathcal F_\gamma$ as a function of $\nu = \sqrt{\frac 9 4 - \left( \frac m H \right)^2}$, in the range of masses below the Higuchi bound: $\nu \in [\frac 1 2, \frac 3 2]$. \label{Fig_F_gamma} }
\end{figure} 
\noindent
The function $\mathcal F_\gamma(\nu)$ is plotted in fig.~\ref{Fig_F_gamma} in the mass range below the Higuchi bound.
The squeezed bispectrum diverges in the limit $\nu \to 3/2$. The situation is analogous to what we have already discussed at the end of sec.~\ref{PPS}. 
The divergence should be trusted up to $m^2/H^2 \lesssim 1/N$.
For smaller masses one must keep the number of e-folds finite.
In the massless limit the computation of $\langle\gamma^{(s)}_{\vec q \to 0} \zeta_{\vec k} \zeta_{-\vec k}\rangle'$ can be done substituting the massless wavefunctions  directly into eqs.~\eqref{spp_bispectrum} and \eqref{sx_corr}. 
\be
\expect{\gamma^{(s)}_{\vec q} \, \zeta_{\vec k} \, \zeta_{-\vec k}}'_{\vec q \to 0} =  \frac34\ \frac{N_q}{c_2^3} \frac{\rho \tilde \rho}{\epsilon \, H^2} \, P_\gamma(q) \ P_\zeta(k) \ \epsilon^{(s)}_{ij}\hat k_i \hat k_j\,.
\ee

\subsubsection{Scalar three-point function in any configuration}

The contribution to the scalar bispectrum, due to a { spinning} particle, can be computed for any configuration of the momenta $\vec k_i$ without much effort if $\s$ is massless and in the limit $c_0\ll1$.
Using the in-in formalism, the expression for $\expect{\zeta\zeta\zeta}$ is given by
\bea
\expect{\zeta_{\vec k_1} \zeta_{\vec k_2} \zeta_{\vec k_3}}' &=& H^3 \int_{-\infty}^{0}d\eta \int_{-\infty}^{\eta}d\tilde\eta \[ \expect{\H(\tilde\eta) \, \H(\eta) \, \pi_{\vec k_1} \, \pi_{\vec k_2} \, \pi_{\vec k_3}} + \expect{ \pi_{\vec k_1} \, \pi_{\vec k_2} \, \pi_{\vec k_3} \, \H(\eta) \, \H(\tilde\eta)} \] \nonumber \\
&&\hspace{4.3cm} - H^3 \int_{-\infty}^{0}d\eta \int_{-\infty}^{0}d\tilde\eta \  \expect{\H(\eta) \, \pi_{\vec k_1} \, \pi_{\vec k_2} \, \pi_{\vec k_3} \, \H(\tilde\eta) } \,,
\eea
where $\H \equiv \H^{(2)} + \H^{(3)}$ denotes the interaction hamiltonian (see eq.~\eqref{intercanonical}).
Performing the time integrals we get
\be
B_\zeta(k_1, k_2, k_3) = - \frac{8\pi^4}{{c_0}^3} \ \frac{\rho \, \tilde \rho}{\ep H^2} \ \frac{\mathcal I (k_1, k_2, k_3)}{{\Delta_\zeta}^4}\,, 
\ee
with $\Delta_\zeta^2 \equiv k^3 P_\zeta(k)/2\pi^2$ and $\mathcal I (k_1, k_2, k_3)$ given by
\be
\mathcal I (k_1, k_2, k_3)  = \frac{k_2 + 2 k_3}{k_1^3 \, k_2 \, k_3 \, (k_2 + k_3)^2} \ \( (\hat k_1 \cdot \hat k_3)^2 - \frac 1 3\) + 5 \ perms\,.
\ee
To analyze the shape of the bispectrum for general momentum configurations it is convenient to define the dimensionless shape function
\be
S(k_1, k_2, k_3) \equiv (k_1 \, k_2 \, k_3)^2 \times \(\mathcal I (k_1, k_2, k_3) + 5 \ perms \)\,.
\ee
The characteristic feature  of this shape is its angular dependence due to the exchange of the higher spin particle: we expect a modulation that approaches the Legendre polynomial $P_2( \vec k_1 \! \cdot \! \vec k_3)$ as we approach the squeezed configuration $k_1\ll k_3$. 
One natural question that arises is how much the triangle has to be squeezed in order to see this behaviour.
figure~\ref{general_bispectrum} shows the shape of the total signal as a function of the angle between the modes $\vec k_1$ and $\vec k_3$, $\theta \equiv \cos^{-1}(\vec k_1 \cdot \vec k_3)$, for a range of momentum configurations with fixed $k_1/k_3$. 
Notice that the signal does not deviate much from the $P_2(\cos \theta)$ in the range $k_1/k_3 \lesssim 0.5$.
As the triangle approaches the equilateral shape ($k_1/k_3 \lesssim 1$) the angular dependence deviates from the pure Legendre behaviour : the peak around $\theta = \ang{180}$ becomes prominent while its width shrinks. 
This happens because for $k_1/k_3 \, \simeq \, 1$ and $\theta \simeq \ang{180}$ the triangle squeezes since $k_2\to0$ making $S(\vec k_1, \vec k_2, \vec k_3)$ diverge.
\begin{figure}[t] 
\centering
\includegraphics[scale=1.35]{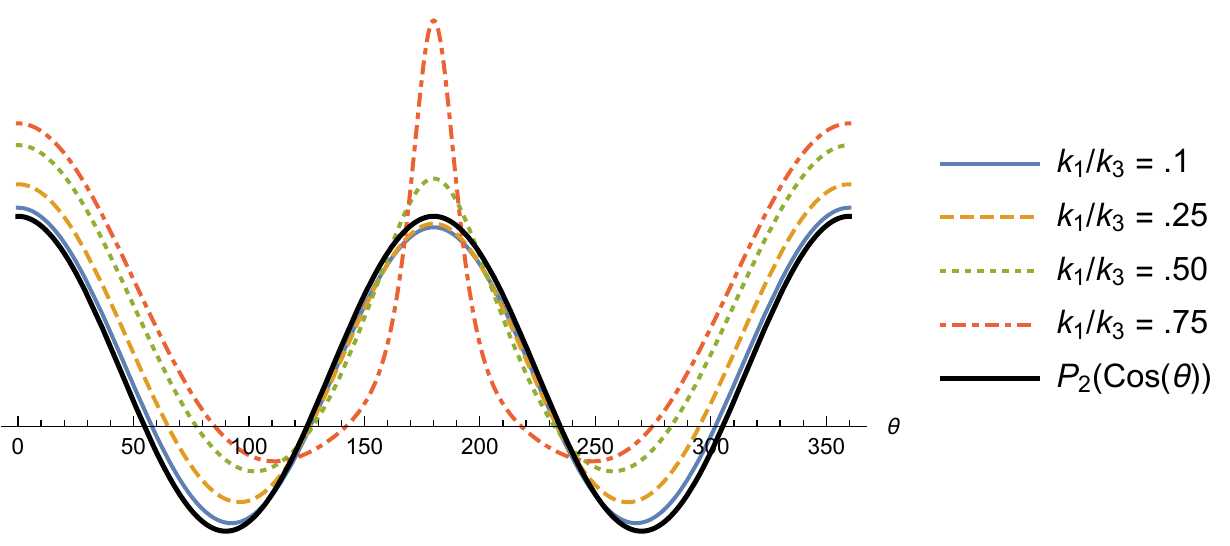}
\caption{\small Bispectrum shape function $S(k_1,k_2,k_3)$ as a function of the angle $\theta = \cos^{-1}(\hat k_1 \! \cdot \! \hat k_3)$ for fixed ratio $k_1/k_3$. For easy comparison, the plot has been normalized such that the height difference between $\theta = \ang 0$ and $\theta = \ang{90}$ of each curve is fixed to $ \frac 3 2$.  }
\label{general_bispectrum}
\end{figure}

\subsection{Trispectrum}

Let us compute the trispectrum $\langle\zeta_{\vec k_1} \zeta_{\vec k_2} \zeta_{\vec k_3} \zeta_{\vec k_4}\rangle$ in the counter-collapsed limit assuming $\rho \ll \tilde \rho$ and $\mu =0$, in such a way that the dominant contribution is given by fig.~\ref{BBB}.
In the counter-collapsed configuration, the four-point function simply expresses the correlation between a pair of two-point functions, which is induced by a low frequency $\sigma^{(s)}$ mode of momentum $q \ll k_1 \approx k_2,\  k_3 \approx k_4$.
This long mode crossed its sound horizon much earlier than any of the $k_a$ modes crossed the Hubble radius and can be considered as a fixed classical background,~\cite{Seery:2008ax}.
This allows us to use eq.~\eqref{modul_curv_power} to compute the trispectrum:
\be\label{trispectrum}
\expect{\zeta_{\vec k_1} \, \zeta_{- \vec k_1-\vec q} \, \zeta_{\vec k_3} \, \zeta_{- \vec k_3 +\vec q}}'_{\vec q \to 0} \simeq H^4 \ \sum_{s=-2}^{+2} \frac{\expect{\sigma^{(s)}_{\vec q} \, \pi_{\vec k_1} \, \pi_{- \vec k_1 - \vec q}}'\expect{\sigma^{(s)}_{- \vec q} \, \pi_{\vec k_3} \, \pi_{- \vec k_3 + \vec q}}' }{P_{\sigma^{(s)}}(q)}\,.
\ee
The squeezed $\langle\sigma^{(s)}\pi\pi\rangle'$ 3-point function is given by, (see eqs.~\eqref{squeezed_sigma_0_pi_pi} and \eqref{squeezed_sigma_2_pi_pi}),
\be\label{squeezed_sigma_pi_pi}
\expect{\sigma^{(s)}_{\vec q} \, \pi_{\vec k} \, \pi_{- \vec k - \vec q}}' _{\vec q \to 0} = \frac{c(\nu) \, \tilde \rho}{\epsilon \, \mpl \, H} \ (-k\,\eta)^{-3/2 + \nu} \ P_{\sigma^{(s)}}(q) \, P_\pi(k) \ \ep^{(s)}_{ij}(\hat q) \hat k_i \hat k_j \,, 
\ee
Plugging eqs.~\eqref{squeezed_sigma_pi_pi} and \eqref{sigma_power_spectrum} into eq.~\eqref{trispectrum}, we get
\bea
\expect{\zeta_{\vec k_1} \, \zeta_{- \vec k_1-\vec q} \, \zeta_{\vec k_3} \, \zeta_{- \vec k_3 +\vec q}}'_{\vec q \to 0}  &=& \sum_{s=-2}^{+2} \frac{\mathcal T (\nu) }{c_s^{2\nu}} \( \frac{\tilde \rho}{H \sqrt \ep} \)^2 \(\frac{q^2}{k_1 k_3}\)^{3/2-\nu} P_\zeta(q) P_\zeta(k_1) P_\zeta(k_3) \\
&& \hspace{4.5 cm} \times \(\ep^{(s)}_{ij}(\hat q) \hat k_{1, \, i} \hat k_{1, \, j}\) \(\ep^{(s)}_{ij}(\hat q) \hat k_{3, \, i} \hat k_{3, \, j}\) \,. \nonumber
\eea
\begin{figure}[t] 
\centering
\includegraphics[scale=.8]{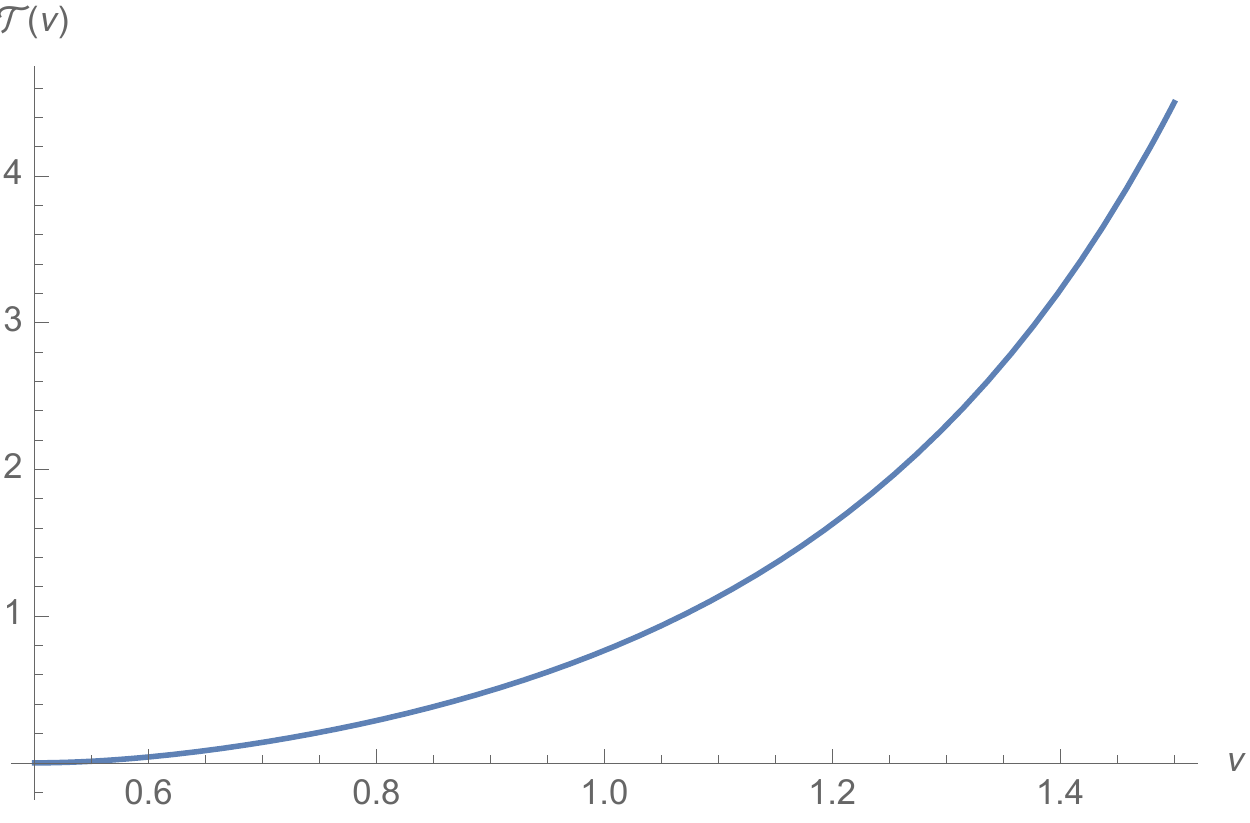}
\caption{\small Trispectrum amplitude $\mathcal T(\nu)$ as a function of $\nu = \sqrt{\frac 9 4 - \left( \frac m H \right)^2}$, in the range of masses below the Higuchi bound: $\nu \in [\frac 1 2, \frac 3 2]$. \label{Fig_T_nu}}
\end{figure} 
\noindent
The function $\mathcal T (\nu)$ is plotted in fig.~\ref{Fig_T_nu} in the mass range below the Higuchi bound and its expression is
\be
\mathcal T(\nu) = \frac{2^{2\nu} \, \Gamma(\nu)^2 \ c(\nu)^2}{\pi} \,,
\ee
with $c(\nu)$ given by eq.~\eqref{c_nu}.
If $\sigma$ is massless, $\nu = 3/2$ then the expression for the trispectrum simplifies to
\be
\expect{\zeta_{\vec k_1} \, \zeta_{- \vec k_1-\vec q} \, \zeta_{\vec k_3} \, \zeta_{- \vec k_3 +\vec q}}'_{\vec q \to 0} = \frac 9 {8 \, c_2^3} \ \(\frac{\tilde \rho}{H \sqrt \ep}\)^2 P_\zeta(q) \, P_\zeta(k_1) \, P_\zeta(k_3) \, \sum_{s=+}^{\times}  \(\ep^{(s)}_{ij}(\hat q) \hat k_{1, \, i} \hat k_{1, \, j}\) \(\ep^{(s)}_{ij}(\hat q) \hat k_{2, \, i} \hat k_{2, \, j}\) \,,
\ee
where we have also assumed that $c_2 \ll c_0 \approx c_1$, we can neglect the contribution from the helicity $0$ and $1$ modes.

\subsubsection{Quadrupolar modulation of the scalar power spectrum}

The same effect that enhances the scalar trispectrum induces, in the presence of super-horizon modes of $\s$, a quadrupolar modulation on the 2-point function of scalar perturbations, see for example \cite{Jeong:2012df}
\be
P_\zeta (\vec k) = P_\zeta(k) \[ 1+ \mathcal Q_{ij} \hat k_i \hat k_j \]\,.
\ee
The matrix $\mathcal Q_{ij}$ can be computed using eqs.~\eqref{modul_curv_power} and \eqref{squeezed_sigma_pi_pi}. We get
\be
\mathcal Q_{ij} \simeq \frac{c(\nu)}{\ep \mpl} \, \frac{\tilde \rho}{H} \, (- k \eta )^{-3/2 + \nu} \ \sum_s \s^{(s)}(\vec q) \, \ep^{(s)}_{ij}(\hat q)\,.
\ee
By averaging over all the super-horizon modes we get the expected (squared) amplitude $\mathcal Q^2$:
\be
\begin{split}
\mathcal Q^2 &= \frac{8\pi}{15} \expect{\mathcal Q_{ij} \mathcal Q_{ij}} \\
&= \frac{16}{15\pi} \mathcal T(\nu) \, \(\frac{\tilde \rho}{\sqrt \ep H}\)^2 \Delta_\zeta^2 \ \int_{q<H_0} \!\!\! dq \, \frac{q^2}{k^3} \(\frac{k}{c_2 \, q}\)^{2\nu} \\
&= \frac{16}{15 \pi} \frac{\mathcal T(\nu)}{3-2\nu}  \(\frac{\tilde \rho}{\sqrt \ep H}\)^2 \frac{\Delta_\zeta^2}{c_2^{2\nu}} \( \frac{H_0}{k} \)^{3-2\nu}\,,
\end{split}
\ee
where in the last two lines we considered, for simplicity, only the contribution of the helicity-2 modes. 
This is justified if $c_2 \ll c_0 \approx c_1$. 
This result can be compared with the experimental limits that are set by the CMB: $\mathcal{Q} \lesssim 10^{-2}$ \cite{Kim:2013gka,Ade:2015hxq}. For future constraints see \cite{Bartolo:2017sbu} (notice that the model studied in this paper differs from ours since the modulation of the scalar power spectrum is quadratic, and not linear, in the higher-spin field).

It is worthwhile stressing that in this paper we never considered the possibility of a large background for $\s_{ij}$.
This is justified provided $\s$ has a non-zero mass squared so that the background anisotropy is diluted away and one is just left with the background due to quantum fluctuations.

\section{Minimal non-Gaussianities and constraints on $\mathbf{c_s}$}\label{sec: radiative constraints}

\noindent
Since the minimal action~\eqref{eq:spins} does not include any mixing between $\sigma$ and $\pi$ there are no tree level contributions to the $\pi$ correlation functions from $\sigma$. 
However, non-Gaussianities in the $\pi$ correlation functions are induced at loop level. 
Due to the smallness of the scalar fluctuation amplitude $\Delta_\zeta$, it is relevant to consider only the lowest order correlation functions---the power spectrum, the bispectrum and the trispectrum. 
The one-loop contributions to the three- and four-point functions of $\pi$ due to the interactions with $\sigma$ are represented by the diagrams on figure~\ref{fig:loops}, where the cubic and quartic vertices are given by the interactions~\eqref{eq:S3} and~\eqref{eq:S4} respectively (while the quintic and sixtic vertices can be easily derived). 
In terms of the canonically normalized scalar field $\pi_c \sim \frac{H^2}{\Delta_\zeta} \pi$ these interactions are proportional to the small factors $\Delta_\zeta$ and $\Delta_\zeta^2$ respectively, up to the factors of $H$ needed to take care of dimensions. The three- and four-point functions generated by these loop diagrams can be, therefore, estimated as $NG_3 \equiv f_{NL} \Delta_\zeta \sim \Delta_\zeta^3$ and $NG_4 \equiv \tau_{NL} \Delta_\zeta^2 \sim \Delta_\zeta^4$. This would make them essentially unobservable\footnote{The normalisation of the scalar perturbations power spectrum is $\Delta_\zeta^2 \approx 2 \cdot 10^{-9}$ and the current observations show the curvature perturbations to be Gaussian with precision $NG_3, NG_4 \lesssim 10^{-3}$ ~\cite{Ade:2015ava}.} unless the speed of propagation of $\sigma$ is parametrically small. In the latter case the loop contributions are enhanced by inverse powers of $c_s$, which could compensate for the smallness of $\Delta_\zeta$ and lead to observably large non-Gaussianities. 
The experimental upper bounds on $NG_3$ and $NG_4$ thus translate to lower bounds on $c_s$. 
We are interested in the lower bound on $c_s$ because the power spectrum of $\sigma$ fluctuations and hence all the observational effects of the extra spinning field are enhanced when $c_s$ is parametrically smaller than unity. 
Perturbativity of the theory and the observed Gaussianity of the scalar perturbations put limits on all other effects from the fields with spin, which we discussed in the previous section. 

\begin{figure}[h] 
\centering
\hspace{\fill}
\subfloat[][\label{fig:3ptloopA}]{\includegraphics[scale=.2, angle=180, origin=c]{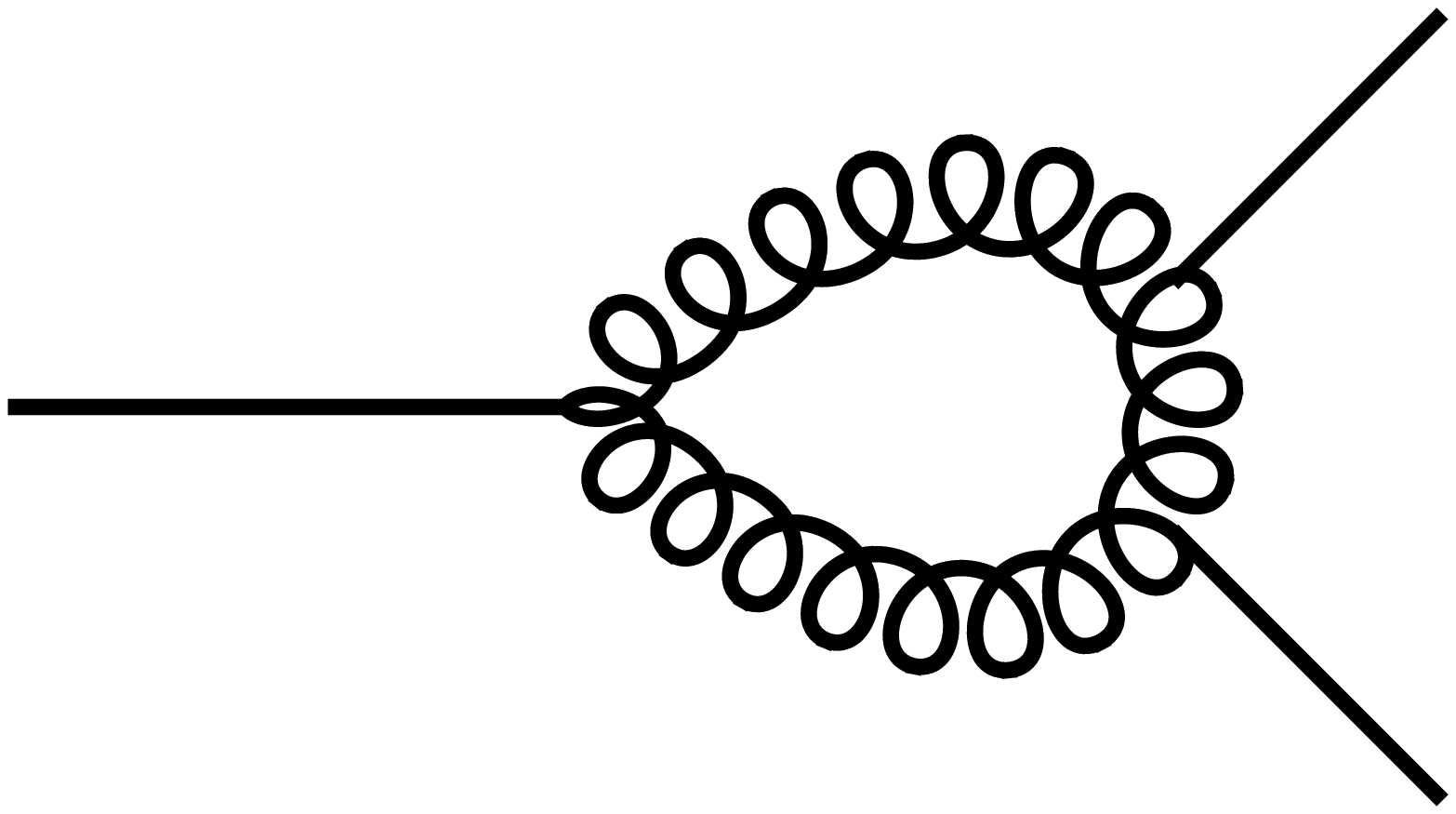}} \hspace{\fill}
\subfloat[][\label{fig:3ptloopB}]{\includegraphics[scale=.2]{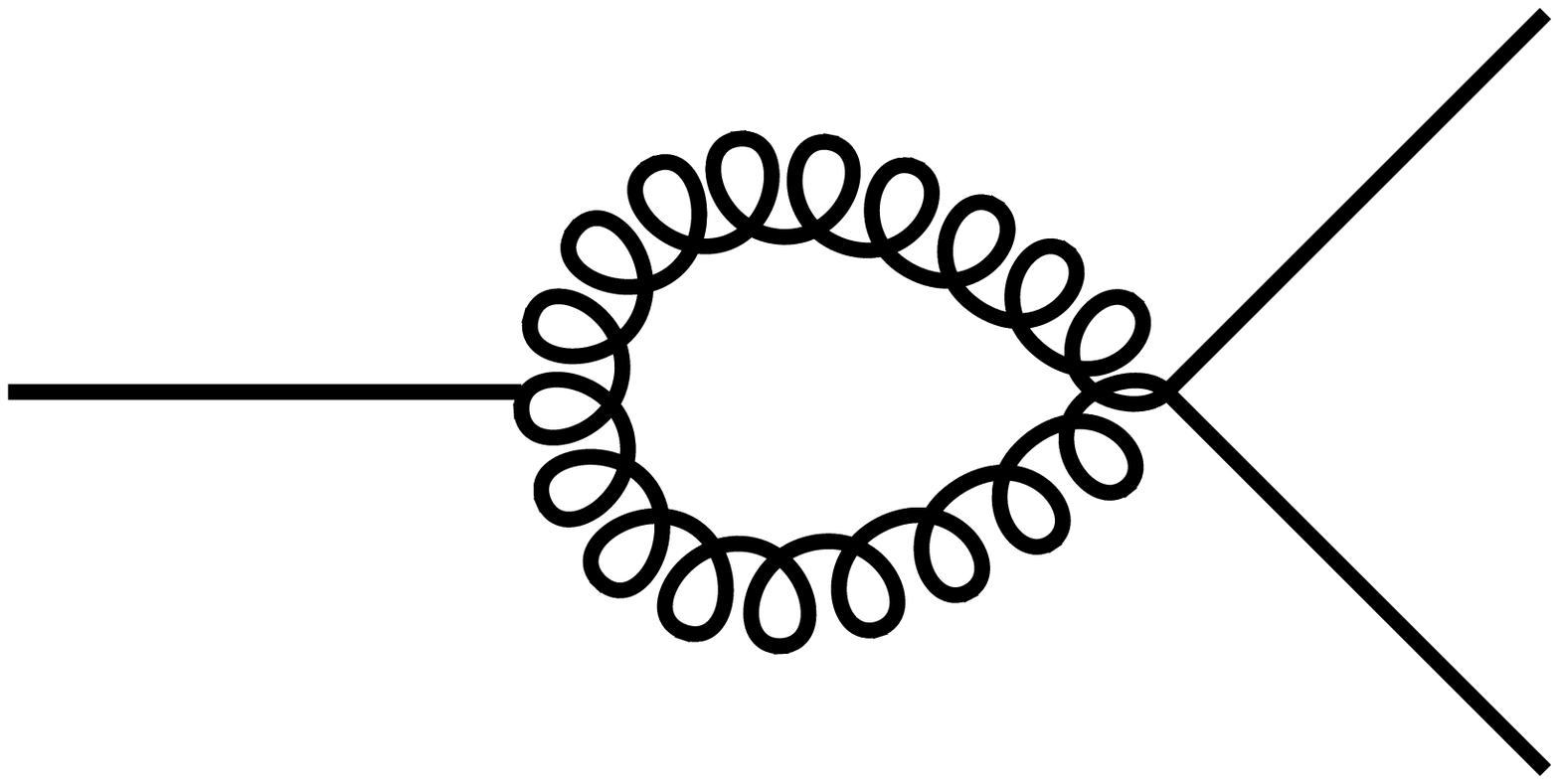}} \hspace{\fill}
\subfloat[][]{\includegraphics[scale=.2]{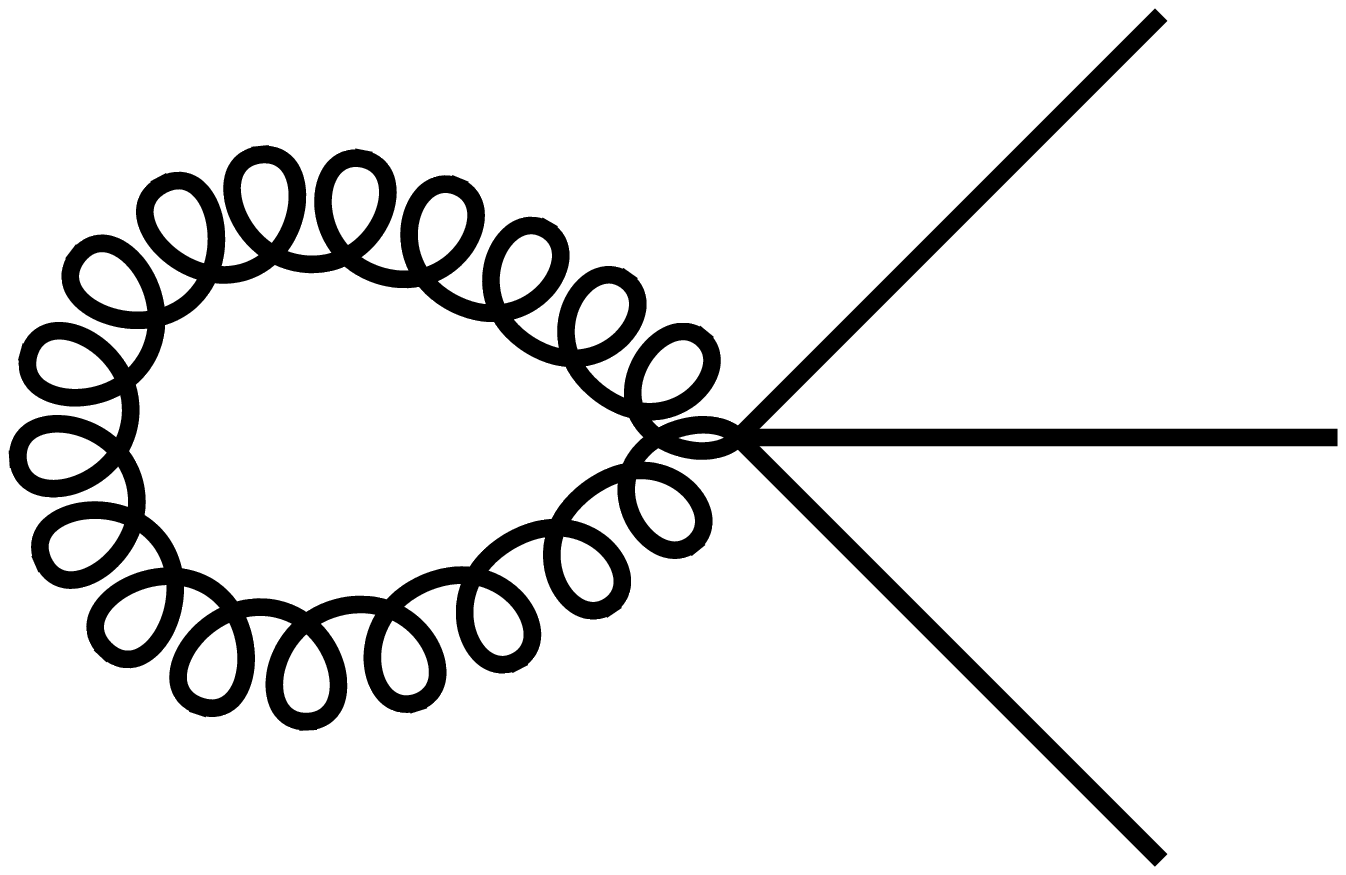}} \hspace{\fill}

\hspace{\fill}
\subfloat[][\label{fig:4ptloopA}]{\includegraphics[scale=.18]{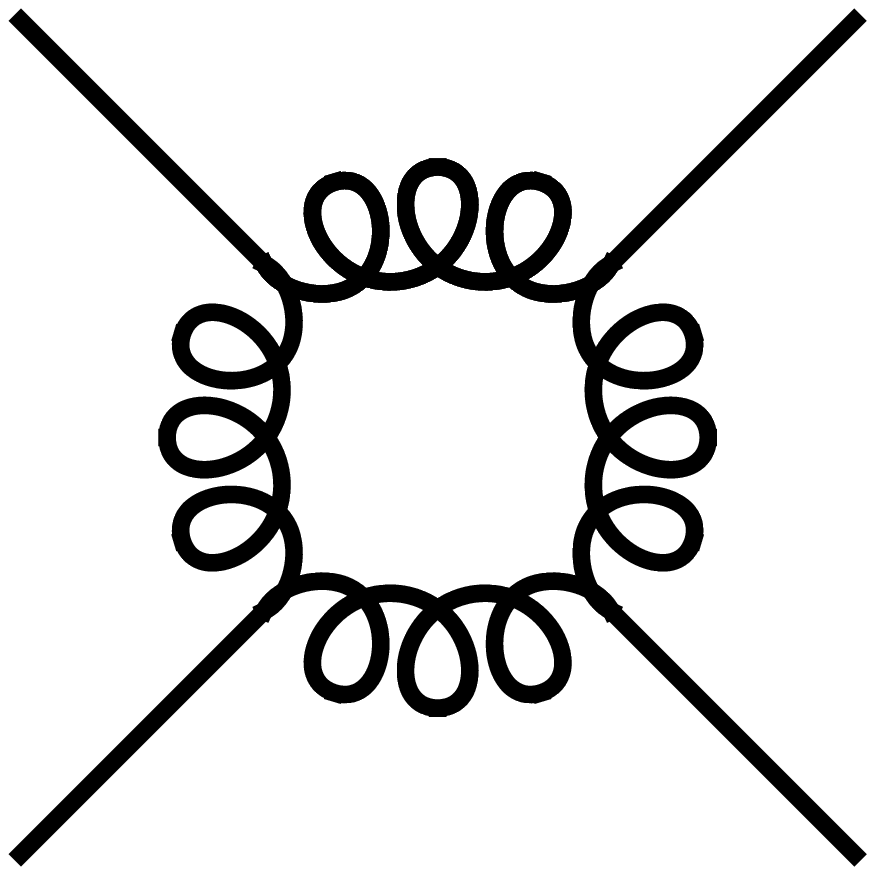}} \hspace{\fill}
\subfloat[][]{\includegraphics[scale=.18]{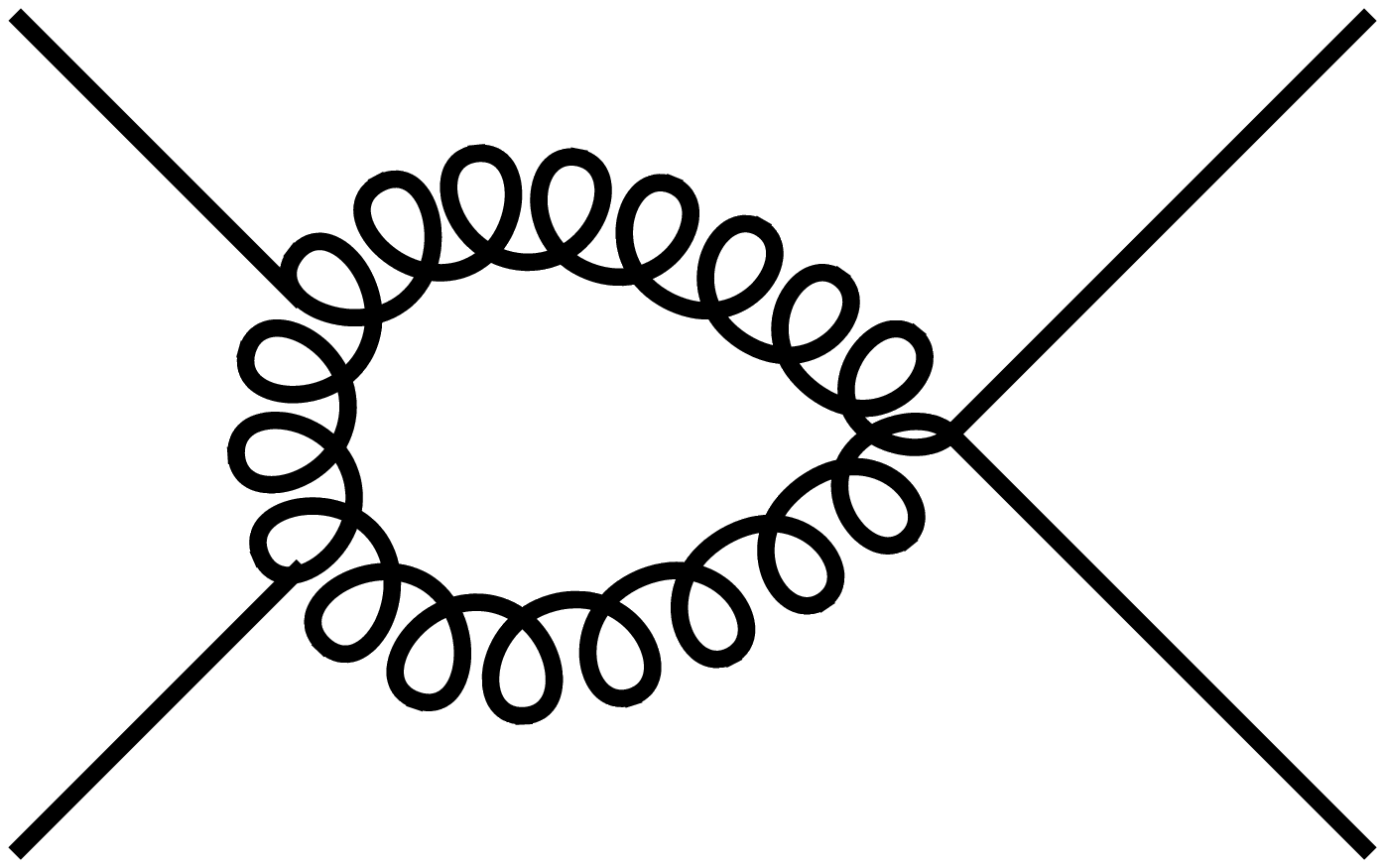}} \hspace{\fill}
\subfloat[][\label{fig:4ptloopC}]{\includegraphics[scale=.18]{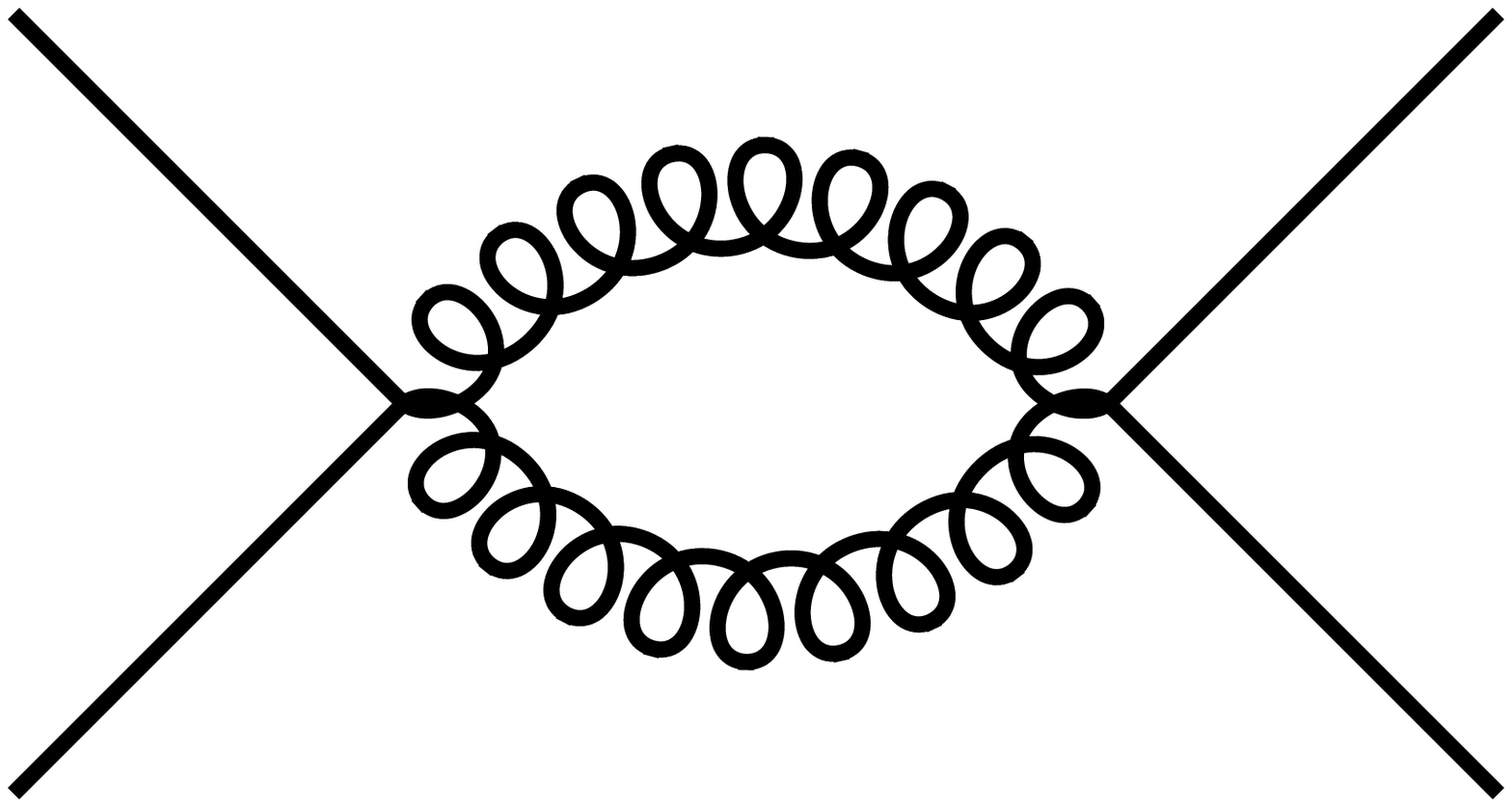}} \hspace{\fill}
\subfloat[][]{\includegraphics[scale=.18]{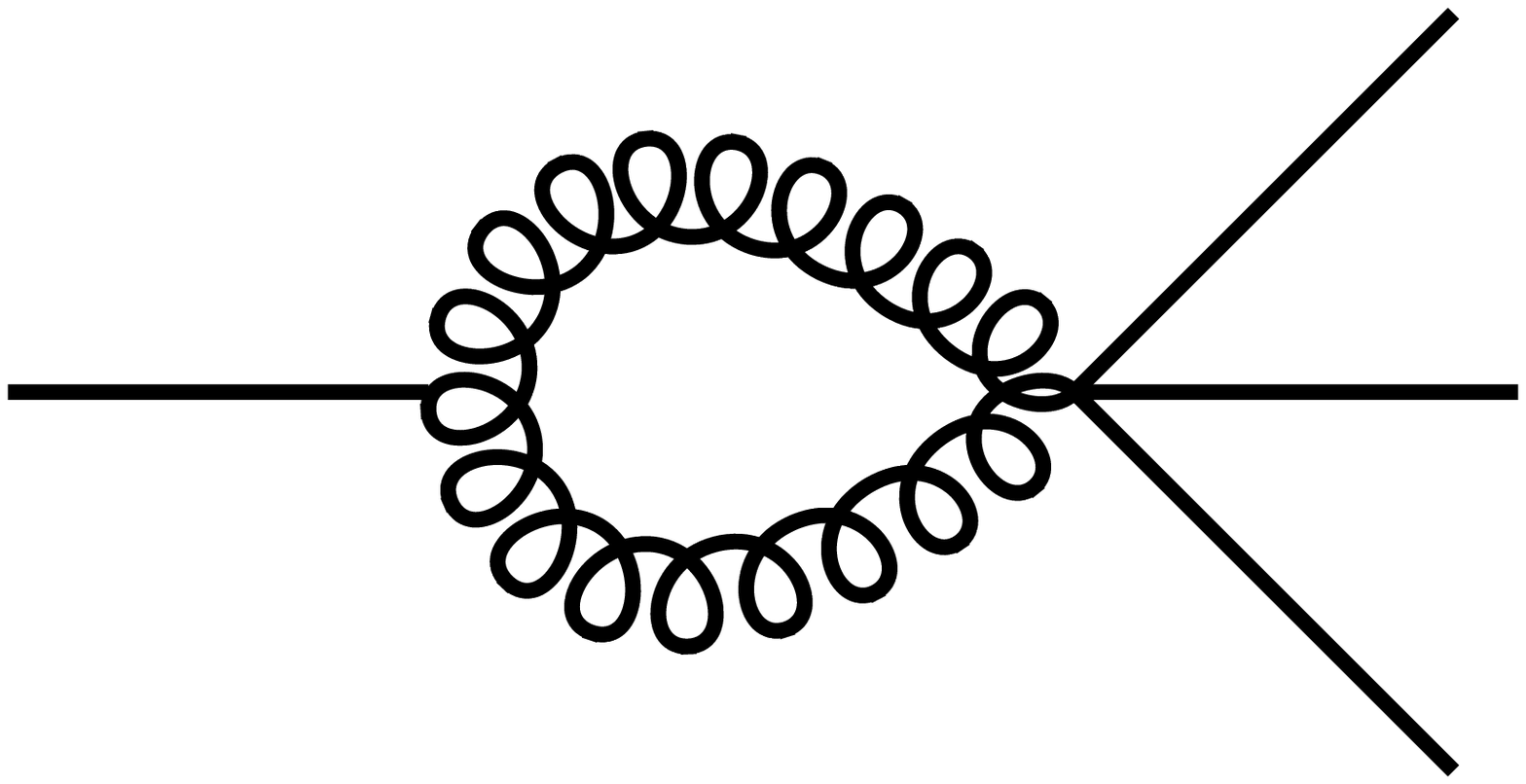}} \hspace{\fill}
\subfloat[][]{\includegraphics[scale=.18]{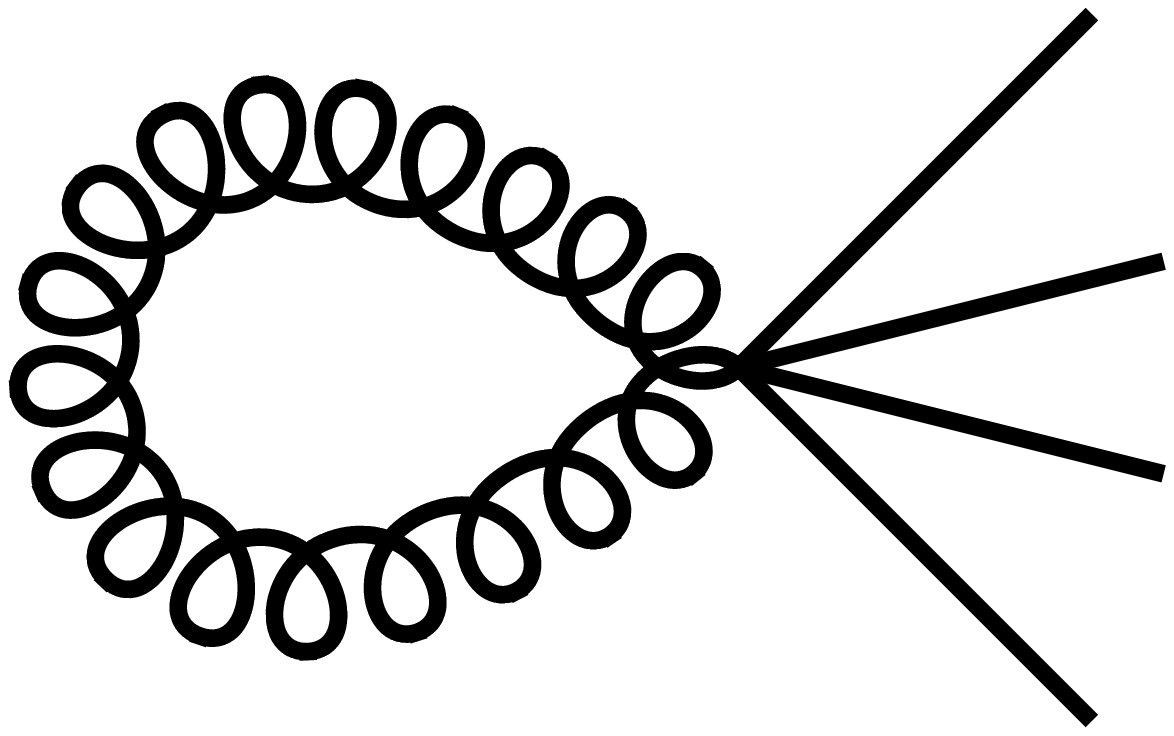}} \hspace{\fill}
\caption{\small One-loop corrections to the three- and four-point function of $\pi$ from the field $\sigma$. The straight lines stand for $\pi$ while the curly ones represent the exchange of a $\sigma$ field. \label{fig:loops}}
\end{figure}

We can split every loop contribution in the IR and UV parts depending on the value of the loop momentum. We start with the UV contributions that correspond to the part of the loop integral where the loop momentum is parametrically higher than the external legs momenta. These contributions are UV divergent and are dominated by the modes with energies at the loop UV cutoff $\Lambda$. For the EFT description to make sense at energies of order Hubble the energy cutoff scale $\Lambda$ has to be larger than $H$. Therefore it should be possible to represent such contributions by local self-interactions of the $\pi$ field in flat space. The coefficients of these self-interaction operators can be estimated by evaluating the divergent part of the loop diagram in Minkowski space and cutting it off at the energy scale $\Lambda$. Given that the value of $\Lambda$ depends on the UV completion of the theory we only demand that the non-Gaussianites induced by the Hubble scale modes do not violate observational constraints, i.e., we evaluate the loop contributions using the lowest possible cut-off, $\Lambda \sim H$. We also require the loop contribution to the power spectrum to be subdominant with respect to the tree level one in order for the loop expansion to make sense.

It is important to note that the minimal action~\eqref{eq:spins} employs only the building blocks that are invariant under reparameterisation of the inflaton field $\psi \mapsto f(\psi)$ and hence possesses an extra symmetry with respect to the standard EFT of Inflation. The subclass of the inflationary theories with this symmetry is called khronon inflation~\cite{1206.1083}. All the $\pi$ operators generated by loops of $\sigma$ have to respect this symmetry and this has important consequences for the loop-induced non-Gaussianities by enforcing cancellations of the leading (and subleading) UV divergencies. Naive dimensional analysis predicts the loop diagrams on figure~\ref{fig:loops} to diverge quartically and they would generate operators like $\dot \pi^3 \, \Lambda^4$ and $\dot \pi^4 \, \Lambda^4$. However all khronon-inflation operators have in total more than one derivative per field. The $\dot \pi^3$ and $\dot \pi^4$ operators with one derivative per $\pi$ field can not arise from these loops and we have checked that the leading quartic divergences do indeed cancel. Furthermore, in khronon inflation the self-interactions of $\pi$ which is leading in the derivative expansion arise from the expansion in $\pi$ of two covariant operators: $M_\alpha^2 \, (\nabla_\mu n^\mu - 3 H)^2 \sim M_\alpha^2 \, (\d \dot \pi)^2$ and $M_\lambda^2 \, n^\mu n^\nu \nabla_\mu n^\rho \nabla_\nu n_\rho \sim M_\lambda^2 \, (\d^2 \pi)^2$. The coefficients $M_\alpha^2$ and $M_\lambda^2$ of these operators have dimensions mass squared, hence these operators can be generated as the remaining quadratic divergence of the loop diagrams. Both these operators start quadratically in the Goldstone field. It means that the coefficients $M_\alpha^2$ and $M_\lambda^2$ of these cubic and quartic operators are fixed in terms of the one-loop corrections to the two-point function of $\pi$.

\begin{figure}[h] 
\centering
\hspace{\fill}
\subfloat[][]{\includegraphics[scale=.22]{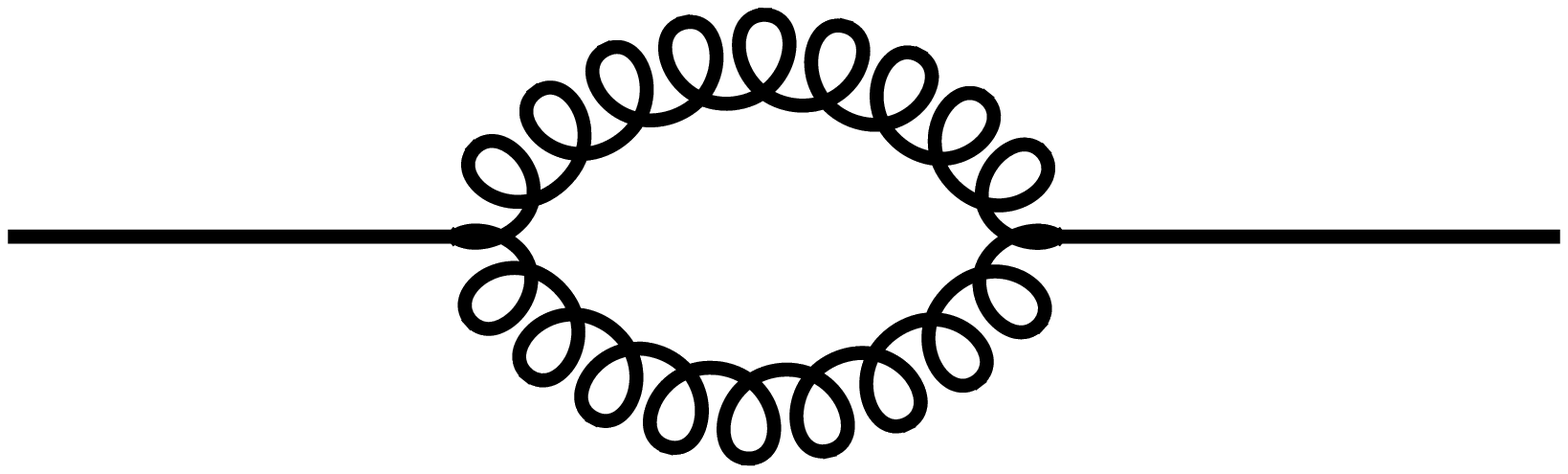}} \hspace{\fill}
\subfloat[][]{\includegraphics[scale=.22]{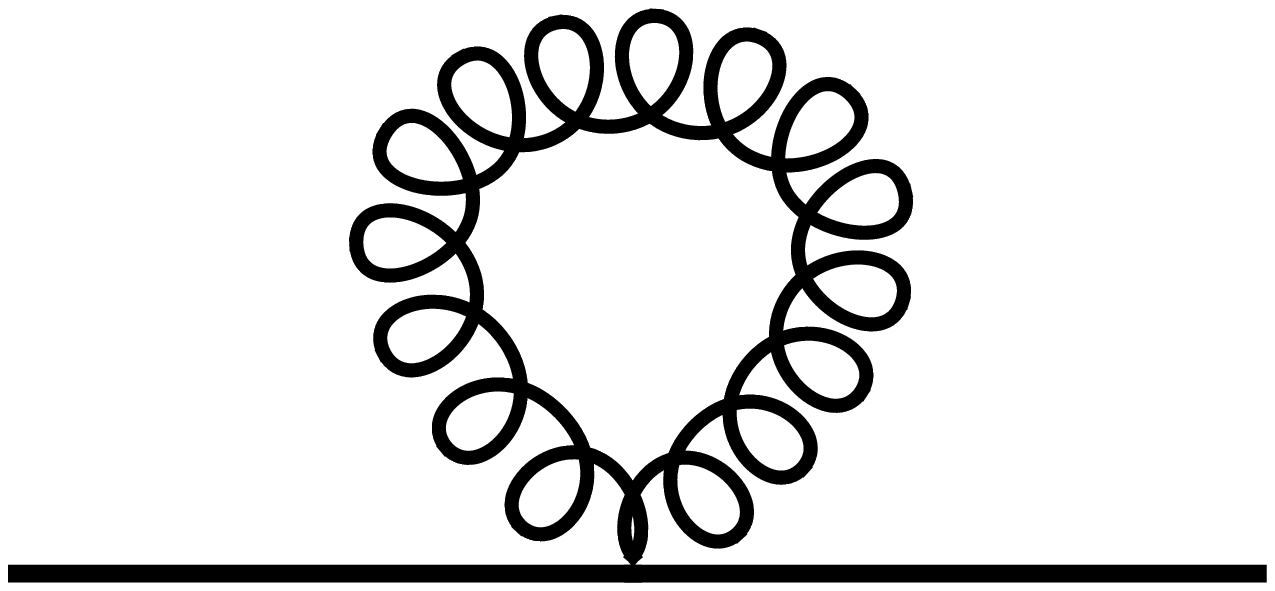}} \hspace{\fill}
\caption{\small One-loop corrections to the two-point function of $\pi$.\label{fig:loops2}}
\end{figure}

The one-loop corrections to the two-point function of $\pi$ come from the two diagrams on figure~\ref{fig:loops2}. The absence of the usual two-derivative kinetic term for $\pi$ in khronon inflation means that the leading quartic divergence in these diagrams cancels out. The remaining quadratically divergent contribution can indeed be represented by the higher derivative khronon operators
\begin{equation}
 \frac{\Lambda^2}{c_s^3}  \, (\d^2 \pi)^2\qquad \text{and} \qquad  \frac{\Lambda^2}{c_s^3} \, (\d \dot \pi)^2 \; ,
\end{equation}
with coefficients that are enhanced by a factor $c_s^{-3}$ in the small $c_s$ limit, $M_\alpha^2 \sim M_\lambda^2 \sim \Lambda^2 / c_s^3$. For the perturbativity of the EFT we require these operators to be subdominant with respect to the tree level kinetic term at horizon crossing, i.e. when the derivatives are of order $H$, at least with $\Lambda \sim H$:
\begin{equation}
\frac{H^2}{c_s^3} \, (\d^2 \pi)^2\Big|_{\d\sim H} \lesssim \frac{H^4}{\Delta_\zeta^2} \, (\d \pi)^2\;.
\end{equation}
Perturbativity imposes a lower bound on the sound speed of the $\sigma$ field
\begin{equation}\label{eq:cspert}
c_s \gtrsim \Delta_\zeta^{2/3} \approx 10^{-3} \; .
\end{equation}
Non-Gaussianities generated in the khronon inflation case were extensively studied in ref.~\cite{1206.1083}. The corresponding cubic and quartic operators can be schematically represented as $\frac{\Lambda^2}{c_s^3} \, (\d^2 \pi)^2 \, \d \pi$ and $\frac{\Lambda^2}{c_s^3} \, (\d^2 \pi)^2 \, (\d \pi)^2$, where some $\d$ can stand for time derivatives. The loop contributions to the three- and four-point functions of $\pi$ from the $\sigma$ modes with energies around $\Lambda \sim H$ can be thus estimated as $NG_3 \sim \frac{\Delta_\zeta^3}{c_s^3}$ and $NG_4 \sim \frac{\Delta_\zeta^4}{c_s^3}$. Using the perturbativity bound~\eqref{eq:cspert} we see that these contributions can be at most of order $f_\text{NL} \sim 1$ and $\tau_\text{NL} \sim 1$ and are below the experimental constrains.

Given that the khronon symmetry constraints the coefficients of the quadratic divergencies, it turns out that the leading UV contribution in the small $c_s$ regime is given by the sub-sub-leading logarithmically divergent terms. The local parts of these terms can be represented by the cubic and quartic $\pi$ self-interactions with seven and eight derivatives respectively, which can be schematically written as
\begin{equation}
\frac{\log{(\Lambda^2/H^2})}{c_s^5}  \, \d^2 \dot \pi \, (\d^2 \dot \pi)^2    \qquad \text{and} \qquad  \frac{\log{(\Lambda^2/H^2)}}{c_s^7}  \, (\d^2 \dot \pi)^4 \; .
\end{equation}
The corresponding contributions to the bi- and tri-spectrum from the modes with frequencies around $H$ are given by
\begin{equation}\label{eq:34ptloop}
NG_3 \sim \frac{\Delta_\zeta^3}{c_s^5} \qquad \text{and} \qquad NG_4 \sim \frac{\Delta_\zeta^4}{c_s^7} \; . 
\end{equation}

Let us turn to the IR contributions to the loop diagrams on figure~\ref{fig:loops}. They can be estimated by taking the loop momentum comparable to the momentum of the external legs. It is easy to count the $c_s$ dependence of the one-loop graphs. Each $\sigma$ line contributes a factor of $c_s^{-3}$ and each time derivative $\dot\sigma$ in the vertices gives an extra factor of $c_s^2$. The extra symmetry of the minimal action~\eqref{eq:spins} also impacts the IR part of the loop contributions. Note that all the cubic interactions in the action~\eqref{eq:S3} have an extra $c_s^2$ suppression either as an explicit factor or coming from the time derivative of $\sigma$. Since this is not the case for the quartic interactions~\eqref{eq:S4} all the diagrams with quartic vertices are enhanced in the small $c_s$ regime with respect to the corresponding cubic diagrams, where one quartic vertex is substituted by two cubic ones. In particular, the leading contribution to the bispectrum is given by diagram~(\ref{fig:3ptloopB}) and can be estimated as
\begin{equation}
NG_3  \sim \frac{\Delta_\zeta^3 c_s^2}{c_s^6} = \frac{\Delta_\zeta^3}{c_s^4} \; .
\end{equation}
Similarly, the leading loop contribution to the four-point function is given by the diagram~(\ref{fig:4ptloopC}) and can be estimated as
\begin{equation}
NG_4 \sim \frac{\Delta_\zeta^4}{c_s^6} \; .
\end{equation}
We see that these IR contributions are smaller than the UV contributions~\eqref{eq:34ptloop} and we shall use the latter in order to compare with the observational limits on the non-Gaussianity. The experimental upper bounds on $NG_3$ and $NG_4$ can be translated into the lower limit on the sound speed of $\sigma$ field:
\begin{equation}\label{eq:csexp}
c_s \gtrsim \left(\frac{\Delta_\zeta^{3}}{NG_3}\right)^{1/5} \qquad \text{and} \qquad c_s \gtrsim \left(\frac{\Delta_\zeta^4}{NG_4}\right)^{1/7} \;.
\end{equation}
These constraints have different parametric behaviour, nevertheless  the resulting numerical bound on $c_s$ is approximately the same (and it is stronger than the bound~\eqref{eq:cspert} from perturbativity alone):
\begin{equation}\label{eq:cs_khronon_num}
c_s \gtrsim 10^{-2} \; .
\end{equation}

Note that the bounds above apply to the propagation speed $c_h$ of every polarisation, where we assumed that there is no splitting of the speeds. In the presence of splitting $\delta c_s^2$ the situation is somewhat more involved. Some of the cubic interactions~\eqref{eq:S3} carry an explicit factor of $c_s^2$ and therefore are not suppressed if $c_s \approx 1$. This implies that for the models with $c_0^2 \lesssim c_s^2 \approx 1$ the non-Gaussianities induced by the helicity-0 mode running in the loops are larger than the estimates above. This translates into a stronger lower bound on $c_0$. We are, however not interested in such a regime, since the phenomenology is not qualitatively different with respect to the case where all $c_h$s are degenerate.
Therefore, we focus on the cases when all the $c_h$ are of the same order or when $c_s$ is the smallest one.

The extra symmetry of the minimal action~\eqref{eq:spins} means that it is possible to describe particles with arbitrary spin, scaling dimension, and propagation speeds in the context of the symmetries of khronon inflation. However, we have no reason to expect this additional symmetry to be present during inflation, and we indeed assume that the action for the Goldstone field $\pi$ does not preserve it.
It is therefore consistent and, in the absence of the khronon reparameterisation symmetry, even compulsory to add more interactions between $\sigma$ and $\pi$ in the EFT. At variance with the cubic and quartic terms in~\eqref{eq:S3} and~\eqref{eq:S4}, the coefficients of these interactions are not predicted by the non-linear realisation of the Lorentz symmetry but are free parameters in the EFT approach. Therefore, comparing the induced non-Gaussianities with the experimental bounds does not allow one to constrain $c_s$, it only leads to constraints on the size of these new couplings. 
However, given the strong impact of the khronon symmetry of the minimal action on the loop-induced three- and four-point functions, one can worry that even reasonably weak $\sigma$-$\pi$ interactions, which break the accidental khronon symmetry, induce loop non-Gaussianities, which are parametrically larger than the estimates~\eqref{eq:34ptloop}. In this case, one should consider such interactions in order to put a realistic lower bound on $c_s$. For example, it is natural to expect that at order $\pi \sigma^2$ there are interactions that are given by the operators already present in the quadratic action~\eqref{eq:S2} multiplied by $\delta  g^{00}$:
\begin{align*}
m^2 \, \delta  g^{00}  \, \Sigma^{\nu_1 \dots \nu_s} \Sigma_{\nu_1 \dots \nu_s} &\sim m^2 \, \dot\pi \, (\dot \sigma^{i_1 \dots i_s})^2 \; , \\
\delta  g^{00}  \, n^\mu n^\lambda \, \nabla_\mu \Sigma^{\nu_1 \dots \nu_s} \nabla_\lambda \Sigma_{\nu_1 \dots \nu_s} &\sim \dot\pi \, (\dot \sigma^{i_1 \dots i_s})^2 \; , \\
\delta  g^{00} \, c_s^2 \, \nabla_\mu \Sigma^{\nu_1 \dots \nu_s} \nabla^\mu \Sigma_{\nu_1 \dots \nu_s} &\sim  c_s^2 \,  \dot\pi \, (\d_j\sigma^{i_1 \dots i_s} )^2 \; .
\end{align*}
 It is straightforward to check that loop diagrams constructed with these cubic vertices contribute in the same way at comparable level when the loop frequency is taken to be of order $H$. They induce a correction to the quadratic action for $\pi$ of the order $H^2 \, c_s^{-3} \, \dot \pi^2$ and three- and four-point functions of size
\begin{equation}
NG_3 \sim \frac{\Delta_\zeta^3}{c_s^3} \qquad \text{and} \qquad NG_4 \sim \frac{\Delta_\zeta^4}{c_s^3} \; .
\end{equation}
The correction to the quadratic action is subleading when the perturbativity constraint~\eqref{eq:cspert} holds, and the induced non-Gaussianities happen to be smaller than the ones induced by the minimal interactions that are given in equation~\eqref{eq:34ptloop}. This shows that despite the cancellations the lower bound~\eqref{eq:cs_khronon_num} is sufficient to render the theory perturbative and to satisfy observational constraints on non-Gaussianity even when the accidental khronon symmetry is broken by the interactions.

We see that in the regime of small $c_s$ the effects of the spinning field $\sigma$ are boosted due to enhancement in the amplitude of its fluctuations. In the previous section we discussed the main observational signatures of $\sigma$. Both the main signatures---changing the tensor mode power spectrum by mixing and inducing spin-dependent angular dependence in the squeezed limits of the scalar correlation functions---are also naturally enhanced by inverse powers of $c_s$. The constraints obtained in this section limit the extent of the parametric range where those observational signatures hold, but leave a rather large one.

Note that for any reasonable UV completion the loop cut-off is somewhat higher than $H$ and the UV part of the loop dominates over the IR part. Hence, if during inflation there was a spinning field with parametrically small $c_s$ the non-Gaussianities induced by its loops will first manifest themselves as the effects of the local self-interactions of $\pi$ and will not carry any distinctive features. In order to see some features of the spinning field $\sigma$ one should introduce a mixing between $\sigma$ and observable tensor and scalar modes $\gamma$ and $\pi$. 
These are the theories whose phenomenology we studied in the previous section.

\section{\label{sec:conclusions}Conclusions and outlook}

In this paper we have shown that the coupling with the inflaton foliation allows to violate the Higuchi bound: one can have particles with spin that do not decay outside the Hubble radius, while preserving the approximate shift symmetry of the inflaton. This opens a window on new phenomenology that we have only started to explore focussing on the case of spin 2. It should be straightforward to extend our analysis in various directions. One could include a non-trivial speed for the inflaton perturbations, consider the case of higher spin, or look at fermions. It would be interesting to understand whether it is possible to distinguish the effect of fermionic fields, although they cannot mix with scalar and tensors perturbations and play a role only at loop level. A more speculative investigation would be to violate the relation spin/statistics using the spontaneous breaking of Lorentz invariance. It would also be interesting to understand how one can observationally distinguish our setup from constructions based on different symmetry patterns \cite{Endlich:2012pz,Bartolo:2015qvr,Ricciardone:2016lym}, where unconventional light spin-2 particles do exist.

In this paper we did not  consider the possibility that higher spin fields have a sizeable background: in this case isotropy is broken as studied in \cite{Franciolini:2018eno,Bartolo:2015dga}. It would be interested to explore this possibility since it does not rely on any mixing with $\zeta$ and $\gamma$, but only on the interactions of $\sigma$ with the foliation required to violate the Higuchi bound.

On the more theoretical side, it would be interesting to explore the possibility of modifying our Lagrangian by adding auxiliary fields that do not change the number of propagating degrees of freedom, but that can change and potentially enlarge the resulting phenomenological consequences. Microcausality, i.e.~the commutativity of fields outside the lightcone, is not automatically preserved after these additions (while it is automatic without them) and it would be nice to have a straightforward procedure to include auxiliary fields.  

\section*{Acknowledgements}
It is a pleasure to thank Mehrdad Mirbabayi for collaboration in the early stages of this work, Riccardo Rattazzi, Sergei Sibiryakov, and Andrew Tolley for useful discussions and Daniel Baumann, Hayden Lee, Azadeh Moradinezhad, Gui Pimentel and Yuko Urakawa for useful comments on the draft.
LS is partially supported by Simons Foundation Origins of the Universe program (Modern Inflationary Cosmology collaboration) and by NSF award 1720397.

\appendix

\section{Explicit calculations}

\subsection{Power Spectra}
\label{power_spectra}

Let us study the quadratic action \eqref{toy_model}. 
In Fourier space, the equations of motion for $\s$ read
\be
{\s_{k}^{(s)}(\eta)}'' - \frac{2}{\eta} {\s_{k}^{(s)}(\eta)}' - \( \frac{m^2}{H^2 \eta^2} +c_s^2 k^2 \)\s_{\vec k}^{(s)} (\eta) =0 \,,
\ee
where $'\equiv\d_\eta$ and the wavefunctions $\s_{k}^{(s)}(\eta)$ are defined by the following expression,
\be
\s_{ij}(\vec k, \eta) \equiv \sum_{s=-2}^2 \s^{(s)}_k (\eta) \, \ep^{(s)}_{ij}(\hat k)\,.
\ee
The polarisation tensors are defined such that $\sum_{i,j}\ep^{(s)}_{ij}(\hat k) \, (\ep^{(s)}_{ij}(\hat k))^*= 2\delta^{s s'}$. 

Let us write an explicit expression for the $\ep^{(s)}_{ij}(\hat k)$.  The helicity-zero polarisation tensor is just given in terms of the direction $\hat k$:
\be
\ep^{(0)}_{ij}(\hat k) = \sqrt 3 \(\hat k_i \hat k_j -\frac{\delta_{ij}}{3}\)\,.
\ee
To define the higher-helicity tensors we introduce the vectors orthogonal to $\hat k$, i.e.~$\hat v$ and $\hat u \equiv \hat k \times \hat v /|\hat k \times \hat v|$. 
Hence the helicity-one polarisation tensors are given by
\be
\ep^{(\pm1)}_{ij}(\hat k) = \frac{\hat k_i \, (\hat v \pm \hat u)_j + (\hat v \pm \hat u)_i \, k_j}{\sqrt 2}\,,
\ee
while the helicity-two tensors are
\be
\ep^{(\pm 2)}_{ij} (\hat k) = \frac{\( \hat u_i \hat u_j - \hat v_i  \hat v_j \) \mp i \( \hat u_i \hat v_j + \hat v_i  \hat u_j \)}{\sqrt 2}\,.
\ee

The solution to the equations of motion are 
\be\label{wavefunction}
\s_k^{(s)}(\eta) = \frac{\sqrt{\pi}}{2} \ H \ (-\eta)^{\frac{3}{2}} H_{\nu}^{(2)}(-c_s k \eta)\,,
\ee
with $\nu = \sqrt{\frac{9}{4}-\frac{m^2}{H^2}}$.
At late times, $\eta \ll 1$, the Hankel function can be expanded in a power series of $c_s k \eta$ and the wavefunction can be approximated with
\be\label{late_sigma}
{\s_q^{(s)}}(\eta\to0) =  \frac{i \ 2^{\nu-1}\ \Gamma(\nu)}{\sqrt \pi}\ H\ \frac{ (-\eta)^{\frac{3}{2}-\nu}}{(c_s \, q)^{\nu}}\,.
\ee

To compute the changes in the scalar and tensor power spectra we need to evaluate integrals of the form  
\be
\mathcal I_1(n, c) \equiv  \int_0^\infty dx \ e^{-i x} \ x^n  \ H_\nu^{(1)}(c \, x)
\ee
and 
\be
\mathcal I_2(n, c) \equiv \int_0^\infty dx \ e^{-i x} \ x^n  \ H_\nu^{(1)}(c \, x) \int_x^\infty e^{-i y} \ y^n  \ H_\nu^{(2)}(c \, y)\,,
\ee
with $c\ll 1$.
To compute them one can slightly rotate the contour of integration in the complex plane {to pick up the interaction vacuum (see for example~\cite{Senatore:2009cf,Senatore:2012ya})}. Then, because of the exponential suppression the integrals take support only if $x,y \lesssim 1$. Since $c \ll 1$, the Hankel functions  can be Taylor expanded. Keeping the first order is enough. The integrals can now be computed easily with Mathematica, giving
\bea
\label{int1}
\mathcal I_1(n, c) &=& - \frac{i \ 2^\nu \ e^{-i \pi (1+n -\nu)/2} \ \Gamma(1+n-\nu) \ \Gamma(\nu)}{\pi \ c^{\,\nu}}\,, \\
\label{int2}
\mathcal I_2(n, c) &=&\frac{4^\nu \ e^{-i \pi (1-\nu)} \ \Gamma(2-2\nu) \ \Gamma(\nu)^2 \ {}_2F_1(2-2\nu,1+n-\nu,2+n-\nu,-1)}{(1+n-\nu) \ \pi^2 \ c^{\,2\nu}} \,.
\eea
Using these results, the contributions to the scalar and tensor power spectra follow straightforwardly. 
For the scalar power spectrum we get
\bea
C_{\zeta,1}(\nu) &=& \frac{2^{2\nu-3} \ (3-2\nu)^2 \ \Gamma(\frac{1}{2}-\nu)^2 \ \Gamma(\nu)^2}{3 \ \pi}\,, \\
C_{\zeta,2}(\nu) &=& - \frac{2^{2\nu-3} \ (3-2\nu)^2 \ \Gamma(\frac{1}{2}-\nu)^2 \ \Gamma(\nu)^2 \ \sin(\pi \, \nu)}{3 \ \pi}\,.
\eea
The changes in the tensor power spectrum are instead given by
\bea
C_{\gamma,1}(\nu) &=& \frac{2^{2\nu-1} \ \Gamma(\frac{1}{2}-\nu)^2 \ \Gamma(\nu)^2}{\pi}\,, \\
C_{\gamma,2}(\nu) &=& - \frac{2^{2\nu - 1} \ \Gamma(\frac{1}{2}-\nu)^2 \ \Gamma(\nu)^2 \ \sin(\pi \, \nu)}{\pi}\,.
\eea

\subsection{Non-perturbative treatment of the $\gamma$-$\sigma$ mixing}
\label{gamma_sigma_model}

In this Appendix we discuss the $\gamma$-$\sigma$ mixing by studying the mode functions of the coupled system.
A very similar analysis can be applied for the $\pi$-$\s$ mixing in quasi-single field inflation,~\cite{Chen:2009we}. 
  The perturbative calculation of the tensor power spectrum presented above leads to a spurious divergence as the mass of $\sigma$ vanishes. The coefficient $\mathcal C_\gamma$ given by expression~\eqref{c_gamma} blows up for small mass as $\mathcal C_\gamma \simeq \frac {9}{m^4}$. In the massless case the time integrals are divergent at late times. It is possible to regulate them by calculating the power spectrum at some finite time $\eta_*$. The result is given in equation~\eqref{eq:p_gamma} and grows as the square of the number of $e$-foldings passed since the horizon crossing for a mode with a given wavenumber $k$, $N_k = \log(- k \eta_*)$. 
{This result} motivates us to study the two-point function of the $\gamma$-$\sigma$ system by finding the solutions to the coupled system. {In fact, as we show shortly, all the modes in the system either remain constant or decay at late times and the power spectrum cannot grow unbounded with $N_k$.} In this appendix we assume the sound speed of the helicity-2 mode to be small $c_2 \ll 1$ and, as usual,  $\rho \ll c_0 \sqrt \ep H \ll H$.

Given the quadratic action~\eqref{toy_model} and the mixing term from the action~\eqref{intercanonical} we can write the system of coupled equations for the modes of given momentum $k$ and polarisation $s = \pm 2$:
\begin{align}
g'' + k^2 \, g - \frac 2 {\eta^2} \, g - \frac{2 \rho}{H \eta} \, s' + \frac{4 \rho}{H \eta^2} \, s &= 0 \; ,\label{eq:geom}\\ 
s'' + c_2^2 \, k^2 \, s + \frac{m^2 - 2 H^2} {H^2 \eta^2} \, s + \frac{\rho}{H \eta} \, g' + \frac{\rho}{H \eta^2} \, g &= 0 \; , \label{eq:seom}
\end{align}
where we have introduced { rescaled fields $g \equiv a(\eta) \, \gamma^{(s)}_c$ and $s \equiv a(\eta) \, \sigma^{(s)}$}. In order to obtain the power spectra of $\gamma$ at the late times we have to find the late time behaviour of the mode functions. At late times $- k \eta \ll 1$ one can neglect the terms proportional to $k^2$ in the equation of motion. Since all the terms scale in the same way under rescaling of $\eta${,} one should look for solutions of the form $g = g_\Delta (- k \eta)^{\Delta-1}$ and $s = s_\Delta (- k \eta)^{\Delta - 1}$, where $g_\Delta$ and  $s_\Delta$ are constant amplitudes and $\Delta$ is the scaling dimension for the original fields $\gamma$ and $\sigma$. Plugging this ansatz in the field equations we obtain the following system:
\begin{equation}
\begin{pmatrix}
\Delta(3 - \Delta) & -\frac{2\rho} H (3 - \Delta) \\
- \frac{\rho}{H} \Delta & \Delta(3 - \Delta) - \frac{m^2}{H^2}
\end{pmatrix} \begin{pmatrix}
g_\Delta \\
s_\Delta
\end{pmatrix} = 0 \; .
\end{equation}
This has non-trivial solution only if the matrix of the coefficients is singular. The solutions correspond to the  eigenvectors with zero eigenvalue. 
The singularity condition fixes the scaling dimension to take one of the four values: $\Delta =$~0, 3, or { $\Delta_\pm = \frac32 \pm \sqrt{\frac94 - \frac{m^2 + 2\rho^2}{H^2}}$}. 
The former pair of scaling dimensions coincide with those of a massless scalar field and the latter pair corresponds to a scalar of mass $  m^2 + 2 \rho^2$. It means that for $\gamma$ and $\sigma$ there are two growing modes{,} one of which is constant and another decays as $  (- k \eta)^{\Delta_-} \simeq (- k \eta)^{\frac{m^2+2 \rho^2}{ 3 H^2}}$ at late times. The general solution for the system at late times is specified by the four constant amplitudes of the corresponding modes and reads
\begin{multline} 
\begin{pmatrix}
g(\eta) \\
s(\eta)
\end{pmatrix} = A_-  (- k \eta)^{-1}
\begin{pmatrix}
1 \\
0
\end{pmatrix} + A_+ (- k \eta)^{2}
\begin{pmatrix}
\frac{m^2}{H^2} \\
- 3 \frac \rho H
\end{pmatrix} \\ 
+ B_- (- k \eta)^{\Delta_- -1}
\begin{pmatrix}
\frac {2\rho} H \Delta_+\\
\frac{m^2 + 2 \rho^2}{H^2}
\end{pmatrix}
+ B_+ (- k \eta)^{\Delta_+ -1}
\begin{pmatrix}
\frac {2\rho} H \\
\Delta_+
\end{pmatrix} \; .
\end{multline}
Here $A_\pm$ and $B_\pm$ are four integration constants. The composition of the modes crucially depends on the relation between the mass and the mixing. If the mass is larger than the mixing, $\rho^2 \ll m^2 \ll H^2$, then the massless modes are in the $\gamma$ direction and the massive modes are in the $\sigma$ direction, just as one would expect:
\begin{multline}
\begin{pmatrix}
g(\eta) \\
s(\eta)
\end{pmatrix} \simeq A_- (- k \eta)^{-1}
\begin{pmatrix}
1 \\
0
\end{pmatrix} + A_+ (- k \eta)^{2}
\begin{pmatrix}
\frac{m^2}{H^2} \\
0
\end{pmatrix}\\
 + B_- (- k \eta)^{\frac{m^2}{3H^2} -1}
\begin{pmatrix}
0\\
\frac{m^2}{H^2}
\end{pmatrix} + B_+ (- k \eta)^{2 - \frac{m^2}{3H^2}}
\begin{pmatrix}
0 \\
3
\end{pmatrix} \; .
\end{multline}
If the mixing dominates, $m^2 \ll \rho^2 \ll H^2$, then both growing modes appear to be in $\gamma$ direction and both decaying modes are mostly in $\sigma$:
\begin{multline}
\begin{pmatrix}
g(\eta) \\
s(\eta)
\end{pmatrix} \simeq A_- (- k \eta)^{-1}
\begin{pmatrix}
1 \\
0
\end{pmatrix} -3 \frac \rho H A_+ (- k \eta)^{2}
\begin{pmatrix}
0 \\
1
\end{pmatrix} \\ 
+ 6 \frac \rho H \, B_- (- k \eta)^{\frac{2\rho^2}{3H^2} -1}
\begin{pmatrix}
1\\
\frac{\rho}{3 H}
\end{pmatrix} + 3 B_+  (- k \eta)^{2 - \frac{2\rho^2}{3H^2}}
\begin{pmatrix}
\frac {2\rho} {3H} \\
1
\end{pmatrix} \; .
\end{multline}
Only the $B_-$ growing mode has a component in the $s$ direction and it is parametrically smaller than its component in $g$ direction. In order to accommodate a growing solution only in {the} $s$ direction at horizon crossing, $ s\big|_{-k\eta = 1} = s_- (-k \eta)^{\frac{2 \rho^2}{3H^2}-1}$ and $g\big|_{-k\eta = 1} = 0$, one has to take the coefficients $ B_- = s_- \frac{H^2}{2 \rho^2}$ and $ A_- = - 6 \frac{\rho}{H} B_-$ to be parametrically large. The solution for $g$ exhibits an exact cancellation between two large $\mathcal O(\frac{H}{\rho})$ terms at horizon crossing
\begin{equation}\label{eq:gmode_approx}
g = - s_- \frac{3 H}{\rho} \left(1 -  (-k \eta)^{\frac{2\rho^2}{3H^2}} \right) (-k \eta)^{-1} \; .
\end{equation}
At late times the second term decays and the amplitude of the growing $g$ mode builds up to be large $g \propto \mathcal O(\frac{H}{\rho})$. Even if the growing $g$ mode is present at  horizon crossing with a comparable amplitude $g\big\rvert_{-k \eta = 1} \sim s_-$ it provides a contribution to $A_- \sim s_-$, which is parametrically smaller. At late times the solution for $g$ is therefore dominated by the mode given in equation~\eqref{eq:gmode_approx}. Similar effect{s} happen for the purely decaying solution in $g$ direction, but we are interested only in the growing modes in $g$ at the late times. For a generic mixing the solution~\eqref{eq:gmode_approx} reads
\begin{equation}\label{eq:gmode}
g \simeq - s_- \frac{6 \rho H}{m^2 + 2\rho^2} \left(1 - (-k \eta)^{\frac{m^2 + 2\rho^2}{3H^2}} \right) (-k \eta)^{-1} = - s_- \frac{6 \rho H}{m^2 + 2\rho^2} \left(1 - e^{-\frac{m^2 + 2\rho^2}{3H^2} N_k} \right) (-k \eta)^{-1} \; .
\end{equation}
The corresponding tensor mode $\gamma_c = (- H \eta) \, g$ grows in time till the massive mode decays and it saturates {at} {a} maximum value. 
The amplitude of the mode depends on the number of $e$-folds $N_k$ between the horizon crossing and the time of the end of inflation. 
Since we require the mixing of $\sigma$ with the scalar perturbations to be small, $\rho^2 \lesssim \epsilon H^2 \lesssim H^2 / N_k$  for the observable modes, {the} $\rho^2$ term is never important in the exponent. Hence, independently of $\rho$ there are two regimes.
If the mass of $\sigma$ is large enough that the $\Delta_-$ mode decays before we observe it, $\frac{m^2} {H^2} N_k \gtrsim 1$, then the amplitude of $g$ is given by the constant term $ g \simeq - s_- \frac{6 \rho H}{m^2} (-k \eta)^{-1}$. 
In the opposite case, when the massive mode does not have enough time to decay, $\frac{m^2} {H^2} N_k \lesssim 1$, one can expand the exponent and obtain the amplitude to be $ g \simeq - s_- \frac{2 \rho N_k}{H} (-k \eta)^{-1}$. 
Both these regimes are captured by the perturbative calculations. The former one corresponds to the case of massive $\sigma$ and asymptotically late times{,} and the latter to the massless $\sigma$ and a finite time cutoff. 
The non-perturbative result~\eqref{eq:gmode} implies that the coefficient of the $m^{-4}$ divergence of the massive $\sigma$ power spectrum~\eqref{p_gamma_c} is $3^2$ times the coefficient of the $N_k^2$ enhancement of the massless $\sigma$ result~\eqref{eq:p_gamma}, which is indeed the case.

We have shown that the late time solution for $\gamma$ is dominated for sufficiently large mixing by the growing mode of $\sigma$ at horizon crossing. Let us now study the solutions at early times and find the amplitude $s_-$ to be matched with the dominant $g$ mode at late times~\eqref{eq:gmode}. At early times, when both $\sigma$ and $\gamma$ modes are inside horizon, $- k \, \eta \gg c_2^{-1} \gg 1$, the mixing is not important and the solutions for $g$ and $s$ are just plane waves. The fields $s$ and $g$ have a canonical time kinetic term in  conformal time $\eta$ (up to a factor $\half$ for $g$) and we can choose two independent positive frequency solutions to be
\begin{equation}\label{eq:gsinit}
\begin{pmatrix} g \\ s \end{pmatrix}_{-c_2 k \eta \gg 1} \simeq \frac{e^{-i k \eta}}{\sqrt k} \begin{pmatrix} 1 \\ 0 \end{pmatrix}\,;
\qquad \text{and} \qquad  
\begin{pmatrix} g \\ s \end{pmatrix}_{-c_2 k \eta \gg 1} \simeq \frac{e^{-i c_2 k \eta}}{\sqrt{2 \,c_2 k}}  \begin{pmatrix} 0 \\ 1 \end{pmatrix} \, .
\end{equation}
The choice of the Minkowski-like vacuum state for the modes deep inside the horizon corresponds to choosing these mode-functions to multiply annihilation operators for the $\gamma$ and $\sigma$ particles respectively. Plugging these early time solutions in the mixing terms of the equations~\eqref{eq:geom} and~\eqref{eq:seom} one can check that for $\rho \ll H$ the back reaction from the mixing does not become important until the $\gamma$ horizon crossing time $- k \eta \sim 1$. { It means that the second solution in equation~\eqref{eq:gsinit}, which correspond to the $\sigma$ mode at early times, has much larger $s$-component $- k \eta \sim 1$, than the first solution, which correspond to the $\gamma$ mode. In addition, its amplitude gets enhanced if $c_2 \ll 1$, and will dominate the $\gamma$ power spectrum at late times.} For this solution the $\sigma$ mode freezes out at its horizon crossing at $-c_2 k \eta \sim 1$ and continues to behave like a free field of mass $ m^2 + 2\rho^2$ till $- k \eta \sim 1$:
\begin{equation}
s\Big\rvert_{1/c_2 \gtrsim - k \eta \gtrsim 1} \simeq \frac{1}{\sqrt{2 \,c_2 k}} \frac1{(-c_2 k \eta)^{1-\Delta_-}}\; .
\end{equation}
Because of the early freeze out, by the moment when $\gamma$ crosses horizon $- k \eta \sim 1$ the derivatives of $\sigma$ are suppressed by the factor $c_2^2$ with respect to its amplitude and field $s$ behaves like a pure growing mode up to the $\mathcal O(c_2^2)$ corrections. Its amplitude is given by { $s_- \simeq s(-k \eta = 1) \simeq  1/{ \sqrt{2 k c_2^\nu}}$, where we have used that at small mixing the mass of $\sigma$ is dominated by $m^2$: $3/2 - \Delta_- = \nu + \mathcal O ({\rho^2}/{H^2})$.}  The fact that the $\sigma$ mode at $- k \eta \sim 1$ is a purely growing mode allows us to use the expression~\eqref{eq:gmode} together with the value of $s_-$ in order to obtain the late time behaviour of $g$. 

{ Recalling the definition of the canonical tensor mode $\gamma_c^{(s)} = \mpl \, \gamma^{(s)} = (- H \eta) \, g$ we can write the late times power spectrum of $\gamma$ as 
\begin{equation}
P_\gamma(k, \eta) \equiv \langle \gamma_{ij, \,\vec k} \, \gamma_{ij, \,- \vec k}\rangle' = 2 \sum_{s = \pm 2}  \langle \gamma_{\vec k}^{(s)} \, \gamma_{- \vec k}^{(s)}\rangle' =  4 \frac {H^2}{\mpl^2} \, \eta^2 \, | g(\eta)|^2 \; .
\end{equation}
Using the solution~\eqref{eq:gmode} with the value of $s_-$ inferred above we obtain the power spectrum of $P_\gamma$ in the case when it is dominated by the mixing with $\sigma$:
\begin{equation} 
P_\gamma = \frac { 4 H^2}{\mpl^2}\frac 1{2 \, c_2^{2\nu} \, k^3} \left(\frac{6 \rho H}{m^2 + 2\rho^2}\right)^2 \left(1 - e^{-\frac{m^2 + 2\rho^2}{3H^2} N_k} \right)^2 \,, \hspace{1cm} c_2 \ll 1\,. 
\end{equation}
It is straightforward to check that this expression coincides with the perturbative results~\eqref{p_gamma_c} and~\eqref{eq:p_gamma} in the regimes $m^2 \gg \rho^2, \ H^2/N_k$ and $m^2 \ll \rho^2 \ll H^2/N_k$ respectively.
}

\subsection{Three-point correlation functions}
\label{bispectra}

Let us start computing the mixing $\langle\pi\s\rangle$, eq.~\eqref{sx_corr}{. We} can use the results of the Appendix \ref{power_spectra}, more in detail eq.~\eqref{int1}. 
For $c_0 \ll 1 $, we thus get
\bea\label{sp_corr}
\expect{\pi_{-\vec q} \, \s^{(0)}_{\vec q}}' \!\!\! &=& \!\!\! \frac{4}{\sqrt 3} \ \rho \, \mpl \ k^2 \ {\rm Re} \left\{ i \(\s^{(0)}_q (\eta_*) \, \pi_q (\eta_*)\) \int_{-\infty}^{\eta_*} \!\! d\eta \, a^2 \, {\sigma^{(0)}_q}^*(\eta) \, {\pi_q}^*(\eta) \right\} =  \nonumber \\
&\simeq& \!\!\! \frac{4}{\sqrt 3} \,  \rho \,  \frac{\mathcal N_\sigma}{\sqrt \epsilon \, H}  \ 
 {\rm Re}\left\{\s^{(0)}_{\vec q}(\eta_*) \ \pi _k (\eta_*) \ \[\mathcal I_1\(-\frac{1}{2},c_0\)+\mathcal I_1\(\frac{1}{2},c_0\)\]\right\} \nonumber \\
 &=& \frac{d_\pi(\nu)}{c_0^{2\nu}} \ \mpl \ \rho \ (-q \, \eta_*)^{3/2-\nu} \ P_\pi (k)\,.
\eea
Here $\eta_*$ is the conformal time at the end of inflation.
The coefficient $d_\pi(\nu)$ is,
\be\label{d_pi_nu}
d_\pi(\nu) = \frac{2^{2 \nu -3/2} (2 \nu -3) \Gamma \(\frac{1}{2} - \nu \) \Gamma(\nu) ^2 \[\cos \left(\frac{\pi  \nu
   }{2}\)-\sin \(\frac{\pi  \nu }{2}\)\]}{ \sqrt{3}\ \pi }\,.
\ee
Now we compute the squeezed $\langle\s^{(0)}\zeta\zeta\rangle$ correlation function.
For generic $\nu$, the integral inside eq.~\eqref{spp_bispectrum} cannot be computed analytically.
However, since we are interested only in the contribution in which $\sigma^{(0)}$ is soft, we can use the late time expansion of the $\sigma^{(0)}$ wavefunction, eq.~\eqref{late_sigma}. 
We then obtain
\be\label{spp_scalar_bispectrum_final}
\expect{\s^{(0)}_{\vec q} \, \pi_{\vec k} \, \pi_{- \vec k}}_{q \to 0} '= \frac{\sqrt 3 \, c(\nu)}{\epsilon \ \mpl} \ \frac{\tilde \rho}{H} \ (-k\,\eta_*)^{-\frac{3}{2}+\nu} \ P_{\sigma^{(0)}}(q) \, P_\pi (k) \((\hat q \cdot \hat k)^2-\frac{1}{3} \)\,,
\ee
where the coefficient $c(\nu)$ is given by
\be\label{c_nu}
c(\nu) \equiv {2^{-\frac{7}{2}+\nu} \ \(2\nu - 9 \) \ \cos\(\frac\pi4(1 + 2\nu)\)\ \Gamma\(\frac{5}{2}-\nu\)}\,,
\ee
and $\eta_*$ is the conformal time at the end of inflation.
One can check that the above result agrees with the exact calculation that can be performed if $\s$ is massless (i.e. $\nu = 3/2$). 
Notice, also, that $c(1/2)=0\,,$ then the leading term of $\langle\sigma^{(s)}\pi\pi\rangle$ vanishes if $\nu = 1/2$.
In fact, using the expression of the wavefunctions for $\nu=1/2$ , one can easily check that the leading contribution to $\langle\sigma^{(s)}\pi\pi\rangle$  is of order $\eta_*^2$ instead of being proportional to $\eta_*$, as one might naively expect.
\begin{center}
***
\end{center}

\noindent
The computation of the squeezed tensor bispectrum $\langle{\gamma\zeta\zeta}\rangle$ follows very closely the one of $\langle{\zeta\zeta\zeta}\rangle$. 
The mixing $\langle \gamma \s \rangle$, eq.~\eqref{sx_corr}, reads
\bea\label{sg_corr}
\expect{\gamma_{-\vec q}^{(s)} \,\s^{(s)}_{-\vec q} }' \!\!\! &=& \!\!\! 2 \, \mpl \ \rho \ {\rm Re} \left\{ i \(\s^{(s)}_q (\eta_*) \, \gamma^{(s)}_q (\eta_*)\) \int_{-\infty}^{\eta_*} \!\! d\eta \ a^3 \, \({\sigma^{(s)}_q}(\eta)\)^* \(\d_{\eta} \, {{{\gamma_q}^{(s)}}(\eta)}\)^* \right\} = \nonumber \\
&\simeq& 2 \, \mathcal N_\sigma \ \frac \rho H  \ {\rm Re}\left\{ \sigma^{(s)}(\eta_*) \ \gamma^{(s)}(0) \ \mathcal I_ 1 \(-\frac{1}{2}, c_2\)\right\}\,. \nonumber \\
&=&  \frac{d_\gamma(\nu)}{c_2^{2\nu}}\ \mpl \ \frac{\rho}{H}\ (-q\,\eta_*)^{\frac{3}{2}-\nu} \ P_\gamma(q)\,.
\eea
Again, the above expression assumes $c_2 \ll 1$. The coefficient $d_\gamma(\nu)$ is
\be \label{d_gamma_nu}
d_\gamma(\nu) = - \frac{2^{2 \nu -7/2} \ \Gamma \(\frac{1}{2}-\nu \) \Gamma (\nu )^2 \[\cos \left(\frac{\pi  \nu
   }{2}\)-\sin \(\frac{\pi  \nu }{2}\)\]}{ \pi } \,.
\ee
The 3-point function $\langle\s^{(\pm2)}\pi\pi\rangle$ in the squeezed limit is given by
\be
\expect{\s^{(\pm 2)}_{\vec q} \, \pi_{\vec k} \, \pi_{- \vec k}} '=  \frac{c(\nu)}{\epsilon \ \mpl} \ \frac{\tilde \rho}{H} \ (-k\,\eta_*)^{-\frac{3}{2}+\nu} \ P_{\sigma^{(2)}}(q) \, P_\pi (k) \epsilon_{ij}^{(\pm2)} \hat k_i \hat k_j \,,
\ee 
with $c(\nu)$ given in eq.~\eqref{c_nu}.

\end{document}